\documentclass[11pt,a4paper]{article} \pdfoutput=1 

\usepackage{jheppub}

\usepackage[T1]{fontenc}

\usepackage[latin9]{inputenc} \usepackage{esint}

\usepackage{hyperref}

\usepackage{booktabs}
\newcommand{\noun}[1]{{\tt #1}}

\newcommand{\POWHEG}{\noun{POWHEG}}
\newcommand{\POWHEGBOX}{\noun{POWHEG BOX}}

\newcommand{\ZNLOPSpPYTHIA}{\noun{Zj-MiNLO}}
\newcommand{\ZNNLOPSpPYTHIA}{\noun{NNLOPS}}

\newcommand{\MINLO}{\noun{MiNLO}}

\newcommand{\ZJMINLO}{\noun{Zj-MiNLO}}
\newcommand{\WJMINLO}{\noun{Wj-MiNLO}}
\newcommand{\VJMINLO}{\noun{Vj-MiNLO}}

\newcommand{\DYNNLO}{\noun{DYNNLO}}
\newcommand{\FEWZ}{\noun{FEWZ}} 
\newcommand{\NNLOPS}{\noun{NNLOPS}}
\newcommand{\NLOPS}{\noun{Zj-MiNLO}}
\newcommand{\PYTHIA}[1]{\noun{Pythia{#1}}}
\newcommand{\MCatNLO}{\noun{MC@NLO}} 

\newcommand{\DYQT}{\noun{DYqT}} 

\newcommand{\ZJ}{\noun{Zj}}
\newcommand{\WJ}{\noun{Wj}}
\newcommand{\VJ}{\noun{Vj}}
\newcommand{\Z}{\noun{Z}}

\newcommand{\JETVHETO}{\noun{JetVHeto}}
\newcommand{\FASTJET}{\noun{FastJet}}
\newcommand{\as}{\alpha_{\scriptscriptstyle \mathrm{S}}}
\newcommand{\Kr}{K_{\scriptscriptstyle \mathrm{R}}}
\newcommand{\Kf}{K_{\scriptscriptstyle \mathrm{F}}}
\newcommand{\mur}{\mu_{\scriptscriptstyle \mathrm{R}}}
\newcommand{\muf}{\mu_{\scriptscriptstyle \mathrm{F}}}
\newcommand{\pt}{p_{\scriptscriptstyle \mathrm{T}}}
\newcommand{\ptz}{p_{\scriptscriptstyle \mathrm{T,Z}}}
\newcommand{\ptw}{p_{\scriptscriptstyle \mathrm{T,W}}}
\newcommand{\ptjone}{p_{\scriptscriptstyle
    \mathrm{T,j_{1}}}}
\newcommand{\kt}{k_{\scriptscriptstyle \mathrm{T}}}
\newcommand{\mtw}{m_{\scriptscriptstyle \mathrm{T, W}}}

\newcommand{\hc}{\beta}
\newcommand{\hgam}{\gamma}

\usepackage[mathscr]{euscript}

\preprint{\\\\CERN-PH-TH/2014-129 \\OUTP-14-12P}

\title{{NNLOPS accurate Drell-Yan production}}

\author[a]{Alexander Karlberg,} \author[a]{Emanuele Re,}
\author[b]{Giulia Zanderighi\footnote{On leave from Rudolf Peierls Centre for Theoretical Physics,
University of Oxford, 1 Keble Road, UK}}

\affiliation[a]{Rudolf Peierls Centre for Theoretical Physics,
University of Oxford\\1 Keble Road, UK} \affiliation[b]{Theory
Division, CERN,\\CH--1211, Geneva 23, Switzerland}

\emailAdd{a.karlberg1@physics.ox.ac.uk}
\emailAdd{e.re1@physics.ox.ac.uk}
\emailAdd{g.zanderighi1@physics.ox.ac.uk}

\abstract{We present a next-to-next-to-leading order accurate
  description of Drell-Yan lepton pair production processes through
  $\gamma^*/Z$ or $W$ exchange that includes consistently parton
  shower effects.  Results are obtained by upgrading the vector-boson
  plus one jet NLO calculation in \POWHEG{} with the \MINLO{}
  procedure and by applying an appropriate reweighting procedure
  making use of the \DYNNLO{} program. We compare to existing data and
  to accurate resummed calculations.  }


\begin{document}
\maketitle \flushbottom

\section{Introduction}

During run I at the LHC, at 7 and then 8 TeV centre of mass energy,
both the ATLAS and CMS collaboration collected almost 30 fb$^{-1}$ of
data. Because of their very large cross-sections and the very small
systematic uncertainties, Drell-Yan production through $W$ and $Z$
exchange are standard candles at the LHC. Given the high-statistics
reached, kinematic distributions can be studied over many orders of
magnitude. With run II at 13-14 TeV even more $W$ and $Z$ events will
be available, in particular it will be possible to study distributions
over an even larger kinematic range. These distributions provide
important input to constrain parton distribution functions. For
example the rapidity distribution and the dilepton invariant mass of
both neutral and charged Drell-Yan data have been used recently to
constrain the photon content of the proton~\cite{Ball:2013hta}.

Higher-order corrections are indispensable for these studies, and
Drell-Yan production is to date the theoretically best described
process at the LHC. The cross-section is known through
next-to-next-to-leading order (NNLO) in QCD including the decay of $W$
and $Z$ bosons to
leptons~\cite{Anastasiou:2003yy,Melnikov:2006kv,Catani:2009sm}.
Two public codes exist (\DYNNLO{}~\cite{Catani:2009sm} and
\FEWZ{}~\cite{Gavin:2010az}) that implement QCD NNLO corrections to
the hadronic $W$ and $Z$ production.
Furthermore electroweak corrections have been the subject of
intensive
studies~\cite{Baur:1997wa,Baur:2001ze,Dittmaier:2001ay,CarloniCalame:2007cd,Arbuzov:2007db,Dittmaier:2009cr}.
Electroweak Sudakov effects become more important at large invariant
mass~\cite{Denner:2000jv,Kuhn:2001hz,CarloniCalame:2007cd,Campbell:2013qaa},
a region that was already interesting at run I and which will be
explored even more during the next LHC run. Version 3 of
\FEWZ{}~\cite{Li:2012wna} implements also NLO EW corrections and the
leading photon initiated processes.
Recently, a framework for the calculation of the mixed QCD-electroweak
${\cal O}(\alpha_s\alpha)$ corrections to Drell-Yan processes in the
resonance region has been developed~\cite{Dittmaier:2014qza}. The
impact of non-factorising (initial-final state) corrections is shown
to be very small. Factorisable ${\cal O}(\alpha_s\alpha_{ew})$
corrections in the pole approximation have been computed. While QCD
corrections are predominantly initial state corrections, and EW
contributions are predominantly final-state corrections, because of
kinematic effects there are sizable differences between the result of
ref.~\cite{Dittmaier:2014qza} and the naive product of NLO QCD and NLO
EW corrections.

While fixed-order predictions provide accurate results for inclusive
distributions, there is an obvious advantage in combining
state-of-the-art fixed-order perturbative calculations with parton
shower generators.
\POWHEG{}~\cite{Nason:2004rx,Alioli:2010xd} and
\MCatNLO~\cite{Frixione:2002ik} generators allow to keep NLO accuracy
and benefit from the exclusive description from a parton shower
(NLO+PS accuracy from now on). Recently, for $W$ and $Z$ production,
QCD and EW corrections have been implemented in the \POWHEGBOX{}
framework~\cite{Bernaciak:2012hj,Barze:2012tt,Barze':2013yca}.

In this paper, we take a step forward in further improving the
accuracy of the Drell-Yan description by detailing and releasing an
NNLO+PS accurate Monte Carlo description of this process. Our
implementation relies on the following inputs:
\begin{itemize}
\item Les Houches events for the $Z$+one jet or $W$+one jet
  process~\cite{Alioli:2010qp} (respectively \ZJ{} and \WJ{} from now
  on), as implemented in \POWHEG{}, upgraded with the improved
  \MINLO{} procedure of ref.~\cite{Hamilton:2012rf} in such a way that
  NLO accuracy is guaranteed for inclusive distributions without any
  jet cut;
\item NNLO accurate distributions, differential in the Born kinematics
  of the leptons, as obtained from \DYNNLO{}~\cite{Catani:2009sm};
\item a local reweighting procedure, described in all details in
  Sec.~\ref{sec:Theoretical-framework}.
\end{itemize}

Compared to the standard NLOPS implementation of $W$ and $Z$ in
\POWHEG{} we find a considerably reduced theoretical uncertainty for
inclusive distributions. This is expected since our predictions have
full NNLO accuracy.
Furthermore the 1-jet region is described at NLO accuracy in our
framework, even at small transverse momentum.

Throughout this work we pay particular attention to the issue of
assigning a theory uncertainty to our predictions. We describe our
procedure in Sec.~\ref{subsec:Estimating-uncertainties}. Unlike the
standard \POWHEG{} approach (without a separation between singular and
finite real contributions), that is known to underestimate the
theoretical uncertainty for the $W$, $Z$ or Higgs boson transverse
momentum~\cite{Nason:2012pr}, we believe that our uncertainties are
more reliable.  Compared to pure NNLO predictions, we find in general
a better description of observables sensitive to multiple emissions,
such as the boson transverse momentum, $\phi^*$, and jet-resolution
variables $d_{i}$ (which just vanish at NNLO starting from
$i=2$). This is both because of the underlying \MINLO{} procedure, and
because of the \POWHEG{} framework.

An important validation of our results comes from comparing with
precise calculations of specific observables. For the $Z$ transverse
momentum distribution we find a good agreement with
\DYQT{}~\cite{Bozzi:2010xn}.\footnote{We thank Giancarlo Ferrera and
  Massimiliano Grazzini for providing us with a preliminary version of
  this code.}
We also compare our results for the $\phi^*$ distribution to the
NLO+NNLL resummation of ref.~\cite{Banfi:2011dx}, and the jet-veto
efficiency with the NNLL+NNLO results of ref.~\cite{Banfi:2012jm}.

An implementation of Drell-Yan lepton pair production at NNLO accuracy
including parton shower effects using the UN$^2$LOPS algorithm in the
event generator \textsc{Sherpa} was recently presented in
ref.~\cite{Hoeche:2014aia}. An important difference of this
implementation, with respect to our approach, is that in
ref.~\cite{Hoeche:2014aia} the pure NNLO correction sits in the bin
where the $Z$ boson has zero transverse momentum. These events do not
undergo any parton showering, populating only the zero-$\pt$
bin. Thus, in the approach of ref.~\cite{Hoeche:2014aia}, the $Z$
boson transverse momentum is insensitive to the NNLO correction.

A general approach to matching NNLO computations with parton showers
was also proposed in ref.~\cite{Alioli:2013hqa}. The \MINLO{}
procedure that we use in this work was discussed by the authors
of ref.~\cite{Alioli:2013hqa} in the context of the general
formulation of their method. It will be interesting to see how our
approach effectively compares with ref.~\cite{Alioli:2013hqa} when the
corresponding implementations for reference processes like Drell-Yan
(and Higgs) production will be available.

The remainder of this paper is structured as follows.  In
Sec.~\ref{sec:Theoretical-framework} we present the theoretical
framework underlying the \NNLOPS{} method.  We start by recalling the
basic formulation in Sec.~\ref{subsec:NNLOPS} and then present few
refinements that we will adopt in the following
Sec.~\ref{subsec:Simple-variations}.  In Sec.~\ref{sec:Results} we
present a selection of results obtained with the \NNLOPS{} method.
First, in Sec.~\ref{subsec:Estimating-uncertainties} we recall our
prescription for the determination of theoretical uncertainties.  In
Sec.~\ref{subsec:practical} we give all details required to reproduce
the results of this work, and in Sec.~\ref{subsec:val} we present some
validation plots. Sec.~\ref{subsec:resum} presents comparisons of our
results with analytical resummations.  We then compare to a number of
existing LHC measurements in Sec.~\ref{sec:compdata}.  We conclude in
Sec.~\ref{sec:Conclusions}. 

\section{NNLO+PS via MiNLO}
\label{sec:Theoretical-framework}
We extend here the method that some of us used recently to achieve
NNLO+PS accuracy in the case of Higgs
production~\cite{Hamilton:2013fea} to the case of a generic particle
that undergoes a two-body decay. For concreteness, we will discuss
explicitly the case of a $V$ boson ($V$=$W$,$Z$) decaying to
leptons. The method relies on the main result that the \VJ{}
generator, improved with the \MINLO{} procedure of
ref.~\cite{Hamilton:2012rf} is NLO accurate both for $V$+1 jet
distributions and for inclusive distributions, without any jet cut. In
this section we describe how this result can be used to reach NNLO+PS
accuracy for $V$ production.

\subsection{The method}
\label{subsec:NNLOPS}
We denote by $d\sigma^{{\scriptscriptstyle \mathrm{MINLO}}}/d\Phi$ the
cross-section obtained from the \VJMINLO{} event generator, fully
differential in the final state phase space, $\Phi$, at the level of
the hardest emission events, i.e. before parton shower.
Because of the properties of \MINLO{}, upon integration, this
distribution reproduces the next-to-leading order accurate,
$\mathcal{O}\left(\as\right)$, leptonic distributions inclusive in all
QCD radiation.
After integration over all QCD radiation only the leptonic system is
left. This can be characterised by three independent variables, for
instance one can choose the invariant mass of the lepton pair, $m_{\rm
  ll}$, the rapidity of the boson before decay $y_V$, and the angle of
the negatively (or positively) charged lepton with respect to the
beam, $\theta_l$.  Other choices are possible, and later on we will
discuss the impact of different choices.  In the following we will use
$\Phi_B$ to denote the ensemble of these three Born variables.

We denote schematically the fixed order NNLO cross-section
differential over $\Phi_B$ by $d\sigma^{{\scriptscriptstyle
    \mathrm{NNLO}}}/d\Phi_B$ and the cross-section obtained from
\VJMINLO{} by $d\sigma^{{\scriptscriptstyle \mathrm{MINLO}}}/d\Phi_B$.
Since these distributions are identical up to terms of
$\mathcal{O}\left(\alpha_{{\scriptscriptstyle \mathrm{S}}}\right)$
terms, their ratio is equal to one up to
$\mathcal{O}\left(\alpha_{{\scriptscriptstyle \mathrm{S}}}^{2}\right)$
terms:
\begin{equation} 
  \label{eq:W}
  \mathcal{W}(\Phi_B) = \frac{\frac{d\sigma^{\scriptscriptstyle
        \mathrm{NNLO}}}{d\Phi_B}}{
    \frac{d\sigma^{\scriptscriptstyle \mathrm{MINLO}}}{d\Phi_B}} =
  \frac{c_{0}+c_{1}\alpha_{{\scriptscriptstyle
        \mathrm{S}}}+c_{2}\alpha_{{\scriptscriptstyle
        \mathrm{S}}}^{2}\phantom{+\ldots}}{c_{0}+c_{1}\alpha_{{\scriptscriptstyle
        \mathrm{S}}}+c_{2}^{\prime}\alpha_{{\scriptscriptstyle
        \mathrm{S}}}^{2}+\ldots} =
  1+\frac{c_{2}-c_{2}^{\prime}}{c_{0}}\,\alpha_{{\scriptscriptstyle
      \mathrm{S}}}^{2}+\ldots\,,
\end{equation} 
where the $c_{i}$ are $\mathcal{O}\left(1\right)$ coefficients.  Since
the \VJMINLO{} generator reproduces the inclusive fixed-order result
up to and including NLO terms, the NLO accuracy of the cross-section
in the presence of one jet (that starts at order $\as$) is maintained
if the cross-section is reweighed by the factor in eq.~\eqref{eq:W}.
This is follows from the simple fact that the reweighting factor
combined with this cross-section yields spurious terms of order
$\as^{3}$ and higher.

It is also obvious that by reweighting \VJMINLO{} distributions with
this ratio, any of the three $\Phi_B$ distributions acquires NNLO
accuracy, and in fact coincides with the NNLO distribution.
We will now argue that the \VJMINLO{} generator reweighed with the
procedure of eq.~\eqref{eq:W} maintains the original NNLO accuracy of
the fixed-order program, used to obtain the
$d\sigma^{\scriptscriptstyle \mathrm{NNLO}}/d\Phi_B$ distribution, for
all observables. The proof trivially extends the proof given in
ref.~\cite{Hamilton:2013fea} by replacing the Higgs rapidity with the
chosen set of three variables $\Phi_B$ associated to the Born phase
space. We find it useful to recall the way the proof works.

The claim is based on the following theorem, that we will prove in the
following.
\begin{quote} 
\textit{A parton level V boson production generator that is accurate
  at ${\cal O}(\alpha_{{\scriptscriptstyle \mathrm{S}}}^{2})$ for all
  IR safe observables that vanish with the transverse momenta of all
  light partons, and that also reaches ${\cal
    O}(\alpha_{{\scriptscriptstyle \mathrm{S}}}^{2})$ accuracy for the
  three Born variables $\Phi_B$ achieves the same level of precision
  for all IR safe observables, i.e. it is fully NNLO accurate.}
\end{quote} 
To this end, we consider a generic observable $F$, including cuts,
that is an infrared safe function of the final state kinematics. $F$
could be for instance be a bin of some distribution. Its value will be
given by
\begin{equation} 
  \label{eq:NNLOPS-F-i}
  \langle F\rangle=\int d\Phi\,\frac{d\sigma}{d\Phi}\,
  F(\Phi)
\end{equation} 
with a sum over final state multiplicities being implicit in the phase
space integral. Infrared safety ensures that $F$ has a smooth limit
when the transverse momenta of the light partons vanish.

Since the Born kinematics is fully specified by the Born kinetic
variables $\Phi_B$, such a limit may only depend upon the value of
Born kinematic variables $\Phi_B$. We generically denote such a limit
by $F_{\Phi_B}$.  The value of $\langle F\rangle$ can be considered as
the sum of two terms: $\left\langle F-F_{\Phi_B}\right\rangle +
\left\langle F_{\phi_B}\right\rangle $.  Since $F-F_{\Phi_B}$ tends to
zero with the transverse momenta of all the light partons, by
hypothesis its value is given with ${\cal
  O}(\alpha_{{\scriptscriptstyle \mathrm{S}}}^{2})$ accuracy by the
parton level generator.  On the other hand,
\begin{equation} \left\langle F_{\Phi_B}\right\rangle =\int d
\Phi_B^{\prime}\,\frac{d\sigma}{d \Phi_B^{\prime}}\,
F_{\Phi_B}\left(\Phi_B^{\prime}\right)\label{eq:NNLOPS-F-ii}\,,
\end{equation} which is also exact at the ${\cal
O}(\alpha_{{\scriptscriptstyle \mathrm{S}}}^{2})$ level by
hypothesis. Thus, $\langle F\rangle=\left\langle
F-F_{\Phi_B}\right\rangle +\left\langle F_{\Phi_B}\right\rangle $ is
accurate at the ${\cal O}(\alpha_{{\scriptscriptstyle
\mathrm{S}}}^{2})$ level. 

The \VJMINLO{} parton level generator (in fact even just \VJ) fulfills
the first condition of the theorem since it predicts any IR safe
observable that vanish when the transverse momentum of the light
partons vanish with $\mathcal{O}(\as^{2})$ accuracy.  The second
hypothesis of the theorem, regarding NNLO accuracy of the Born
variables, is simply realised by augmenting the \VJMINLO{} generator
by the reweighting procedure described above. We note that \MINLO{} is
crucial to preserve the first property after rescaling.  The proof of
$\mathcal{O}(\as^{2})$ accuracy for these observables thus corresponds
to the general proof of NLO accuracy of the \POWHEG{} procedure, given
in refs.~\cite{Nason:2004rx,Frixione:2007vw}.

Observe, also, that for observables of the type $\left\langle
  F-F_{\Phi_B}\right\rangle$, adding the full shower development does
not alter the $\mathcal{O}(\as^{2})$ accuracy of the algorithm, for
the same reasons as in the case of the regular \POWHEG{} method.  The
only remaining worry one can have concerns the possibility that the
inclusive Born variable distributions are modified by the parton
shower evolution at the level of $\mathcal{O}(\as^{2})$
terms. However, our algorithm already controls the two hardest
emissions with the required $\as^{2}$ accuracy. A further emission
from the shower is thus bound to lead to corrections of higher order
in $\as$. This concludes then our proof.

\subsection{Variant schemes}
\label{subsec:Simple-variations}
As discussed in detail in ref.~\cite{Hamilton:2013fea} the reweighting
in eq.~\eqref{eq:W} treats low and high transverse momentum
distributions equally, i.e. it spans the virtual correction over the
full transverse momentum range considered. On the other hand, if one
considers the transverse momentum distribution of the vector boson $V$
or of the leading jet, in the high transverse-momentum region
\VJMINLO{} and the NNLO calculation have formally the same (NLO)
accuracy. Hence, while it is not wrong to do so, there is no need to
``correct'' the \VJMINLO{} result in that region. It is therefore
natural to introduce a function that determines how the two-loop
virtual correction is distributed over the whole transverse momentum
region,
\begin{equation} h(\pt)=\frac{(\hc\, m_{{\scriptscriptstyle
        \mathrm{V}}})^{\hgam}}{(\hc\, m_{{\scriptscriptstyle
        \mathrm{V}}})^{\hgam}+\pt^{\hgam}},\label{eq:hfact}
\end{equation} where $\hc$ and $\hgam$ are constant parameters, and $m_{{\scriptscriptstyle
    \mathrm{V}}}$ is the mass of the vector boson. This function has the
property that when $\pt$ goes to zero it tends to one, while
when $\pt$ becomes very large, it vanishes.

The function $h(\pt)$ can therefore be used to split the cross-section
according to
\begin{eqnarray} 
d\sigma_{\phantom{0}} & = &
d\sigma_{A}+d\sigma_{B}\,,\label{eq:NNLOPS-dsig-eq-dsig0-plus-dsig1}
\\ d\sigma_{A} & = & d\sigma\, h\left(\pt\right)\,,\\ d\sigma_{B} & =
& d\sigma\,\left(1-h\left(\pt\right)\right)\,. \label{eq:NNLOPS-dsig1}
\end{eqnarray} 
We then reweight the \VJMINLO{} prediction using the following factor

\begin{eqnarray} 
  \mathcal{W}\left(\Phi_B,\, p_{{\scriptscriptstyle
      \mathrm{T}}}\right)&=&h\left(\pt\right)\,\frac{\smallint
    d\sigma^{{\scriptscriptstyle
        \mathrm{NNLO\phantom{i}}}}\,\delta\left(\Phi_B-\Phi_B\left(\Phi\right)\right)-\smallint
    d\sigma_{B}^{{\scriptscriptstyle
        \mathrm{MINLO}}}\,\delta\left(\Phi_B-\Phi_B\left(\Phi\right)\right)}{\smallint
    d\sigma_{A}^{{\scriptscriptstyle
        \mathrm{MINLO}}}\,\delta\left(\Phi_B-\Phi_B\left(\Phi\right)\right)}\nonumber\\ &+&\left(1-h\left(\pt\right)\right)\,,\label{eq:NNLOPS-overall-rwgt-factor-1}
\end{eqnarray} 
which preserves the exact value of the NNLO cross-section
\begin{eqnarray}
  \left(\frac{d\sigma}{d\Phi_B}\right)^{{\scriptscriptstyle
      \mathrm{NNLOPS}}} & = &
  \left(\frac{d\sigma}{d\Phi_B}\right)^{{\scriptscriptstyle
      \mathrm{NNLO}}}\,.\label{eq:NNLOPS-NNLOPS-eq-NNLO_0+MINLO_1-1}
\end{eqnarray}

The proof of NNLO accuracy for this rescaling scheme is completely
analogous to the proof given above, so we omit it here.

\section{Phenomenological analysis}
\label{sec:Results}
In this section we will present some validation plots and comparisons
with resummed predictions. Before doing so, we will define the
procedure we use to estimate theoretical uncertainties and we will
give all details about our practical implementation of the formulae in
Sec.~\ref{sec:Theoretical-framework}.

\subsection{Estimating uncertainties}
\label{subsec:Estimating-uncertainties}

We detail here the method that we use to estimate the uncertainties in
our NNLO event generator.
As is standard, the uncertainties in the \VJMINLO{} generator are
obtained by varying by a factor 2 up and down independently all
renormalisation scales appearing in the \MINLO{} procedure by $\Kr$
(simultaneously) and the factorisation scale by $\Kf$, keeping $1/2
\le \Kr/\Kf \le 2$. This leads to 7 different scale choices given by
\begin{equation} (K_{{\scriptscriptstyle
      \mathrm{R}}},K_{{\scriptscriptstyle
      \mathrm{F}}})=(0.5,0.5),(1,0.5),(0.5,1),(1,1),(2,1),(1,2),(2,2)\,.\label{eq:Errors-KR-KF-list}
\end{equation} 
We will consider the variation in our results induced by the above
procedure. The seven scale variation combinations have been obtained
by using the reweighting feature of the \POWHEGBOX{}.

For the pure NNLO results, the uncertainty band is the envelope of the
7-scale variation obtained by varying the renormalisation and
factorization scale by a factor 2 around the central value $M_V$
keeping $1/2 \le \Kr/\Kf \le 2$.\footnote{Since in the case of $Z$
  production a symmetric scale variation with $\Kr=\Kf$ (3-point
  variation) gives very similar results to the 7-scale variation, in
  the case of $W$ production we used only a 3-point scale variation at
  NNLO.}

For the NNLOPS results, we have first generated a single of \ZJMINLO{}
event file with all the weights needed to compute the integrals
$d\sigma_{A/B}^{\scriptscriptstyle \mathrm{MINLO}}/d\Phi_B$, in
eq.~(\ref{eq:NNLOPS-overall-rwgt-factor-1}), for all 7 scale choices.
The differential cross-section $d\sigma^{\scriptscriptstyle
  \mathrm{NNLO}}/d\Phi$ was tabulated for each of the three scale
variation points corresponding to $\Kr'=\Kf'$. 
The analysis is then performed by processing the
\MINLO{} event for given values of $(\Kr,\Kf)$, and multiplying its
weight with the factor
\begin{equation} h\left(\pt\right)\times\,\frac{\smallint
d\sigma_{(\Kr',\Kf')}^{{\scriptscriptstyle
\mathrm{NNLO\phantom{i}}}}\,\delta\left(\Phi_B-\Phi_B(\Phi)\right)-\smallint
d\sigma_{B,(\Kr,\Kf)}^{{\scriptscriptstyle
\mathrm{MINLO}}}\,\delta\left(\Phi_B-\Phi_B(\Phi)\right)}{\smallint
d\sigma_{A,(\Kr,\Kf)}^{{\scriptscriptstyle
\mathrm{MINLO}}}\,\delta\left(\Phi_B-\Phi_B(\Phi)\right)}+\left(1-h\left(\pt\right)\right)\,.\label{eq:master}
\end{equation} 
The central value is obtained by setting $(\Kr,\Kf)$ and $(\Kr',\Kf')$
equal to one, while to obtain the uncertainty band we apply this
formula for all the seven $(K_{{\scriptscriptstyle
    \mathrm{R}}},K_{{\scriptscriptstyle \mathrm{F}}})$ and three
$(K_{{\scriptscriptstyle \mathrm{R}}}^{\prime},K_{{\scriptscriptstyle
    \mathrm{F}}}^{\prime})$ choices. This yields 21 variations at
\NNLOPS{} level.\footnote{We have checked that performing instead a
  49-point variation, i.e. doing a 7-point scale variation both
  for \MINLO{} and the NNLO result, does not lead to appreciable
  differences in any of the distributions.}

The reasoning behind varying scales in the NNLO and \ZJMINLO{} results
independently is that we regard uncertainties in the overall
normalisation of distributions, as being independent of the respective
uncertainties in the shapes. This is consistent with the recently
introduced efficiency method~\cite{Banfi:2012yh}, used for estimating
errors on cross-sections in the presence of cuts.

\subsection{Practical implementation}
\label{subsec:practical}
There is some degree of freedom in the way the reweighting procedure
described above is carried out in practice. 
In particular, one is free to choose the three Born variables with
respect to which one performs the reweighting, as well as the form of
the damping factor $h$ in eq.~\eqref{eq:hfact}. Our choice of the Born
variables is driven by the fact that one wants to populate all bins in
the three-dimensional histograms sufficiently well. To produce the
results presented in the following we used the rapidity of the $Z$
boson, $y_Z$, a variable directly related to the dilepton invariant
mass, $a_{\rm mll} = {\rm \arctan}((m_{\rm ll}^2 - M_Z^2)/(M_Z
\Gamma_Z))$ and $\theta^*_l$, where the latter is the angle between
the beam and a charged lepton in the frame where the boson has no
longitudinal momentum.\footnote{For the $W$ boson we always pick the
  charged lepton, for the $Z$ boson we always pick the negatively
  charged one.}

In the following we will use the values of $\hc = 1$ and $\hgam = 2$
in eq.~\eqref{eq:hfact}.  The choice for $\hc$ is motivated by the
fact that typical resummation scale for Drell-Yan production is set to
the boson mass. The second choice is partly driven by the fact that in
the case of Higgs production, with this choice the \NNLOPS{} Monte
Carlo agrees well with the Higgs transverse momentum and leading jet
NNLL+NNLO resumed results.
We stress however that while $\hc$ and $\hgam$ are arbitrary, the
dependence on $\hc$ and $\hgam$ is formally ${\cal O}(\as^3)$ (or
exactly zero in the case of inclusive quantities).  $\pt$ in
eq.~\eqref{eq:hfact} denotes the transverse momentum that is used to
decide how to distribute the virtual corrections. One could for
instance choose the $V$ transverse momentum, the leading jet
transverse momentum, or the total transverse momentum of the event. In
the following, we will adopt the choice of
ref.~\cite{Hamilton:2013fea}, namely to use the leading jet transverse
momentum when clustering events according to the inclusive
$\kt$-algorithm with $R=0.7$~\cite{Catani:1993hr}.  This choice
ensures that $h$ goes to one when no radiation is present, since in
that case the leading jet transverse momentum vanishes. On the
contrary, the $V$ transverse momentum can vanish also in the presence
of radiation.

Finally we note that for the reweighting we used 25 bins per variable,
meaning that our 3-dimensional distributions involve a total of
$15625$ equal bins.
Results presented in the following are based on generating 20 Million
events.
Even if we perform high-statistics runs, there might be bins that are
not well-populated. This can give rise to a rescaling factor that
is unphysically large just because too few events ended up in one
bin, and this in turn can give rise to spikes in kinematical
distributions. To avoid these occurrences, instead of using the local
reweighting factor, we use the global $K_{\rm NNLO}/K_{\rm NLO}$
factor to perform the reweighting whenever the local reweighting
factor (which is formally $1+{\cal O}(\alpha_s)^2$) exceeds 5. This
happens rarely, in about 0.3\% of the points. We checked that this
procedure has no visible systematic effect on distributions, other
than that of removing unphysical spikes.  Other workarounds could of
course also be adopted.

Before showing validation plots, we list here the settings used for
the results obtained in this paper.  We used the code
\DYNNLO{}~\cite{Catani:2009sm} to obtain NNLO
predictions.\footnote{Even with high-statistics runs with
  \FEWZ{}~\cite{Gavin:2010az}, we did not obtain high-quality triple
  differential distributions, as required here.}
Throughout this work we consider
the MSTW2008NNLO parton distribution functions~\cite{Martin:2009iq}
and set $M_Z = 91.1876$ GeV, $\Gamma_Z = 2.49595$ GeV, $M_W = 80.398 $
GeV and $\Gamma_W = 2.008872$ GeV. We choose to use $\alpha_{\rm em} =
1/128.94$ and $\sin^2\theta_W = 0.22264585$.
Jets have been constructed using
\FASTJET{}~\cite{Cacciari:2005hq,Cacciari:2011ma}. To compute the
$h(\pt)$ factor we use the $\kt$
algorithm~\cite{Catani:1993hr,Ellis:1993tq} with $R=0.7$.
To shower partonic events, we have used both
\PYTHIA{8}~\cite{Sjostrand:2007gs} (version 8.185) with the ``Monash
2013''~\cite{Skands:2014pea} tune and \PYTHIA{6} (version 6.4.28) with
the MSTW2008LO variation of the ``Perugia'' tune (``Perugia
P12-M8LO'', tune 378). 
We have generated $Z$ and $W$ events with decays into electrons and
positrons, and always switched off QED radiation off leptons and quarks
in the showering stage.
Moreover, in the following, in order to define
the leptons from the boson decays we will always use the Monte Carlo
truth, \emph{i.e.}~we disregard complications due to the fact that
there might be other leptons in the event.

To obtain the results shown in the following, we have switched on the
``doublefsr'' option introduced in ref.~\cite{Nason:2013uba} and used
a standard driver for \PYTHIA{6} and the \PYTHIA{8} driver suggested
by the \PYTHIA{} authors for showering events generated with
\POWHEG{}.  Although we have also explored the effect of using the
alternative prescription first introduced in section 4 of
ref.~\cite{Nason:2013uba} to compute the scale used by the parton
shower to veto hard emissions (which is also available as an option in
the \PYTHIA{8} driver),~\footnote{A factor 2 is missing in eq.~(4) of
  ref.~\cite{Nason:2013uba}, but not in the practical implementation.}
the plots shown throughout the paper have been obtained keeping the
veto scale equal to the default \POWHEG{} prescription, both for
\PYTHIA{6} and \PYTHIA{8}.

\subsection{Validation plots}
\label{subsec:val}

In this section we consider the case of inclusive $Z\to e^+e^-$
production at 14 TeV, with no cuts on the final state other than
requiring the dilepton invariant mass to be in the range $66 \mbox{
  GeV} \le m_{\rm ll} \le 116 \mbox{ GeV}$. In order to validate our
results, we will show comparisons to NNLO predictions and to
\MINLO{}-improved \ZJ{}-\POWHEG{} results (henceforth denoted as
\ZNLOPSpPYTHIA{}). Since we compare to \DYNNLO{}, for the
\ZNLOPSpPYTHIA{} and \NNLOPS{} results it is useful to consider here
pure parton level results, before hadronization (with underlying event
and multiparton interaction switched off). Unless otherwise stated,
for \DYNNLO{} we have set $\muf=\mur=M_Z$, and the associated
uncertainty bands are obtained from a 7-points scale variation
envelope. Also unless stated otherwise, we will shower events with
\PYTHIA{8}.

\begin{figure}[htb]
  \begin{centering}
    \includegraphics[clip,width=0.9\textwidth]{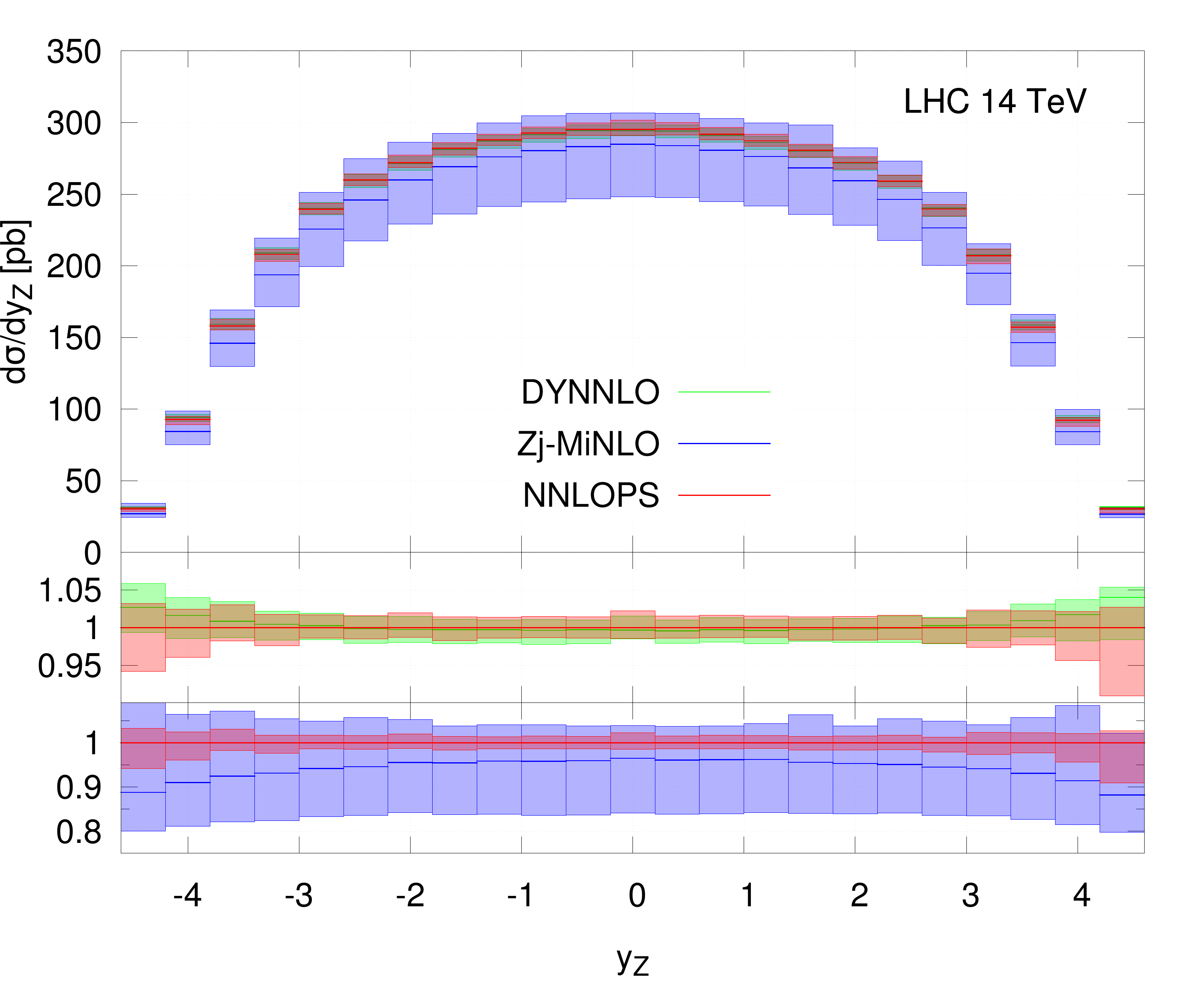}
    \par\end{centering}
  \caption{Comparison of the \ZNNLOPSpPYTHIA{} (red), \ZNLOPSpPYTHIA{}
    (blue) and \DYNNLO{} (green) results for the $Z$ boson fully
    inclusive rapidity distribution at the LHC running at 14 TeV.  The
    \DYNNLO{} central scale is $\muf=\mur=M_Z$, and its error band is
    the 7-point scale variation envelope.  For \ZNNLOPSpPYTHIA{} and
    \ZNLOPSpPYTHIA{} the procedure to define the scale uncertainty is
    described in detail in Sec.~\ref{subsec:Estimating-uncertainties}.
    The two lower panels show the ratio of \DYNNLO{} and
    \ZNLOPSpPYTHIA{} predictions with respect to \ZNNLOPSpPYTHIA{}
    obtained with its central scale choice.}
    \label{fig:val-yz}
\end{figure}

In Fig.~\ref{fig:val-yz} 
we show the $Z$ boson rapidity distribution $y_Z$ as predicted at NNLO
(green), with \ZNLOPSpPYTHIA{} (blue), and at \ZNNLOPSpPYTHIA{}
(red). As expected \ZNNLOPSpPYTHIA{} agrees very well with \DYNNLO{}
over the whole rapidity range, both for the central value and the
uncertainty band, defined as detailed in
Sec.~\ref{subsec:Estimating-uncertainties}.  We also note that as
expected the uncertainty band of the \ZNNLOPSpPYTHIA{} result is
considerably reduced compared to the one of \ZNLOPSpPYTHIA{}, which is
NLO accurate. In the central region the uncertainty decreases from
about ($+$5:$-$15) \% to about ($+$2:$-$2) \%. We finally note that
because of the positive NNLO corrections, the central value of
\ZNNLOPSpPYTHIA{} lies about 5 \% above the one of \ZNLOPSpPYTHIA{},
while no considerably difference in shape is observed in the central
rapidity region. Moderate but slightly more pronounced shape
differences can be seen at large rapidity.

\begin{figure}[htb]
  \begin{centering}
    \includegraphics[clip,width=0.49\textwidth]{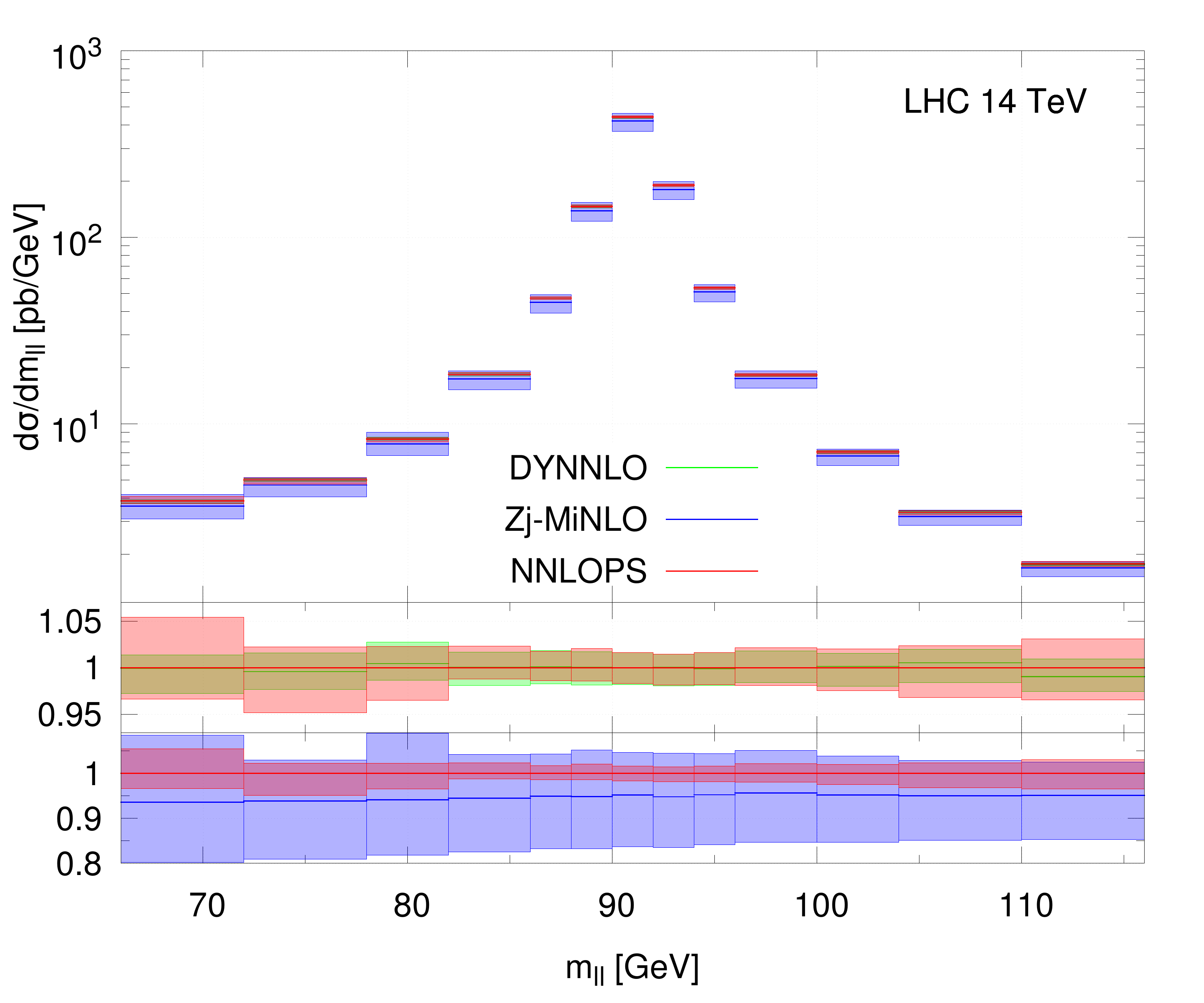}
    \hfill{}\includegraphics[clip,width=0.49\textwidth]{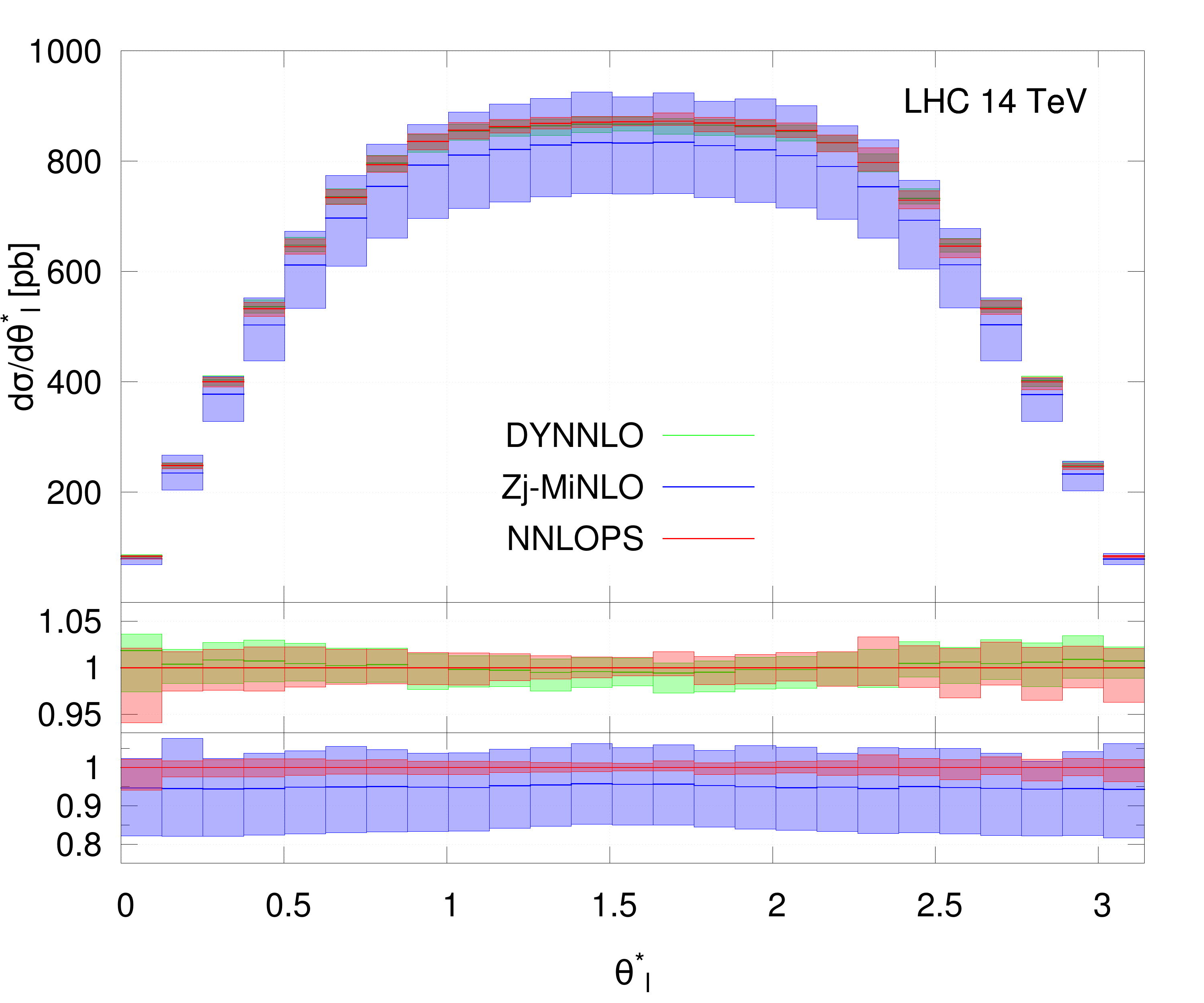}
    \par\end{centering}
    \caption{As for Fig.~\ref{fig:val-yz} but for $m_{\rm ll}$ (left) and
      $\theta_l^*$ (right).}
    \label{fig:val-mll-theta}
\end{figure}

We proceed by examining the other two distributions that have been used
in the reweighting procedure, namely $a_{\rm m_{\rm ll}}$ and
$\theta^*_l$. Instead of showing $a_{\rm m_{\rm ll}}$ we plot in Fig~\ref{fig:val-mll-theta}
the invariant mass of the dilepton system $m_{\rm ll}$ which is
directly related to $a_{\rm m_{\rm ll}}$. We notice that the same
features observed above hold: \ZNNLOPSpPYTHIA{} agrees well with
\DYNNLO{}, it tends to be about 5$-$10\% higher than \ZNLOPSpPYTHIA{}
and the uncertainty band is reduced by about a factor 4. No sizable
difference in shape is observed in these two distributions, when
comparing \ZNLOPSpPYTHIA{} and \ZNNLOPSpPYTHIA{}.

\begin{figure}[htb]
  \begin{centering}
    \includegraphics[clip,width=0.49\textwidth]{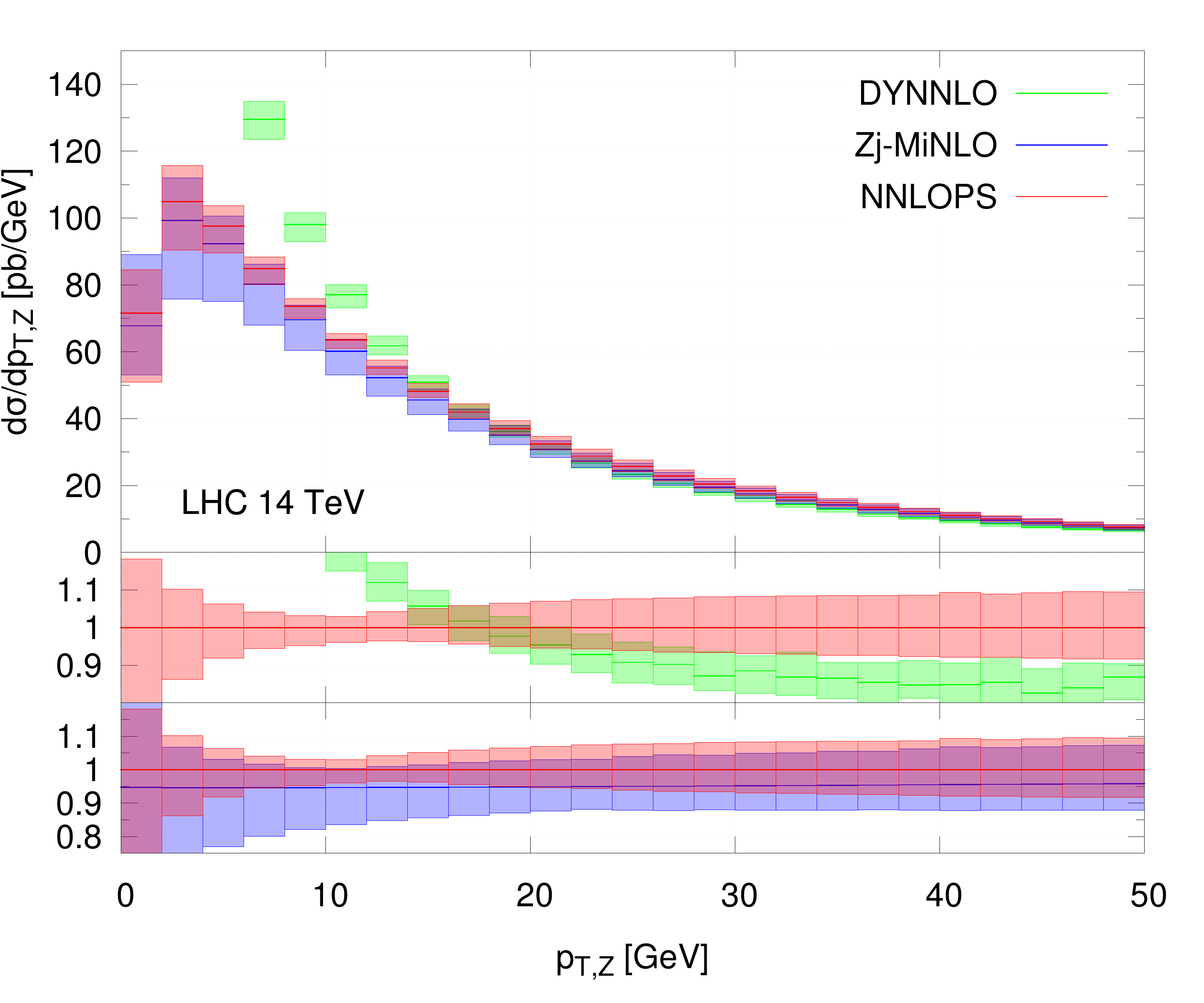}
    \hfill{}\includegraphics[clip,width=0.49\textwidth]{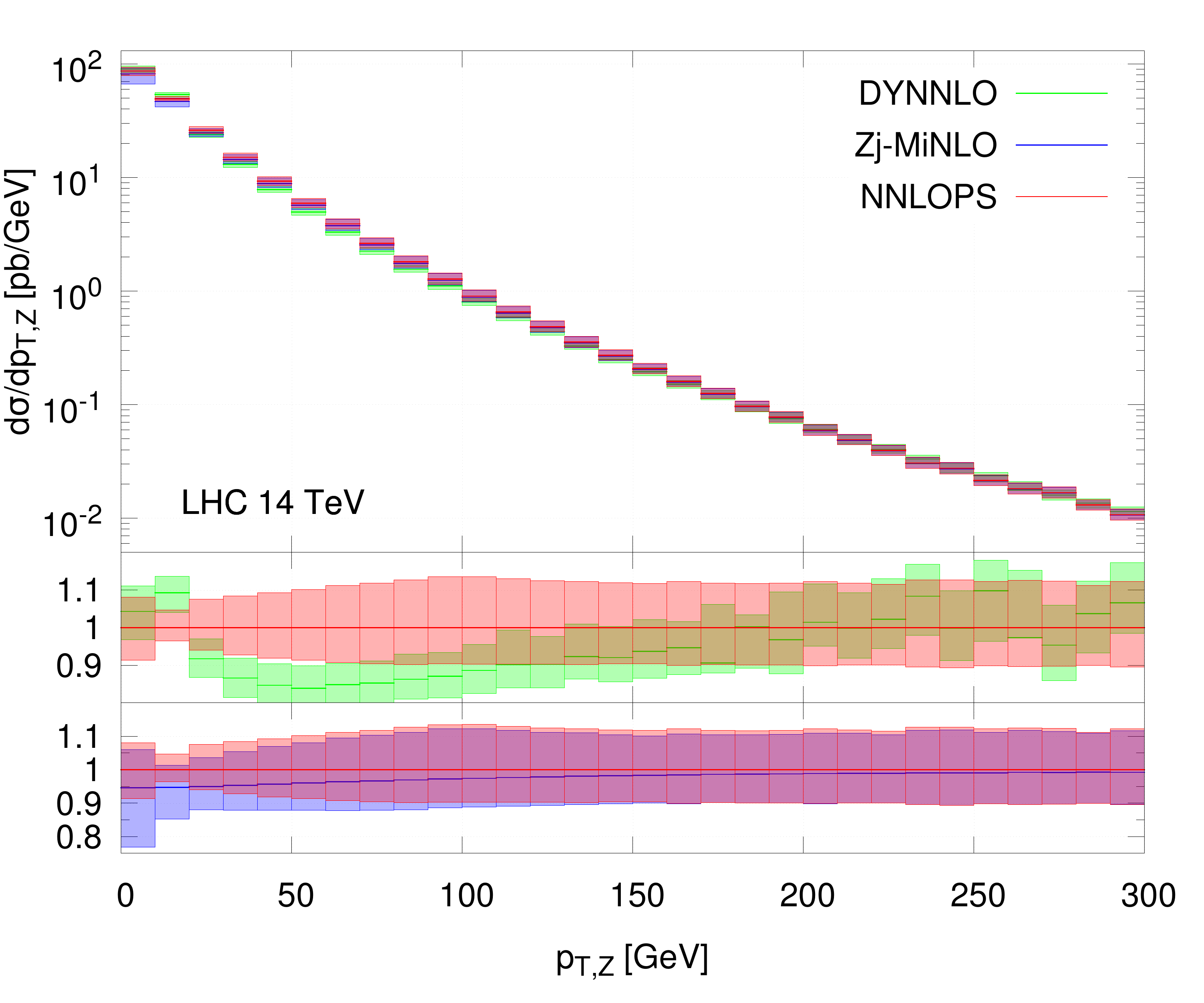}
    \par\end{centering}
    \caption{As for Fig.~\ref{fig:val-yz} but for the $Z$-boson transverse
      momentum for two different ranges in $\ptz$.}
    \label{fig:val-ptz}
\end{figure}

We now show in Fig.~\ref{fig:val-ptz} 
the $Z$ boson transverse momentum in two different ranges. At finite
values of $\ptz$ this quantity is described at NLO accuracy only by
all three codes, \DYNNLO{}, \ZNNLOPSpPYTHIA{} and \ZNLOPSpPYTHIA{}.
In fact, at higher value of $\ptz$ the bands of \ZNNLOPSpPYTHIA{} and
\ZNLOPSpPYTHIA{} overlap and are very similar in size.

At small transverse momenta, while \DYNNLO{} diverges,
\ZNNLOPSpPYTHIA{} remains finite because of the Sudakov damping. The
difference in shape observed between \DYNNLO{} and \NNLOPS{} at finite
$\ptz$ has to do with the fact that in that region the fixed-order
calculation has to compensate for the divergent behaviour at small
$\ptz$.  We also note that the uncertainty band in \DYNNLO{} is far
too small when approaching the divergence at $\ptz = 0$.  The
uncertainty band of \ZNNLOPSpPYTHIA{} instead tends to increase at
very low transverse momenta, reflecting the fact that one is
approaching a non-perturbative region. One can however also note that
the uncertainty tends to shrink at about $\ptz = 10$ GeV. We have
checked that this is not an artifact due to having used a 21-point
scale variation as opposed to a 49-point one.
We attribute this feature to the fact that the uncertainty band of the
fixed-order result shrinks in this region. This is true both for the
3-point and the 7-point scale variation in the fixed order, although
in the latter case this effect is slightly less pronounced.  We also
observe that our \ZNLOPSpPYTHIA{} result that uses the \MINLO{} scale
prescription does not show this feature. When we upgrade
\ZNLOPSpPYTHIA{} to NNLO accuracy, we necessarily inherit this feature
from the NNLO results we are using as input. 
It is also worth mentioning that this feature has been already
observed in several studies where an analytic resummation matched with
fixed-order results was performed for this
observable~\cite{Bozzi:2010xn,Becher:2010tm,Banfi:2012du}.

We end our discussion on neutral Drell-Yan production by looking
briefly into the effects of including non-perturbative contributions
in the \NNLOPS{} simulation, by turning on hadronization, underlying
event and multiparton interaction (MPI). In particular, given the
small perturbative uncertainties found with $\ptz$, it is interesting
to see how much non-perturbative corrections affect the 5-15 GeV
region.

\begin{figure}[htb]
  \begin{centering}
    \includegraphics[clip,width=0.49\textwidth]{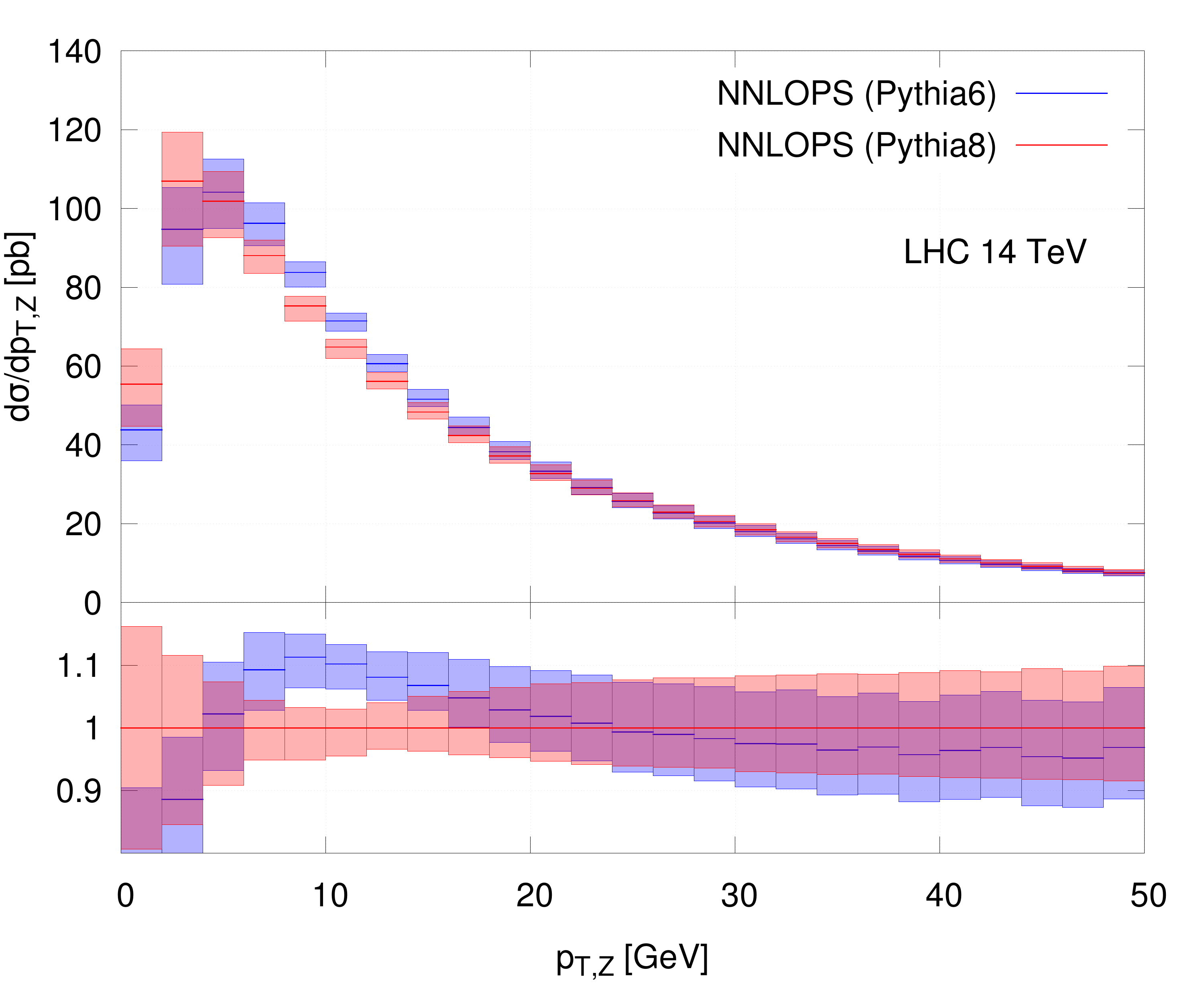}
    \hfill{}\includegraphics[clip,width=0.49\textwidth]{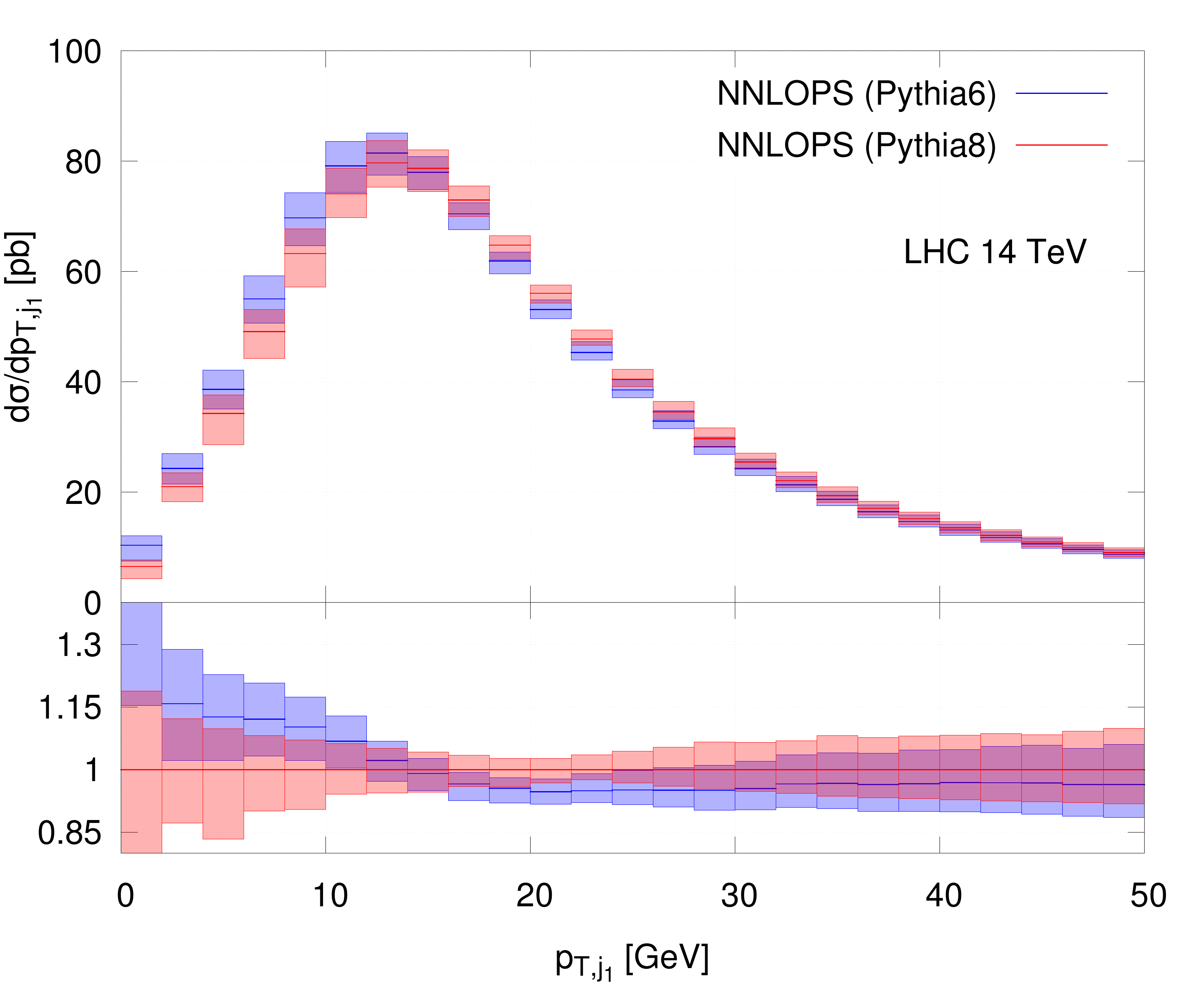}
    \par\end{centering}
  \caption{\NNLOPS{} predictions for the transverse momentum of the
    reconstructed $Z$ boson (left) and leading jet (right) obtained
    using \PYTHIA{6} (blue) and \PYTHIA{8} (red). Non-perturbative
    effects have been included here, by using the tunes mentioned in
    section.~\ref{subsec:practical}.}
    \label{fig:par-ptz}
\end{figure}

In Fig.~\ref{fig:par-ptz}  
we show $\ptz$ (left) and the leading-jet transverse momentum, defined
according to the anti-$\kt$~\cite{Cacciari:2008gp} algorithm with
$R=0.7$ (right), after all non-perturbative stages are included, with
\PYTHIA{6} (blue) and \PYTHIA{8} (red).  We observe sizable
differences between the two results, in particular for the $Z$ boson
transverse momentum at $\ptz < 15$ GeV. This is not surprising since
this is a region dominated by soft effects, hence the details of the
modelling of non-perturbative effects are expected to matter. For the
jet-transverse momentum the difference between the two shower models
is slightly smaller. This can probably be attributed to the fact that
the Sudakov peak is at larger values of the transverse momentum,
compared to the $Z$-boson transverse momentum.

\begin{figure}[htb]
  \begin{centering}
    \includegraphics[clip,width=0.49\textwidth]{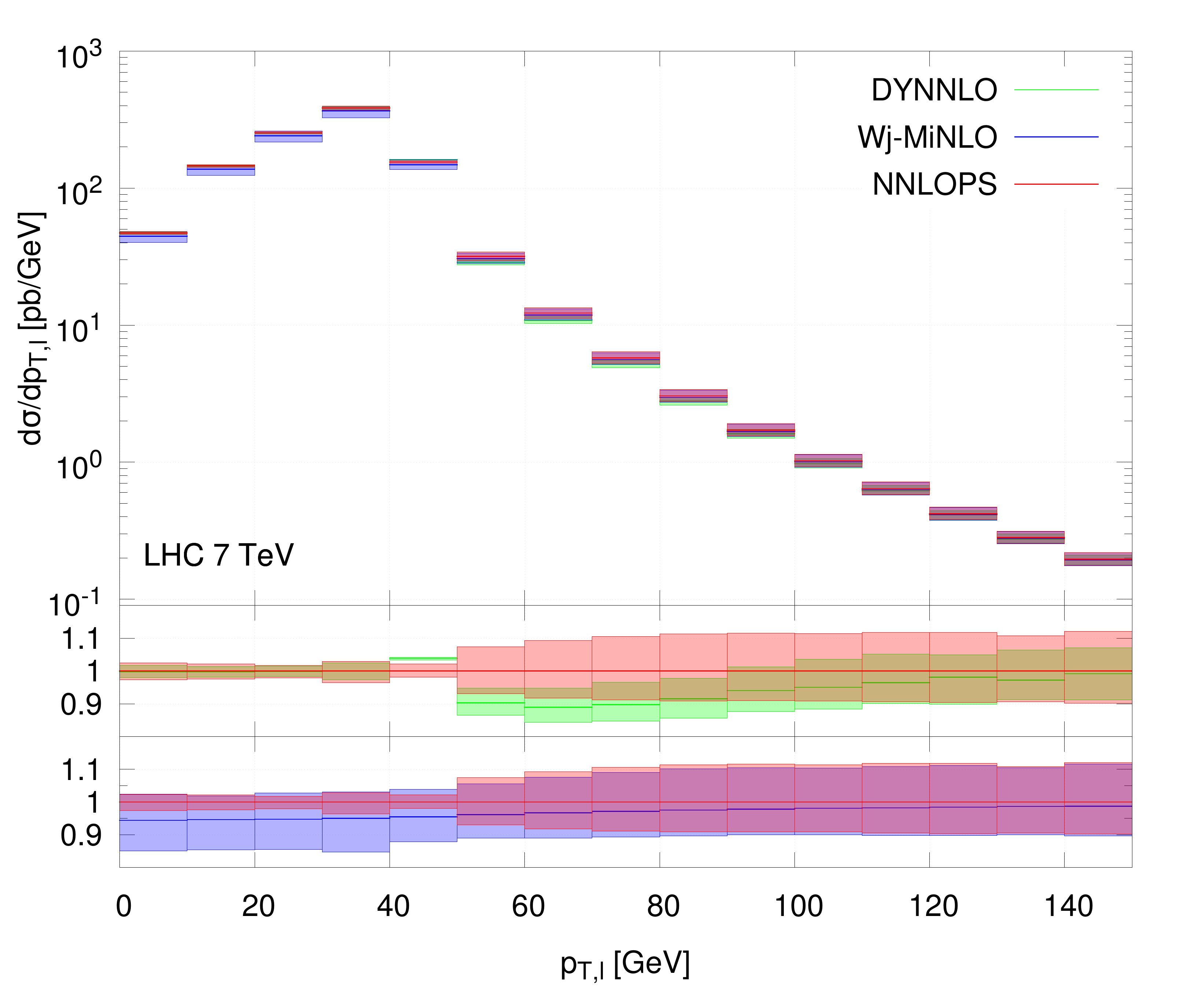}
    \hfill{}\includegraphics[clip,width=0.49\textwidth]{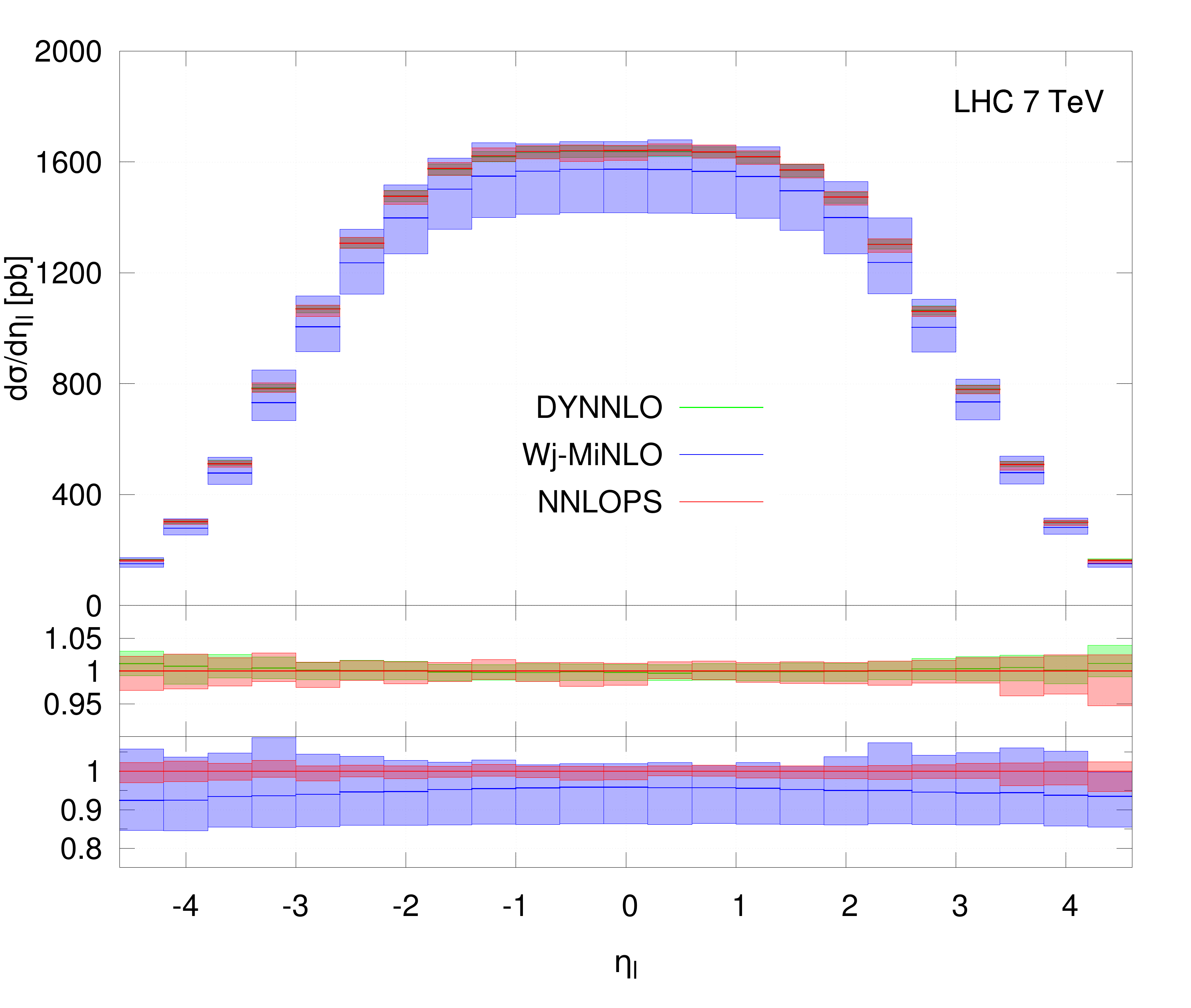}
    \par\end{centering}
    \caption{Comparison of the \NNLOPS{} (red), \NLOPS{} (blue) and
      \DYNNLO{} (green) 
      results for the transverse momentum (left) and rapidity distributions
      of the charged lepton (right) in $W$ production at the LHC running at
      7 TeV.  The \DYNNLO{} central scale is $\muf=\mur=M_W$, and its
      error band is the 3-point scale variation
      envelope. For
      \NNLOPS{} and \NLOPS{} the procedure to define the scale
      uncertainty is described in detail in
      Sec.~\ref{subsec:Estimating-uncertainties}.  The lower panels show
      the ratio with respect to the \NNLOPS{} prediction obtained with
      its central scale choice.}
    \label{fig:val-lepplus}
\end{figure}

Finally, for illustrative purposes, we conclude this section by
showing in Fig.~\ref{fig:val-lepplus} 
predictions for the transverse momentum (left) and pseudorapidity
(right) of the charged electron or positron in $W$ production at 7 TeV
(combining the $W^+$ and $W^-$ samples).
It is interesting to look at these leptonic observables since they
don't coincide with the quantities we are using to perform the NNLO
reweighting. Nevertheless, we should recover NNLO accuracy in the
regions where the lepton kinematics probes the fully-inclusive phase
space: this is precisely what we have found, as illustrated in
Fig.~\ref{fig:val-lepplus}. Here we include no cuts on the final state
other than requiring the transverse $W$ mass $\mtw = \sqrt{2
  (p_{{\scriptscriptstyle \mathrm{T,l}}}\, p_{{\scriptscriptstyle
      \mathrm{T}},miss} -\vec p_{{\scriptscriptstyle \mathrm{T,l}}}
  \,\cdot \vec p_{{\scriptscriptstyle \mathrm{T}},miss}) }$ to be
larger than $40 \mbox{ GeV}$.  All parameters and settings are the
same as those used for the $Z$ production case and \DYNNLO{} results
have been obtained choosing $\mur=\muf=M_W$ as central scale and
performing a 3-point scale variation.

As expected, in the left panel of Fig.~\ref{fig:val-lepplus} we
observe a much more narrow uncertainty on the charged lepton
transverse momentum for values of $p_{\rm T, l}$ smaller than $M_W/2$,
and a very good agreement with \DYNNLO{} in this region, both in the
absolute value of the cross-section as well as in the size of the
theoretical uncertainty band.
When $p_{\rm T, l}$ is larger than $M_W/2$ all distributions have NLO
accuracy only, since in this region the lepton kinematics requires
non-vanishing values for $\ptw$. We observe that in fact all
uncertainty bands are larger in this region. Our \NNLOPS{} result
reproduces the \ZNLOPSpPYTHIA{} one well, while there is some
difference between \DYNNLO{} and \ZNLOPSpPYTHIA{} predictions. This is
expected since the scales used in the two calculations are effectively
different: \DYNNLO{} always uses the mass of the $W$ boson, whereas in
\MINLO{} the transverse momentum of the $W$ boson is used. Therefore
when the lepton is just slightly harder than $M_W/2$ we are probing
phase-space regions where the bulk of the cross-section typically has
$0\lesssim \ptw \lesssim M_W$, and hence \DYNNLO{} yields smaller
cross-sections.

It is also worth mentioning that both the \NNLOPS{} and
\ZNLOPSpPYTHIA{} plots exhibit a smooth behaviour in proximity of the
jacobian peak $p_{\rm T, l}\simeq M_W/2$, also when thinner bins (not
shown) are used. This smooth behaviour is due both to parton-shower
effects and to the \MINLO{} Sudakov form factor.
On the contrary, the NNLO prediction displays the typical numerical
instability of fixed-order predictions due to the numerical
cancellation between real and virtual corrections close to the
kinematical boundary. Furthermore, the NNLO prediction has spuriously
small uncertainties in this region.

Finally, in the right panel of Fig.~\ref{fig:val-lepplus} we plot the
rapidity distribution of the charged lepton. Since in each bin of this
distribution we are fully inclusive with respect to QCD radiation, we
observe the expected good agreement with the NNLO prediction over the
whole range, as well as a quite narrow uncertainty band.

\subsection{Comparison to analytic resummations}
\label{subsec:resum}

The \MINLO{} method at the core of the results presented in this paper
works by including NLL and (some) NNLL terms in the Sudakov form
factors used to improve the validity and accuracy of the underlying
NLO computation at hand.  Although the formal logarithmic accuracy
achieved by \MINLO{}-improved \POWHEG{} simulations has not been
addressed (yet), it is interesting to compare \NNLOPS{} predictions
against results obtained with higher-order analytic resummation, for
observables where the latter are available. In this subsection we will
show results for Z production 
and focus on three quantities for which NNLL resummation has been
performed.

\begin{figure}[htb]
  \begin{centering}
    \includegraphics[clip,width=0.49\textwidth]{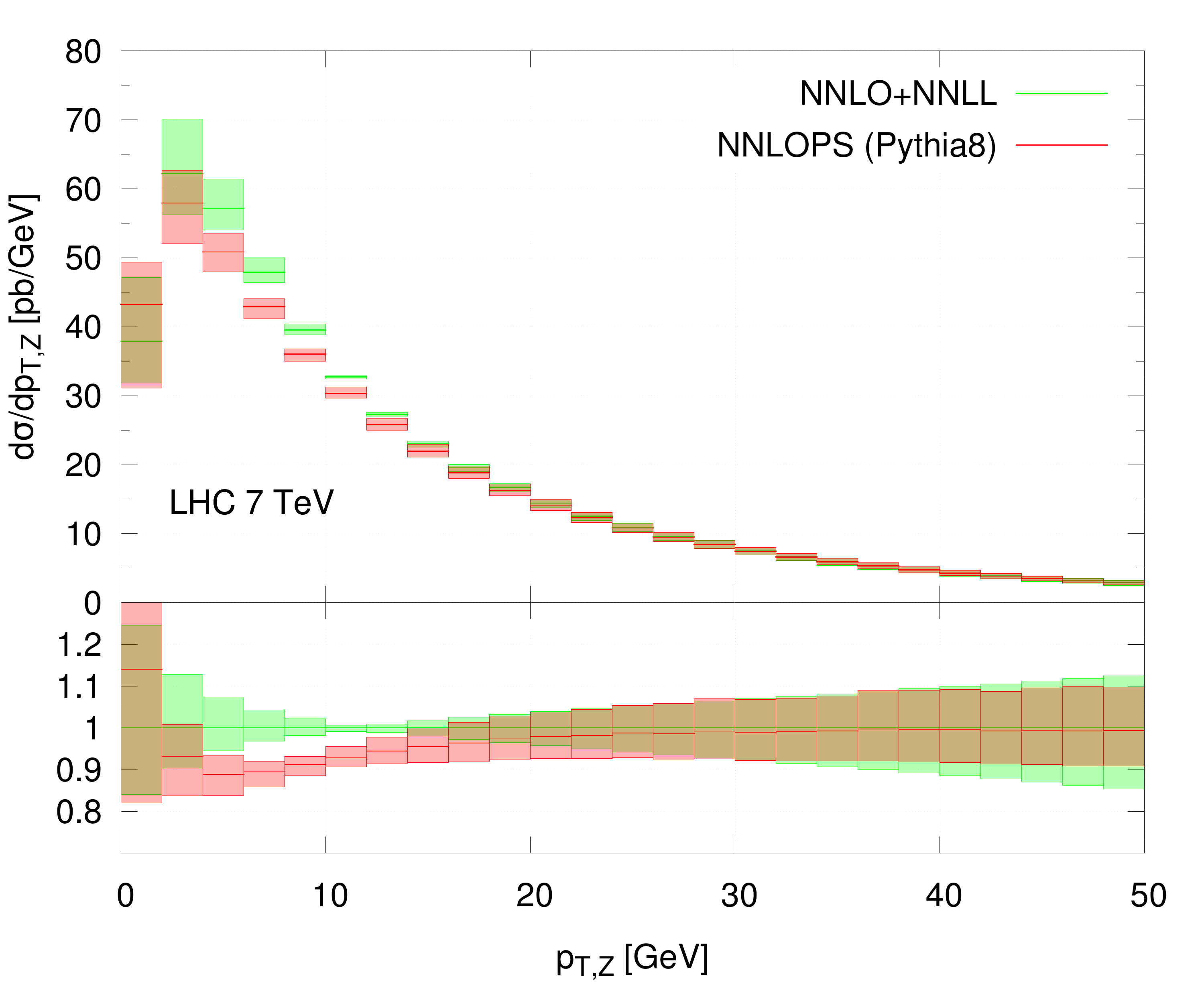}
    \includegraphics[clip,width=0.49\textwidth]{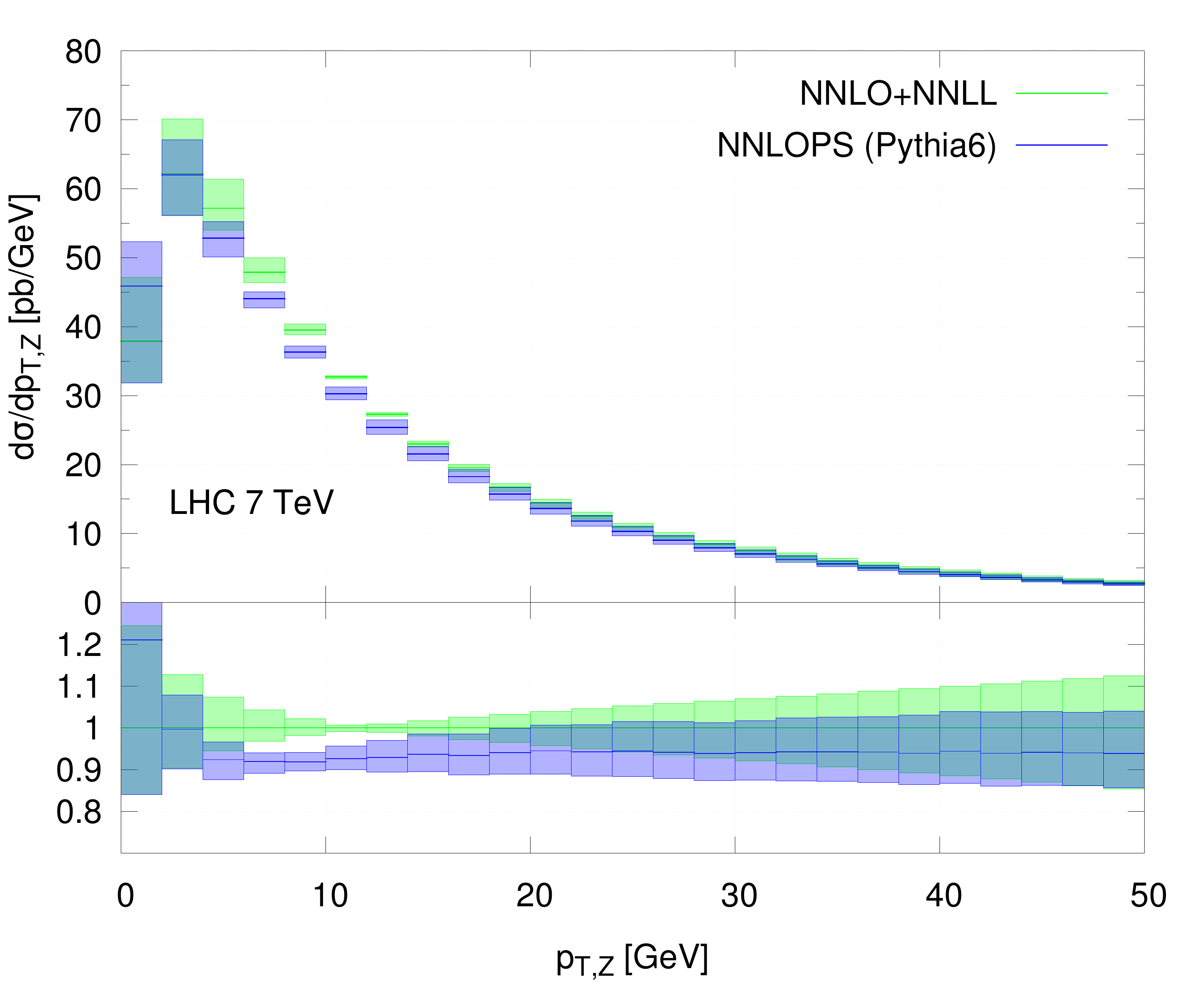}
    \par\end{centering}
    \caption{Comparison of \NNLOPS{} prediction and NNLO+NNLL
      resummation for $\ptz$ at the 7 TeV LHC. The \NNLOPS{}
      prediction is shown at parton level with parton showering
      performed with \PYTHIA{8} (left, red) and \PYTHIA{6} (right, blue).
      The resummed result is shown in green in both panels. The lower
      panels show the ratio to the NNLO+NNLL resummation.}
    \label{fig:res-ptZ}
\end{figure}

The classical observable to consider in Drell-Yan production to study
the effects of soft-collinear radiation is the transverse momentum of
the dilepton-pair system.
This observable has been extensively studied in the past and is now
known to NNLO+NNLL level~\cite{Bozzi:2010xn,Becher:2010tm}.
In Fig.~\ref{fig:res-ptZ} 
we show a comparison between the NNLO+NNLL resummation obtained using
\DYQT{} and our NNLOPS result obtained with \PYTHIA{8} (left) and
\PYTHIA{6} (right), switching off all non-perturbative effects
(\emph{i.e.}~hadronization and MPI are switched off, primordial-$\kt$
is set to zero). As usual, the uncertainty band for our results has
been obtained as the envelope of a 21-pts scale variation. \DYQT{}
uses as resummation scale $m_{\rm ll}$, and the associated band has
been obtained varying $\mur$ and $\muf$ among the usual 7
combinations. On top of this, for the central value of $\mur=\muf$ we
varied the resummation scale by a factor 2 up and down. This gives 9
combinations, the envelope of which is used here to define the \DYQT{}
uncertainty.  The ratio plot shows a pattern quite similar to what was
observed in Figs. 4 and 5 of ref.~\cite{Hamilton:2013fea}, namely
differences of up to $\mathcal{O}(10-12\%)$ between analytic
resummation and \NNLOPS{}, with a slightly more marked difference in
the very first bin. In the region $\ptz \sim 5-15$ GeV, the
uncertainty bands do not overlap, mainly because of the very narrow
uncertainty bands in both predictions, in particular in the NNLO+NNLL
result. In the case of Higgs production, instead, uncertainty bands
are wider, hence the predictions are more compatible.
Changing the $\beta$ parameter might improve this agreement, although we recall that the \NNLOPS{} prediction does not have NNLL
accuracy in this region.
By comparing the two \NNLOPS{} results shown in the two panels of
fig.~\ref{fig:res-ptZ}, we also observe that the spectra obtained with
\PYTHIA{8} are typically $\sim 5$ \% harder than those with
\PYTHIA{6}, a feature that was already noticeable in
Fig.~\ref{fig:par-ptz}, and which will be present also in other
distributions where \NNLOPS{} results are ``only'' NLO
accurate. Few percent differences between different NLO+PS results in
these kinematic regions can be due to subleading effects, such as
differences in details of the two parton-shower algorithms, as well as
the use of different tunes. A comprehensive assessment of these issues
goes beyond the purpose of this work, and is therefore left for future
work. 

Another interesting observable to consider is the $\phi^*$
distribution which is a measure of angular correlations in Drell-Yan
lepton pairs~\cite{Banfi:2010cf}. This observable is defined
as~\cite{Banfi:2012du}
\begin{equation} \phi^* = \tan \left(\frac{\pi - \Delta
\phi}{2}\right)\sin \theta^*\,,
\end{equation} where $\Delta \phi$ is the azimuthal angle between the
two leptons and $\theta^*$ is the scattering angle of the electron
with respect to the beam, as computed in the boosted frame where the $Z$
boson is at rest.
We note that ATLAS uses a slightly different definition of the angle
$\theta^*$, and defines it as
\begin{equation} \cos \theta^* = {\rm
tanh}\left(\frac{y_{l^-}-y_{l^+}}{2}\right)\,.
\end{equation}
Since we will compare to ATLAS data in Sec.~\ref{subsec:Wdata}, we
will use the latter definition throughout this work.
\begin{figure}[htb]
  \begin{centering}
    \includegraphics[clip,width=0.49\textwidth]{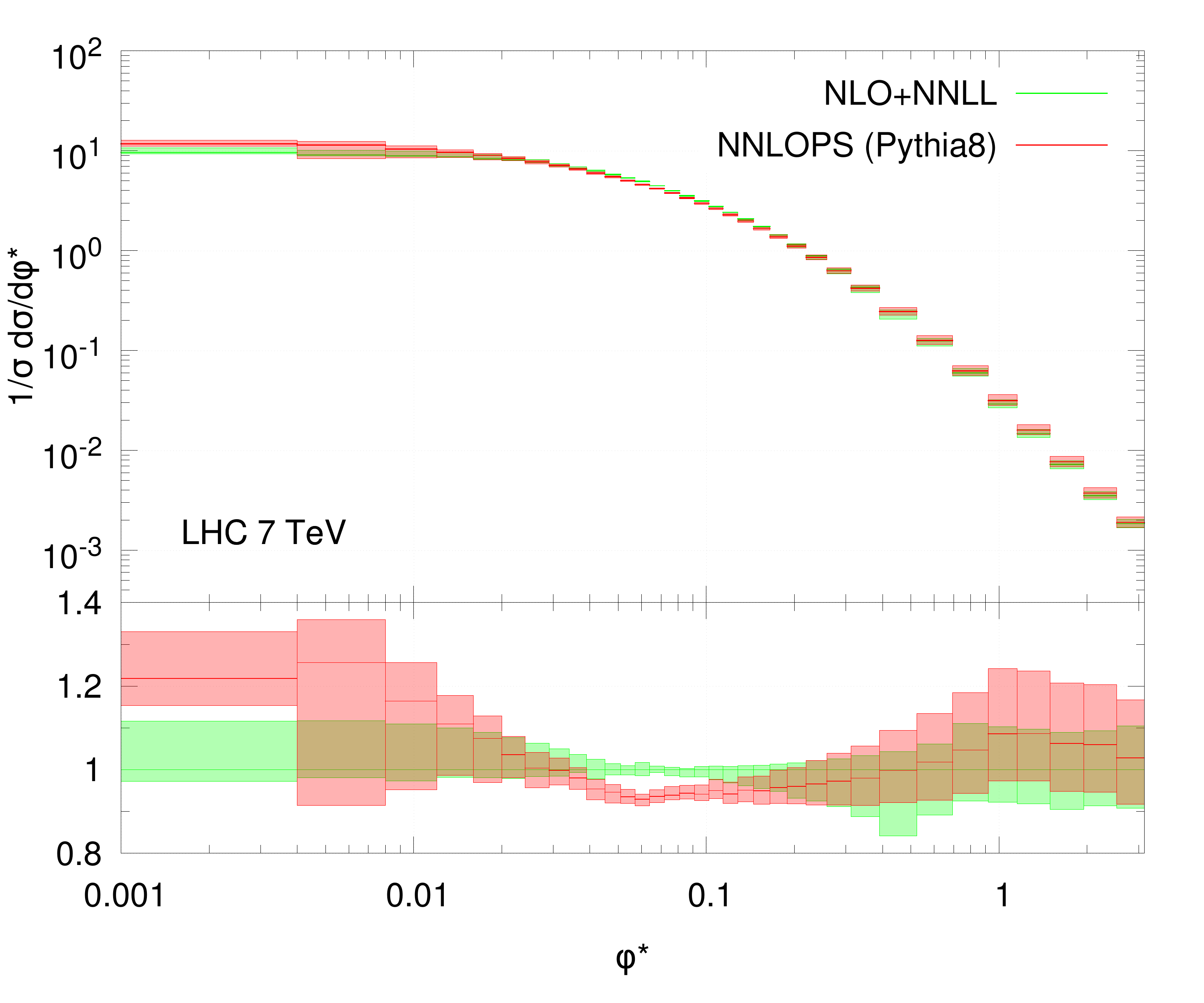}
    \includegraphics[clip,width=0.49\textwidth]{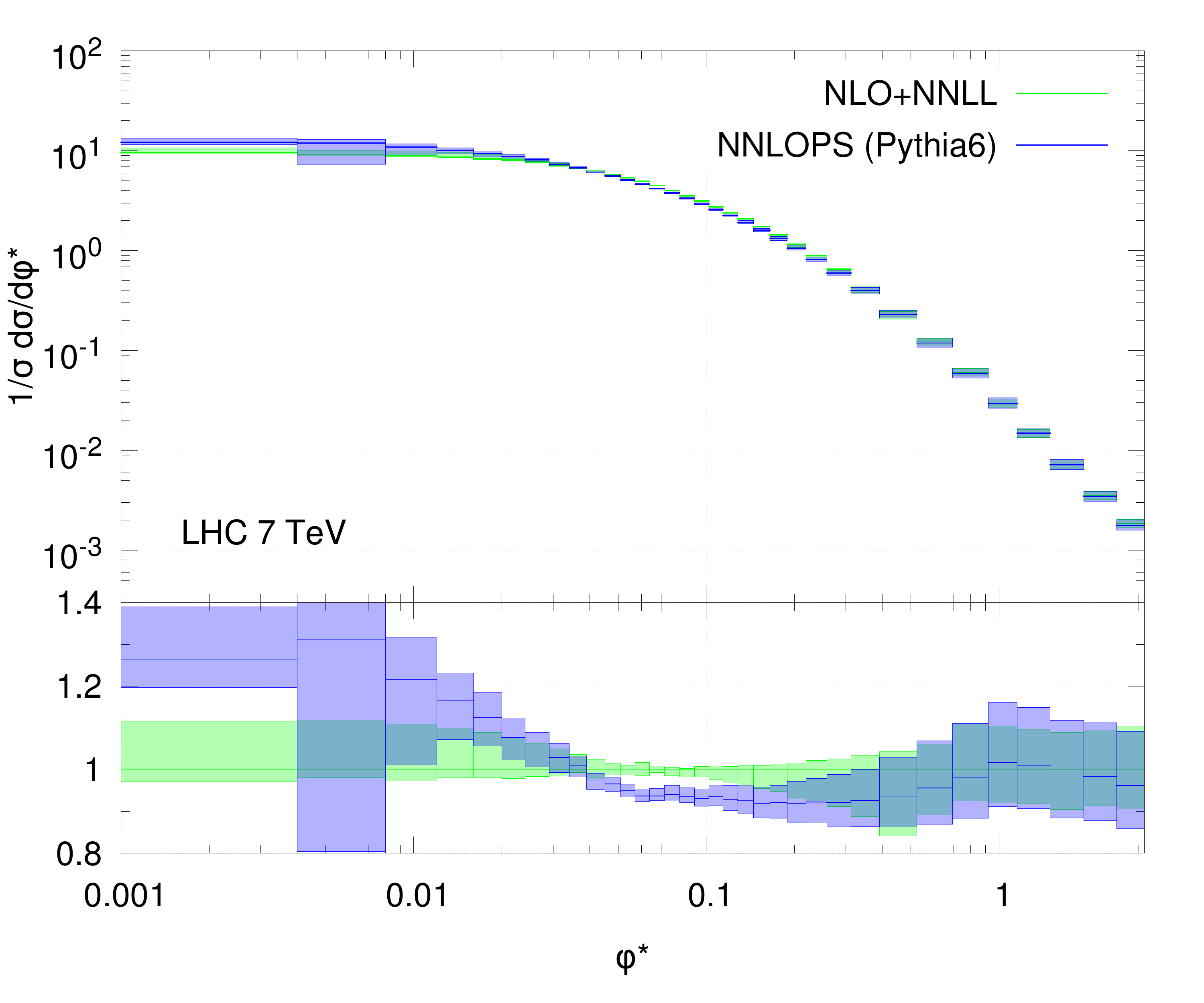}
    \par\end{centering}
    \caption{Comparison of \NNLOPS{} prediction and NLO+NNLL
      resummation for $\phi^*$ in $Z\to e^+e^-$ production at 7 TeV
      LHC. The \NNLOPS{} prediction is shown at parton level with
      parton showering performed with \PYTHIA{8} (left, red) and
      \PYTHIA{6} (right, blue).  The resummed result is shown in green
      in both panels.}
    \label{fig:res-phistar}
\end{figure}
In Fig.~\ref{fig:res-phistar} we compare our \NNLOPS{} simulation with
the NLO+NNLL resummation of ref.~\cite{Banfi:2012du}.\footnote{We
  thank Andrea Banfi and Lee Tomlinson for providing us with their
  resummed results.} From the definition of $\phi^*$, it is clear that
large values of $\phi^*$ correspond to events where the $Z$ boson tends
to be boosted, while for low values the $Z$
boson is almost at rest. We see that the two predictions agree
reasonably well for $\phi^* \gtrsim 0.2$, in particular when
\PYTHIA{6} is used, while the uncertainty bands to not overlap below
that point.\footnote{We remark that the resummation of
  ref.~\cite{Banfi:2012du} uses CTEQ6M parton distribution functions.}
For high $\phi^*$, \PYTHIA{8} is slightly harder than
\PYTHIA{6}. Since for large values of $\phi^*$ the probed phase space
regions are dominated by large values of $\ptz$, this difference is
expected, in view of the discussion at the end of the previous
paragraph.
We will show a comparison to data for this observable in
Sec.~\ref{subsec:Zdata}, where we will also comment on the impact of
non-perturbative corrections.

Finally we consider the jet-veto efficiency which is defined as
\begin{equation}
\epsilon(p_{\rm T,veto}) \equiv \frac{1}{\sigma}\int_0^{p_{\rm
      T,veto}} d\ptjone \frac{d\sigma(\ptjone)}{d\ptjone}\,.  
\end{equation}
\begin{figure}[htb]
  \begin{centering}
    \includegraphics[clip,width=0.49\textwidth]{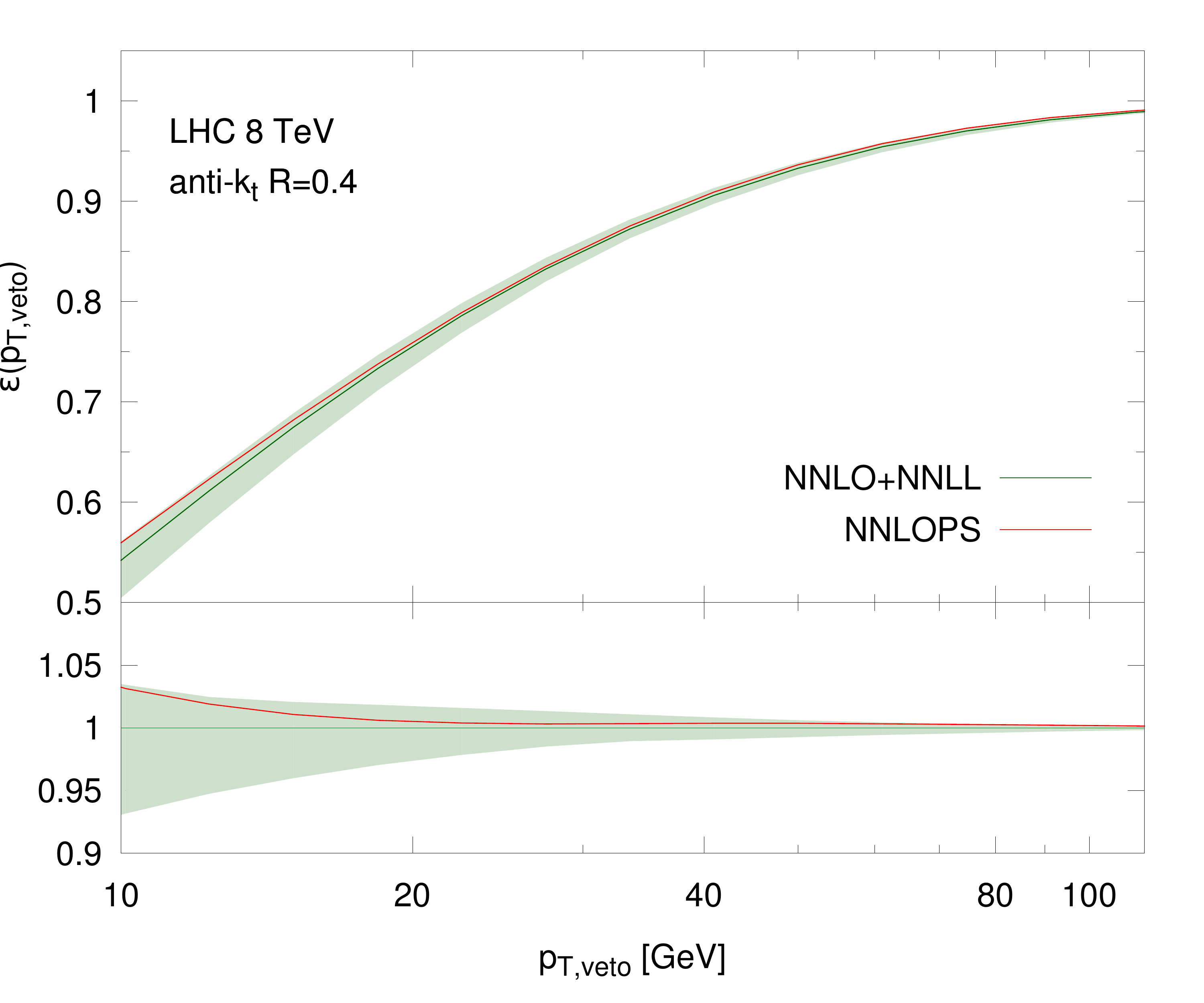}
    \includegraphics[clip,width=0.49\textwidth]{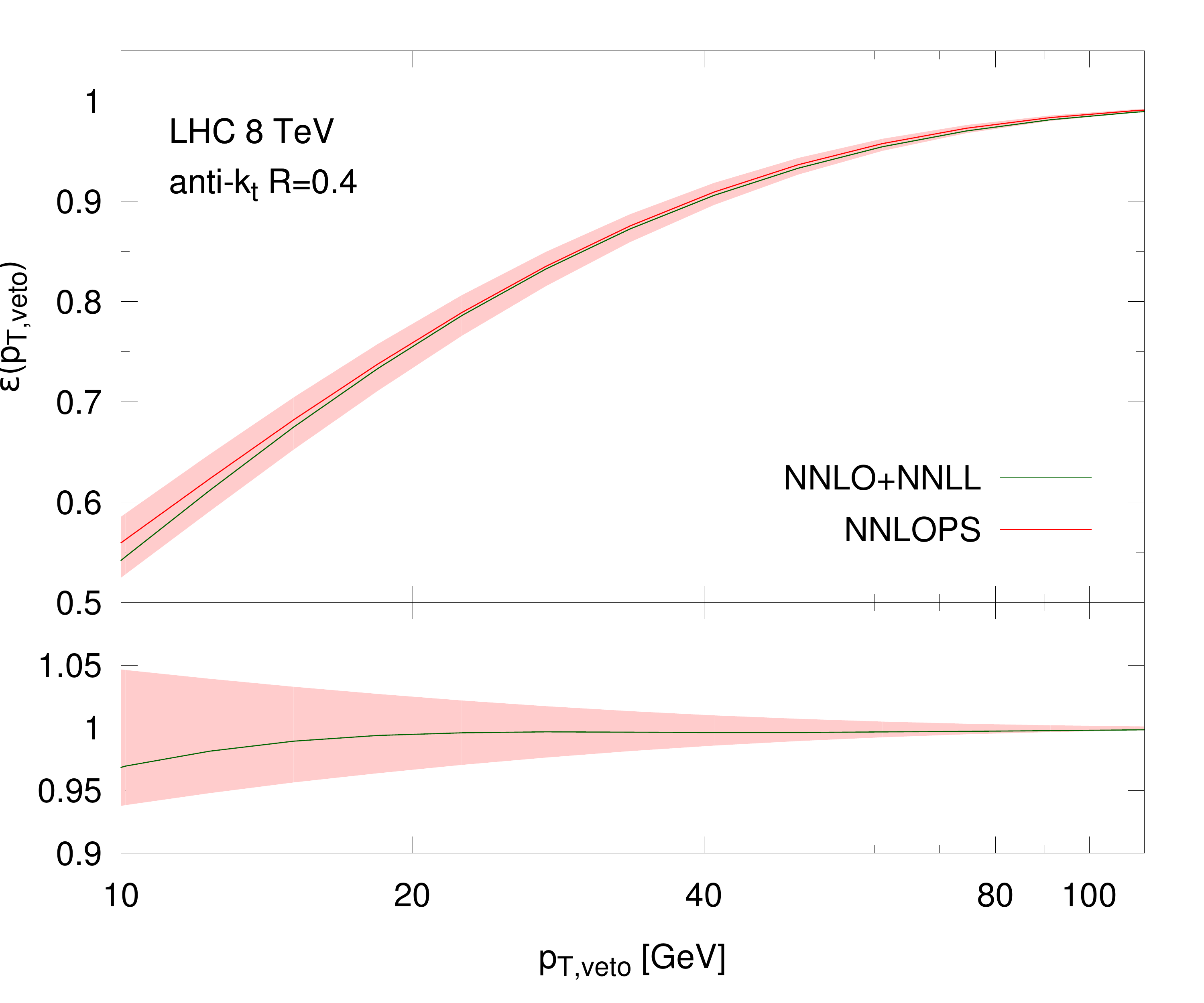}\\
    \includegraphics[clip,width=0.49\textwidth]{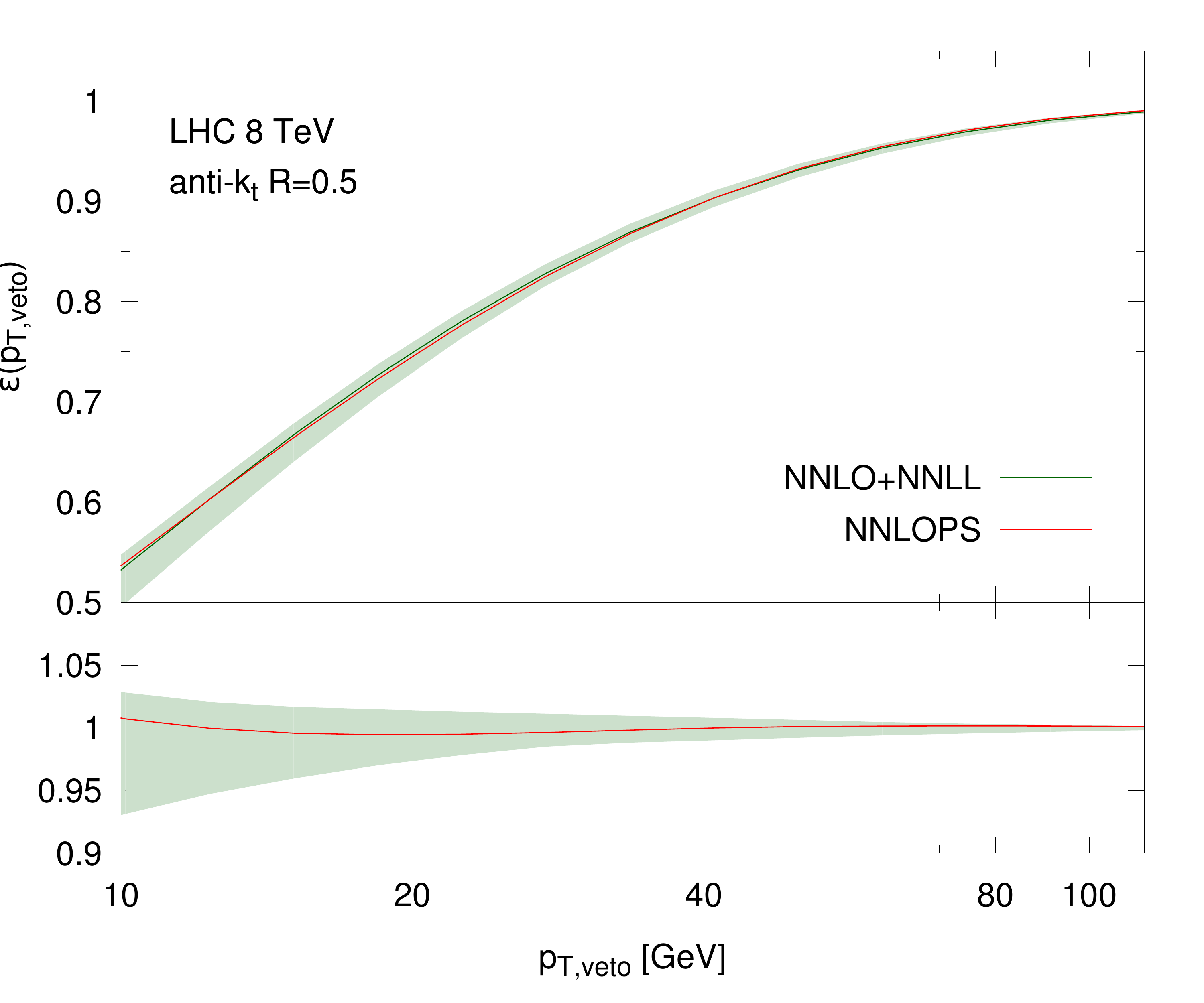}
    \includegraphics[clip,width=0.49\textwidth]{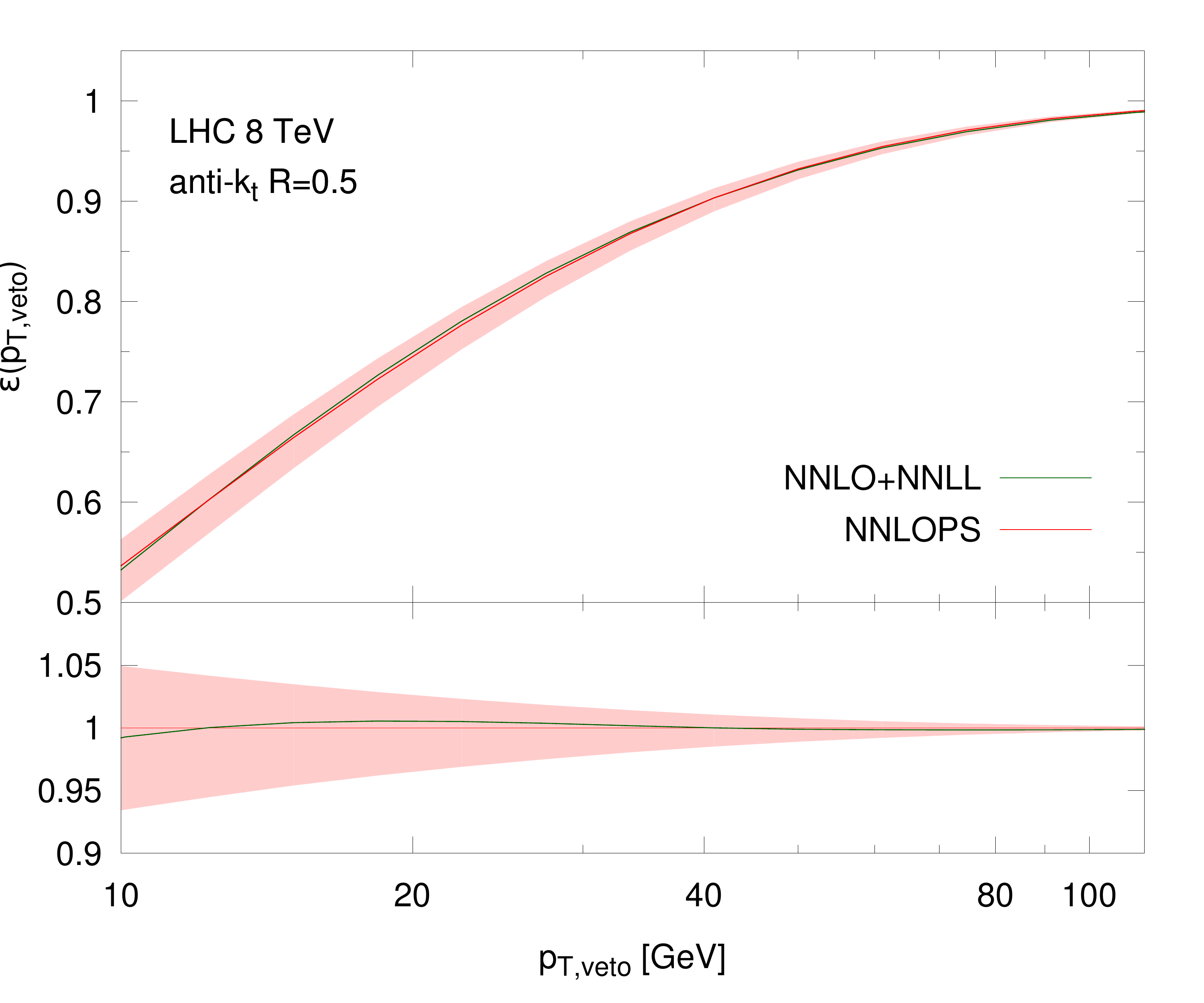}\\
    \includegraphics[clip,width=0.49\textwidth]{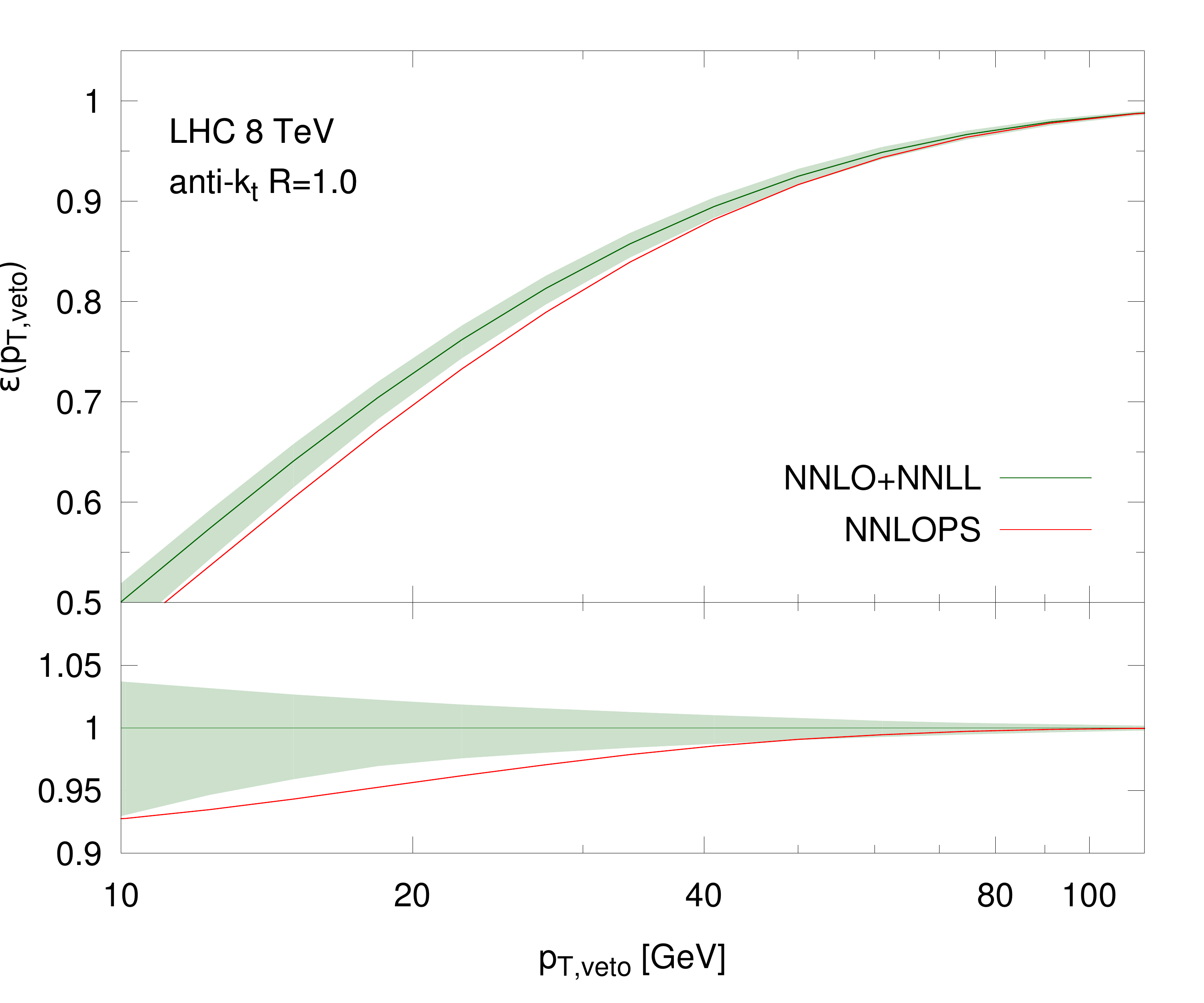}
    \includegraphics[clip,width=0.49\textwidth]{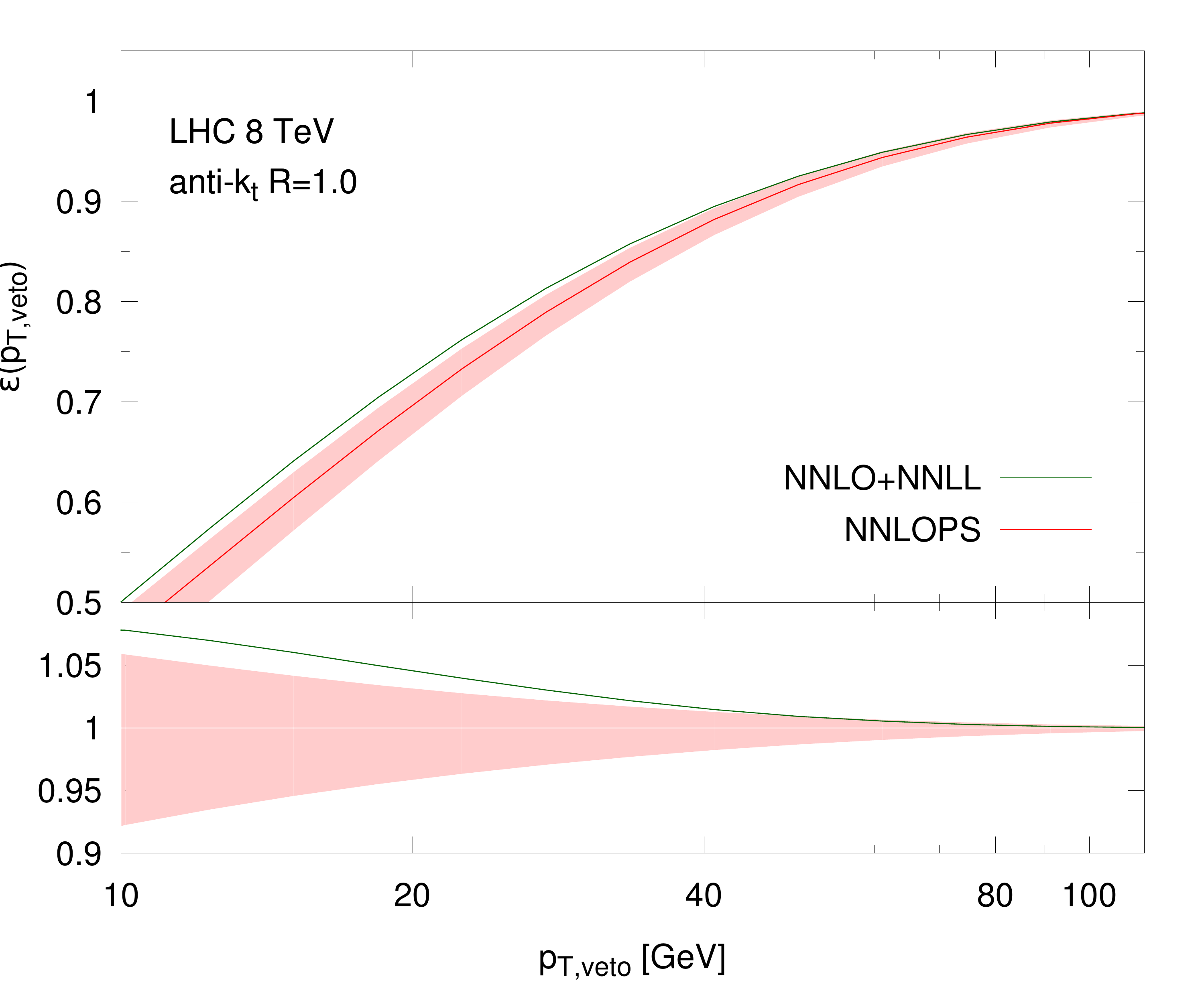}
    \par\end{centering}
    \caption{Comparison of \NNLOPS{} (red) prediction and NNLO+NNLL
      resummation (green) for the jet-veto efficiency for $Z$ production
      at 8 TeV LHC for three different values of the jet-radius. The
      \NNLOPS{} prediction is shown at parton level with parton
      showering performed with \PYTHIA{8}.}
    \label{fig:res-jetveto}
\end{figure}

Here results have been obtained for the LHC at 8 TeV, using the
anti-$k_t$ algorithm to construct jets.  We restrict ourselves to show
\NNLOPS{} results obtained with \PYTHIA{8}. Analytically resummed
results have been obtained using \JETVHETO{}~\cite{Banfi:2012jm}, and
have NNLL+NNLO accuracy. 
The \JETVHETO{} results have been obtained using its default setting,
i.e. using as a central value for the renormalisation, factorisation
and resummation scale $M_Z/2$.\footnote{Since we use $\beta=1$ in the
  definition of $h(\pt)$ for the \NNLOPS{} results, it is interesting
  also to examine the \JETVHETO{} results using as a central value for
  renormalisation, factorisation and resummation scale $M_Z$. We have
  done so and find that there are minimal differences for $R=0.4$ and
  $R=0.5$. For $R=1.0$ we find a slightly worse agreement.}

As is recommended in ref.~\cite{Banfi:2012jm}, for the \JETVHETO{}
results we have obtained the uncertainty band as an envelope of eleven
curves: we have varied the renormalization and factorization scales
independently giving rise to the usual 7-scale choices, additionally,
for central renormalization and factorization scales we have varied
the resummation scale up and down by a factor 2 and looked at two
different additional schemes to compute the efficiency. 

We observe a very good agreement between the two approaches
for $R=0.4$ and $0.5$, whereas for $R=1$ differences are more marked.
Few comments are in order here: first, the pattern shown in the plots
is consistent with what was already observed in the Higgs case (fig. 7
of ref.~\cite{Hamilton:2013fea}), namely differences up to few
percents, and good band overlap, for smaller values of $R$, and larger
differences for $R=1$. For very large values of $R$, the leading-jet
momentum will balance against the transverse momentum of the vector
boson. Given what we observed for $\ptz$, it is therefore no surprise
that, when $R=1$, we have $\mathcal{O}(3-5) \%$ differences with
respect to the resummed result for values of \mbox{$p_{\rm T,veto}\sim
  25-30$ GeV}, as used currently by ATLAS and CMS in Higgs production.

\section{Comparison to data}
\label{sec:compdata}

In this section we compare our predictions with a number of available
data from ATLAS, both for $Z$ and for $W$ production.

\subsection{Z production}
\label{subsec:Zdata}
We show here a comparison to a number of measurements performed by
ATLAS at 7 TeV~\cite{Aad:2011dm,Aad:2011gj,Aad:2012wfa}. ATLAS applied
the following cuts: they consider the leptonic decay of the $Z$ boson
to electrons or muons and require an electron (muon) and a positron
(anti-muon) with $\pt > 20$ GeV and rapidity $|y| < 2.4$. The
invariant mass of the di-lepton pair should lie in the window 66 GeV <
$m_{\rm ll}$ < 116 GeV.

\begin{figure}[htb]
\begin{centering}
\includegraphics[clip,width=0.49\textwidth]{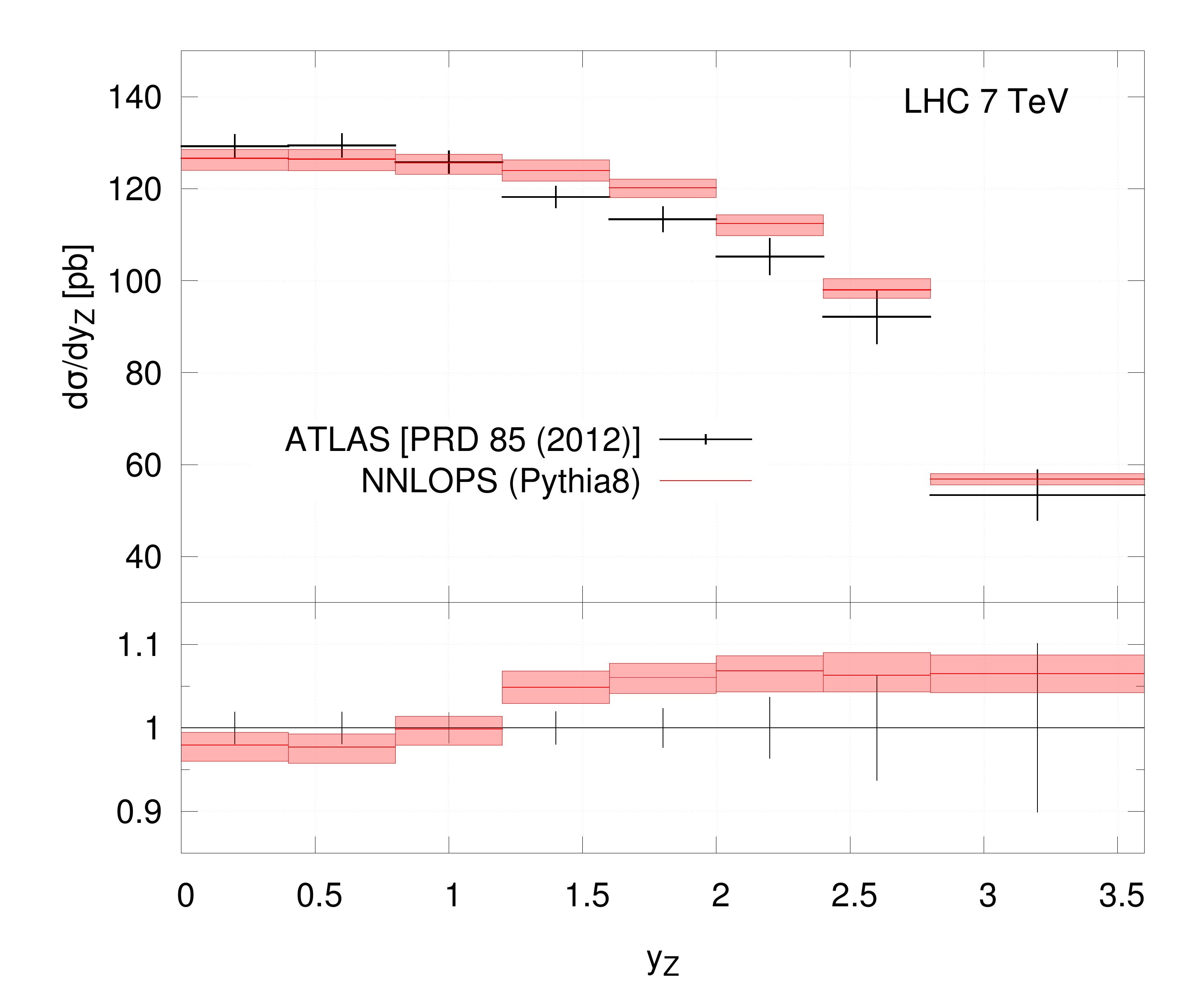}
\includegraphics[clip,width=0.49\textwidth]{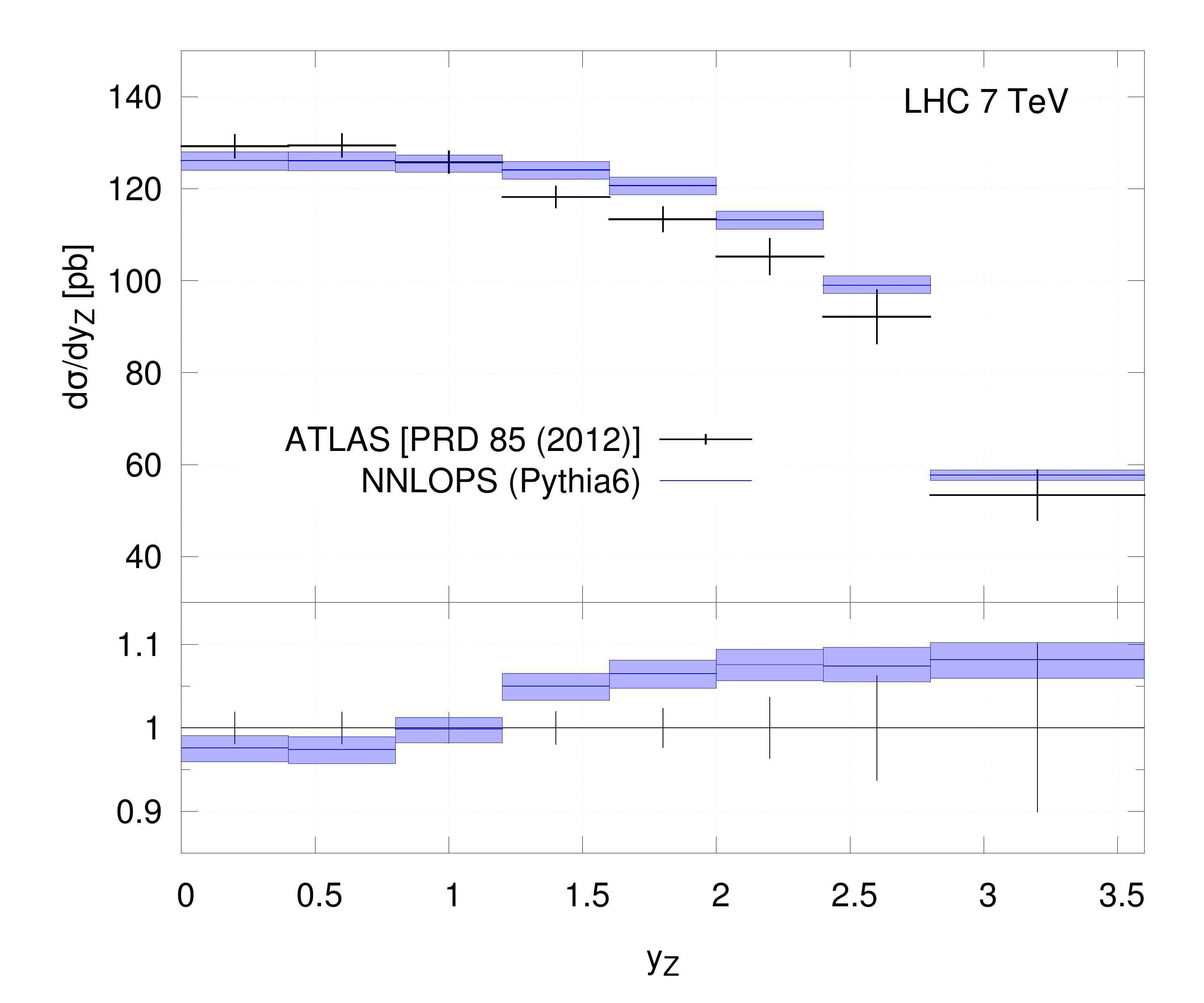}
\par\end{centering}
\caption{Comparison of \NNLOPS{} prediction obtained with \PYTHIA{8}
  (left) and \PYTHIA{6} (right) to data (black) from
  ref.~\cite{Aad:2011dm} for the $Z$ boson rapidity distribution at 7
  TeV LHC.}
\label{fig:data-yz}
\end{figure}

We begin by showing in Fig.~\ref{fig:data-yz}
a comparison of \NNLOPS{} results (with two versions of \PYTHIA{}) to
data from ref.~\cite{Aad:2011dm} for the $Z$ boson rapidity
distribution. As expected, our result displays a quite narrow
uncertainty band, due to having included NNLO corrections. Since this
is a fully inclusive observable, the absolute value of the
cross-section and the size of uncertainty band will be driven by the
NNLO reweighting: hence, \PYTHIA{6} and \PYTHIA{8} results are almost
indistinguishable, as expected.  We also observe that we agree with
data within the errors for central rapidities. At high rapidity,
however, there seems to be a tension between data and our
results. This discrepancy between data and pure NNLO was already
observed in the original ATLAS paper, although the NNLO results shown
in ref.~\cite{Aad:2011dm} have a slightly larger uncertainty band
since they also contain PDF uncertainties.  We note that, at the
moment, the dominant error is coming from data. We therefore expect
the agreement to improve, as more data become available, although
systematic errors are non-negligible~\cite{Aad:2011dm}.

\begin{figure}[htb]
\begin{centering}
\includegraphics[clip,width=0.49\textwidth]{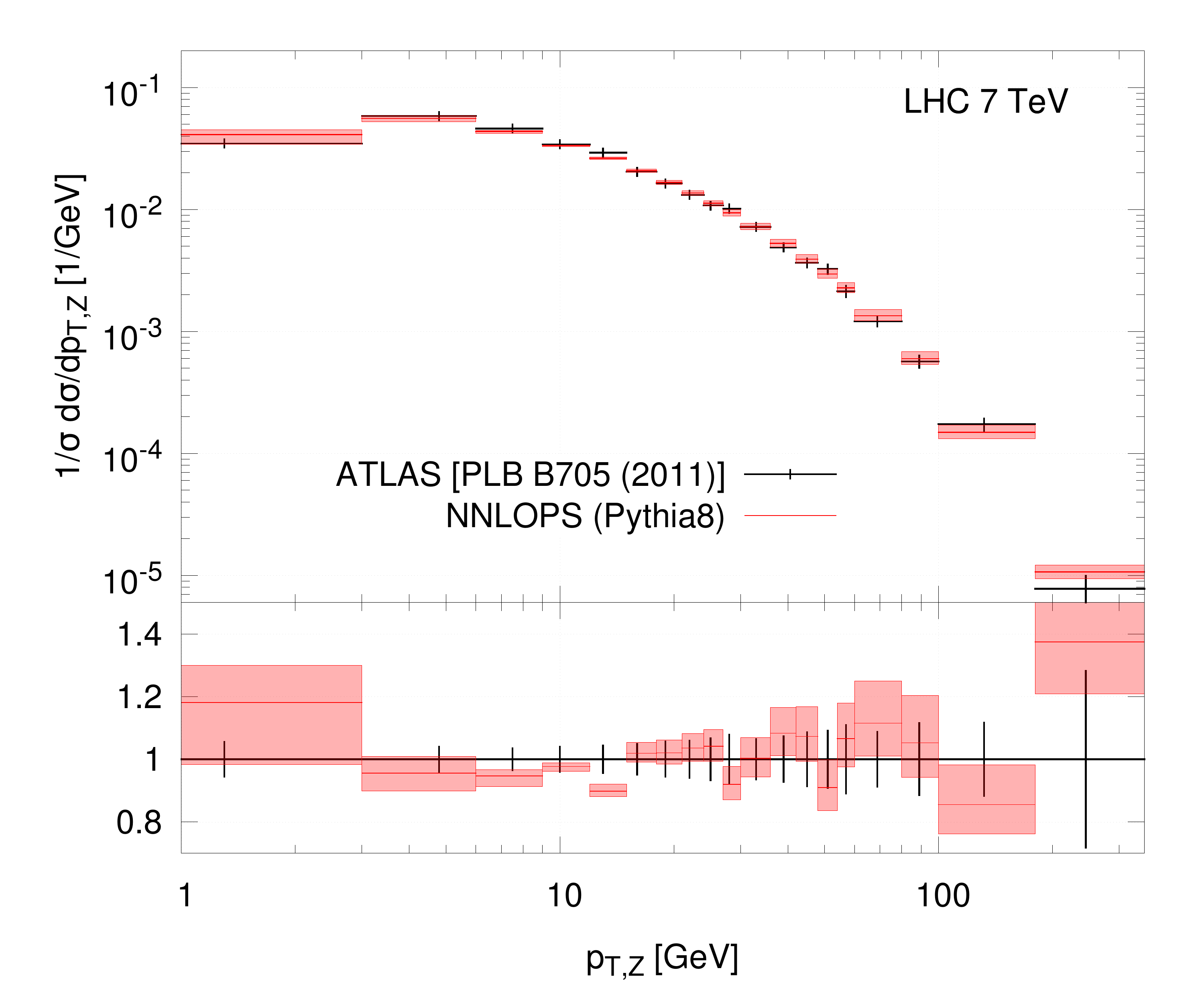}
\includegraphics[clip,width=0.49\textwidth]{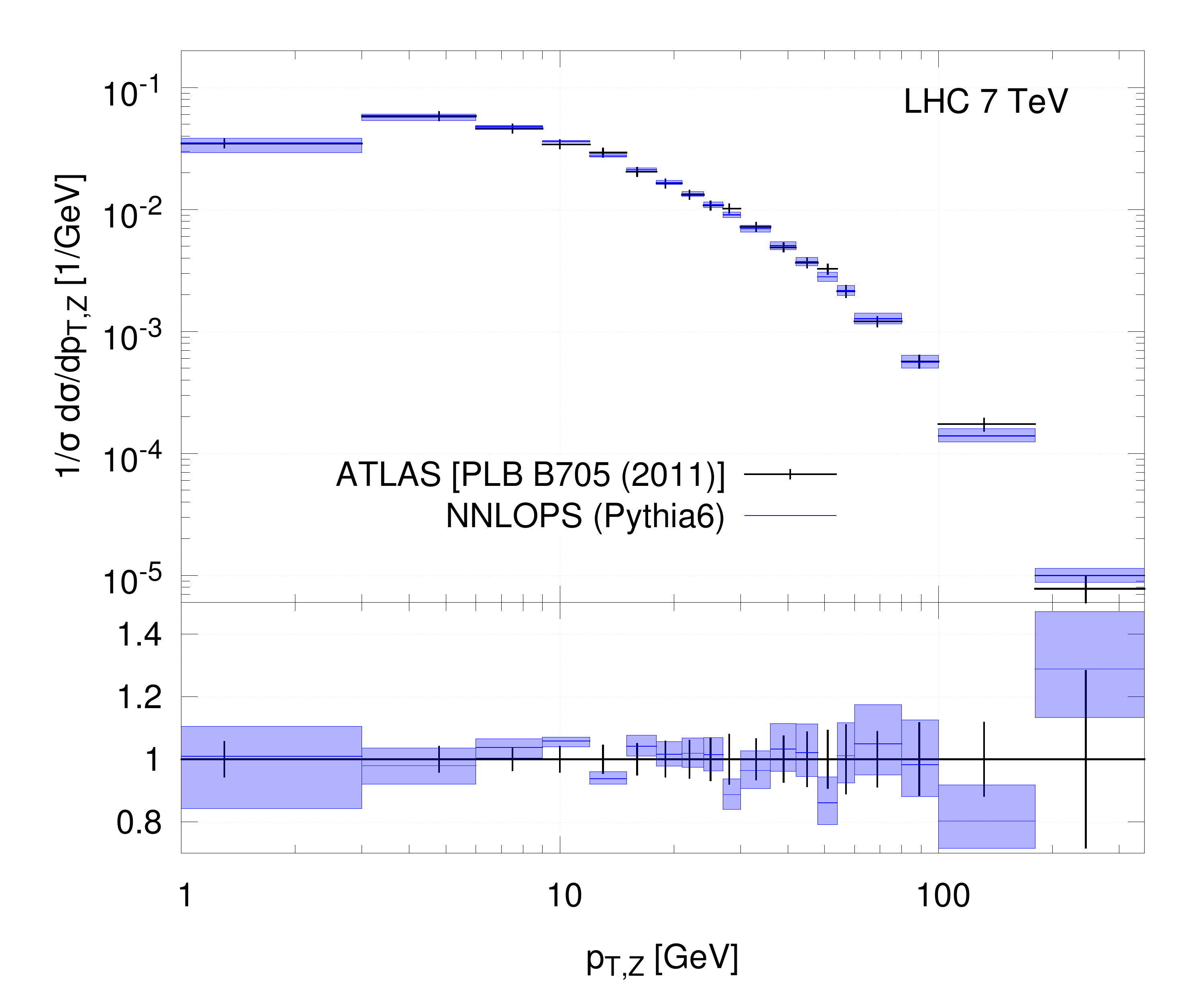}
\par\end{centering}
\caption{Comparison to data from ref.~\cite{Aad:2011gj} for the $Z$
  boson transverse distribution at 7 TeV LHC. Normalised data compared to \NNLOPS{}
  showered with \PYTHIA{8} (left plot, red) and \PYTHIA6{} (right
  plot, blue). Uncertainty bands for the theoretical predictions are
  obtained by first normalising all scale choices, as described in
  Sec.~\ref{subsec:Estimating-uncertainties} and then taking the
  associated envelope of these normalised distributions.}
\label{fig:data-ptz}
\end{figure}

In Fig.~\ref{fig:data-ptz} 
\begin{figure}[htb]
\begin{centering}
\includegraphics[clip,width=0.49\textwidth]{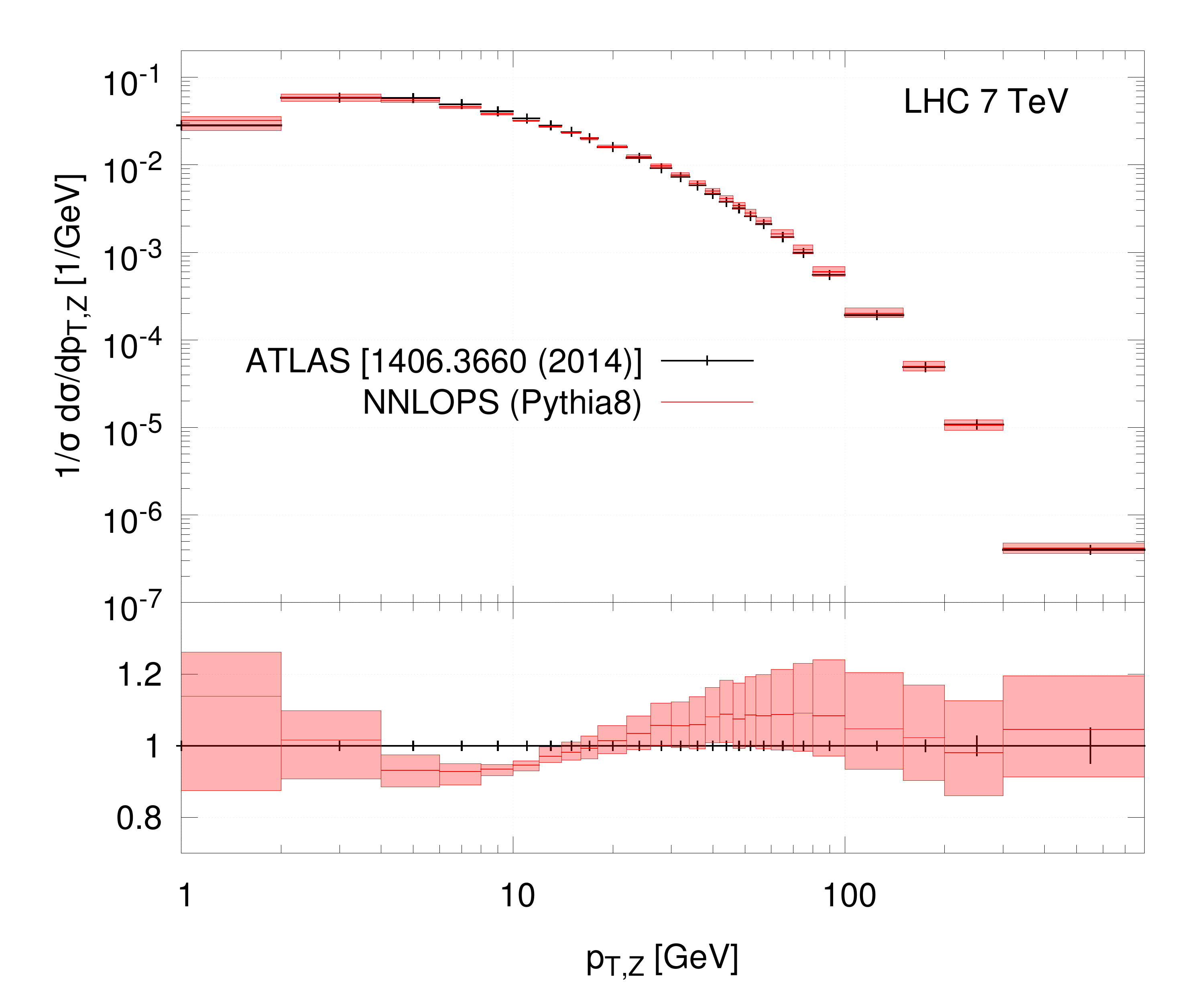}
\includegraphics[clip,width=0.49\textwidth]{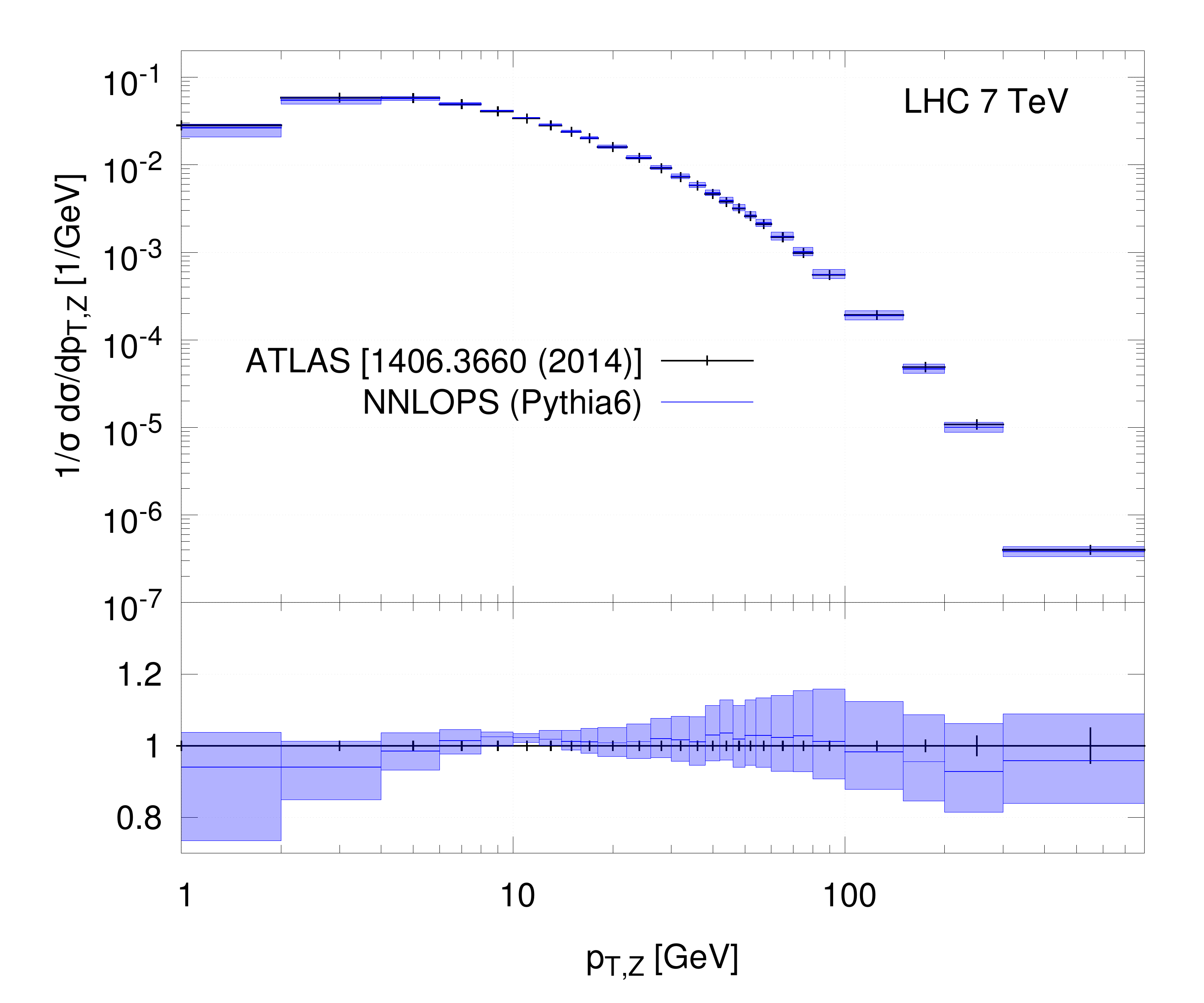}
\par\end{centering}
\caption{As in previous figure, but with more luminosity, thinner
  binnings, and up to larger values of $\ptz$. Data are now from taken
  from ref.~\cite{Aad:2014xaa}. }
\label{fig:data-ptz-new}
\end{figure}
we now show the same comparison for the $Z$ boson transverse momentum
against data from ref.~\cite{Aad:2011gj}. In the left panel we use
\PYTHIA{8} to shower events and in the right panel \PYTHIA{6}. We see
that there is a very good agreement between our \NNLOPS{} prediction
over the whole $\pt$ range.  Only in the very first bin we observe a
slight tension with data in the case of \PYTHIA{8}, which can be due
to a different modelling of non-perturbative effects.  Given the fact
that we use two different parton-showers, with different tunes (and
even different PDFs for the tunes), an $\mathcal{O}(20\%)$ difference
in the $1-3$ GeV region can be expected. It is foreseeable that once a
tuning of \PYTHIA{8} is done in conjunction with \POWHEG{}, such
differences will go away. Fig.~\ref{fig:data-ptz-new} shows again
$\ptz$, based now on 4.7 fb$^{-1}$ of data from
ATLAS~\cite{Aad:2014xaa}. Due to the thinner binning, it is now
possible to appreciate clearly the differences between \PYTHIA{6} and
\PYTHIA{8}: the \NNLOPS{} result obtained with \PYTHIA{6} shows a
remarkable agreement with data across all the $\ptz$ range, whereas
with \PYTHIA{8} we can observe differences of up to 10\% for $
5\lesssim \ptz \lesssim 100$ GeV. What we observe from these plots is
consistent with fig.~\ref{fig:data-ptz}, the difference being that
here the improved precision in data allows to conclude that, for the
setups and tunes we are using, the best description is obtained when
using \PYTHIA{6}.

\begin{figure}[htb]
\begin{centering}
\includegraphics[clip,width=0.49\textwidth]{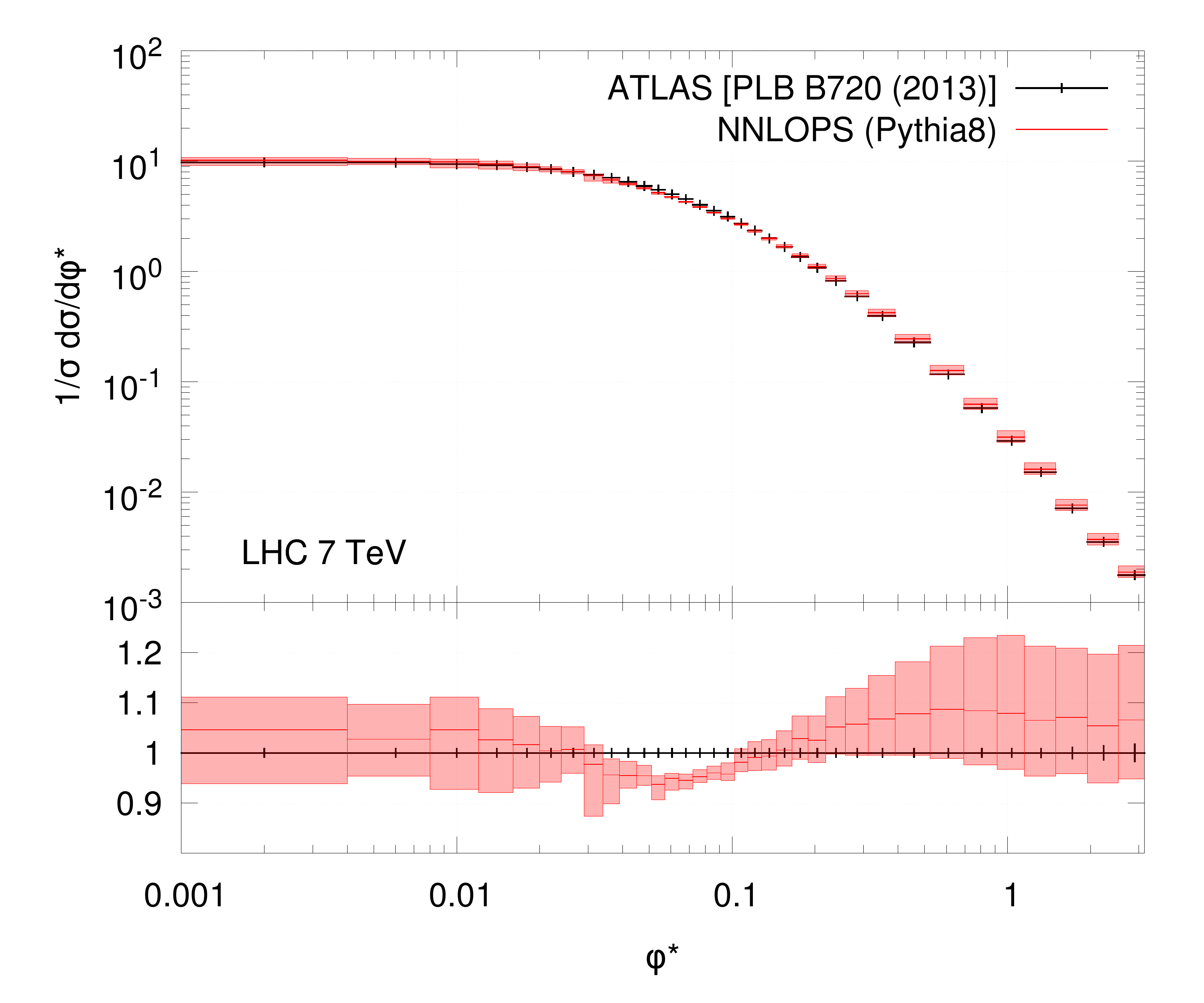}
\includegraphics[clip,width=0.49\textwidth]{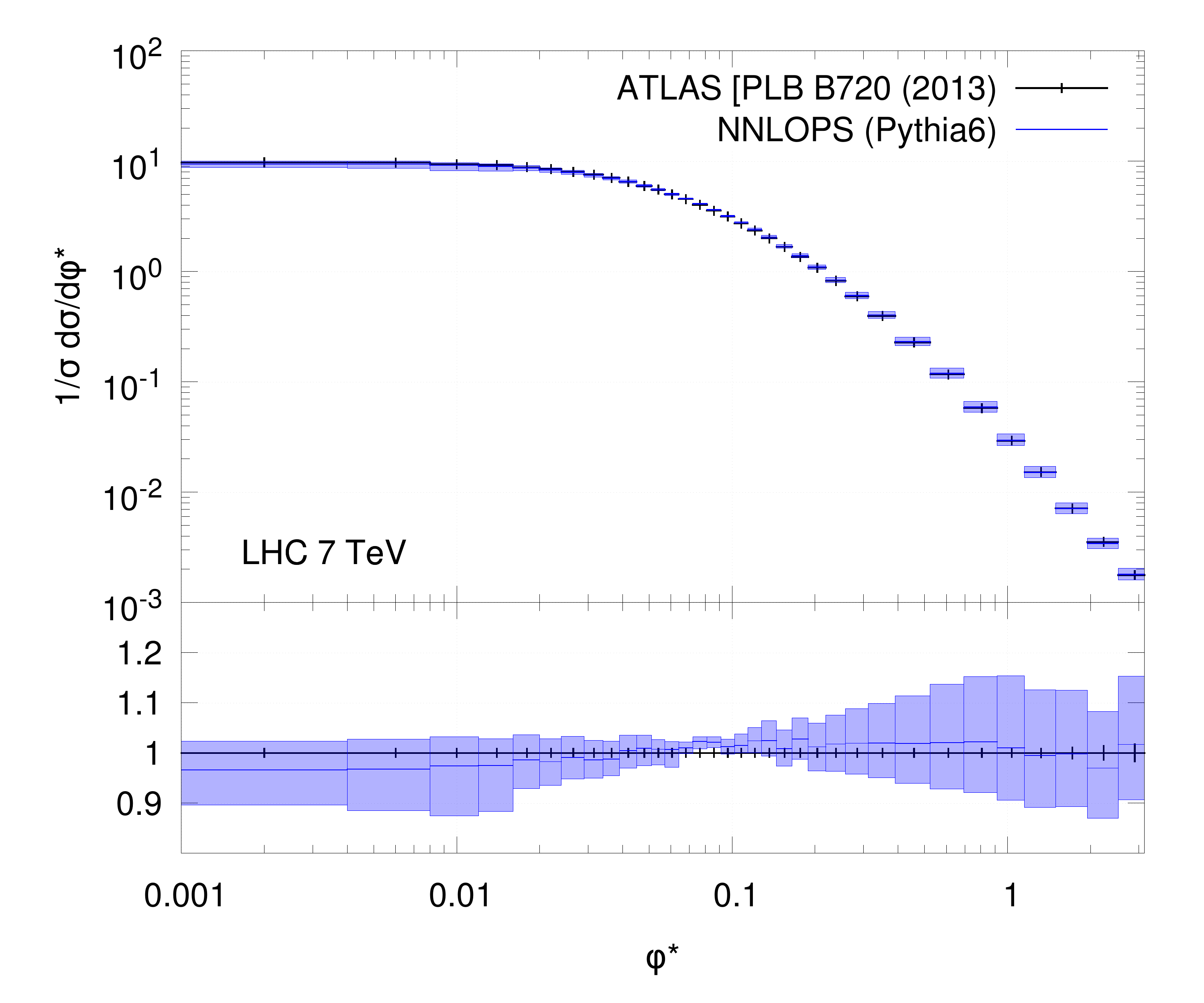}
\par\end{centering}
\caption{As in Fig.~\ref{fig:data-ptz} for the $\phi^*$ distribution
  in $Z$ boson production. Data taken from ref.~\cite{Aad:2012wfa}. }
\label{fig:data-phistar}
\end{figure}

Next, in Fig.~\ref{fig:data-phistar},
we consider the comparison to data for the $\phi^*$ distribution.
Although both our predictions are consistent with the ATLAS measure,
it is clear that events showered with \PYTHIA{6} agree better with
data, whereas \NNLOPS{} showered with \PYTHIA{8} has a slightly
different shape, exhibiting a dip in the theory prediction compared to
data, at around $\phi^*=0.06$.  By comparing with our predictions
before the inclusion of non-perturbative effects, we have checked that
these effects play a sizable role in the region below $\phi^*<0.1$,
therefore it is not completely unexpected that \PYTHIA{6} and
\PYTHIA{8} give slightly different shapes. The fact that
non-perturbative effects introduce non-trivial changes on the shape of
this distribution was also noted in ref.~\cite{Banfi:2011dx} (for
predictions at the Tevatron). We also observe the same pattern shown
in ref.~\cite{Banfi:2011dx}, namely a moderate increase at $\phi^*\sim
0.05$ and a more pronounced decrease for very low values of $\phi^*$
when non-perturbative effects are included.  Finally, it is worth
mentioning that the results we have obtained (especially with
\PYTHIA{6}) clearly show a better agreement with data than what was
found in ref.~\cite{Aad:2011gj}, where different tunes were used, both
for \PYTHIA{6} and \PYTHIA{8}. It is difficult to draw a solid
conclusion, since in ref.~\cite{Aad:2011gj} \POWHEG{}-\Z{} (as opposed
to \ZJMINLO{}) was used, and events were also reweighted using
\noun{ResBos}~\cite{Balazs:1997xd}. Nevertheless it seems clear that
the best predictions are obtained when higher-order perturbative
corrections are included and modern tunes are used.

\subsection{W production} 
\label{subsec:Wdata}
In this section we compare our predictions to results of
refs.~\cite{Aad:2011fp,Aad:2013ueu}, and in particular we use the
combined decay of the $W$ to electrons and muons. The charged lepton
is required to have $\pt > 20$ GeV and rapidity $|y| < 2.4$. The event
must have a missing energy $p_{\rm T, miss} > 25$ GeV and the
transverse mass of the $W$ boson defined as $\mtw = \sqrt{2
  (p_{{\scriptscriptstyle \mathrm{T}},l}\, p_{{\scriptscriptstyle
      \mathrm{T}},miss} -\vec p_{{\scriptscriptstyle \mathrm{T}},l}
  \,\cdot \vec p_{{\scriptscriptstyle \mathrm{T}},miss}) }$ must be
above 40 GeV. As was the case for the Z studies, ATLAS only provides
normalised distributions for the standard observables we will show
here.

\begin{figure}[htb]
\begin{centering}
\includegraphics[clip,width=0.49\textwidth]{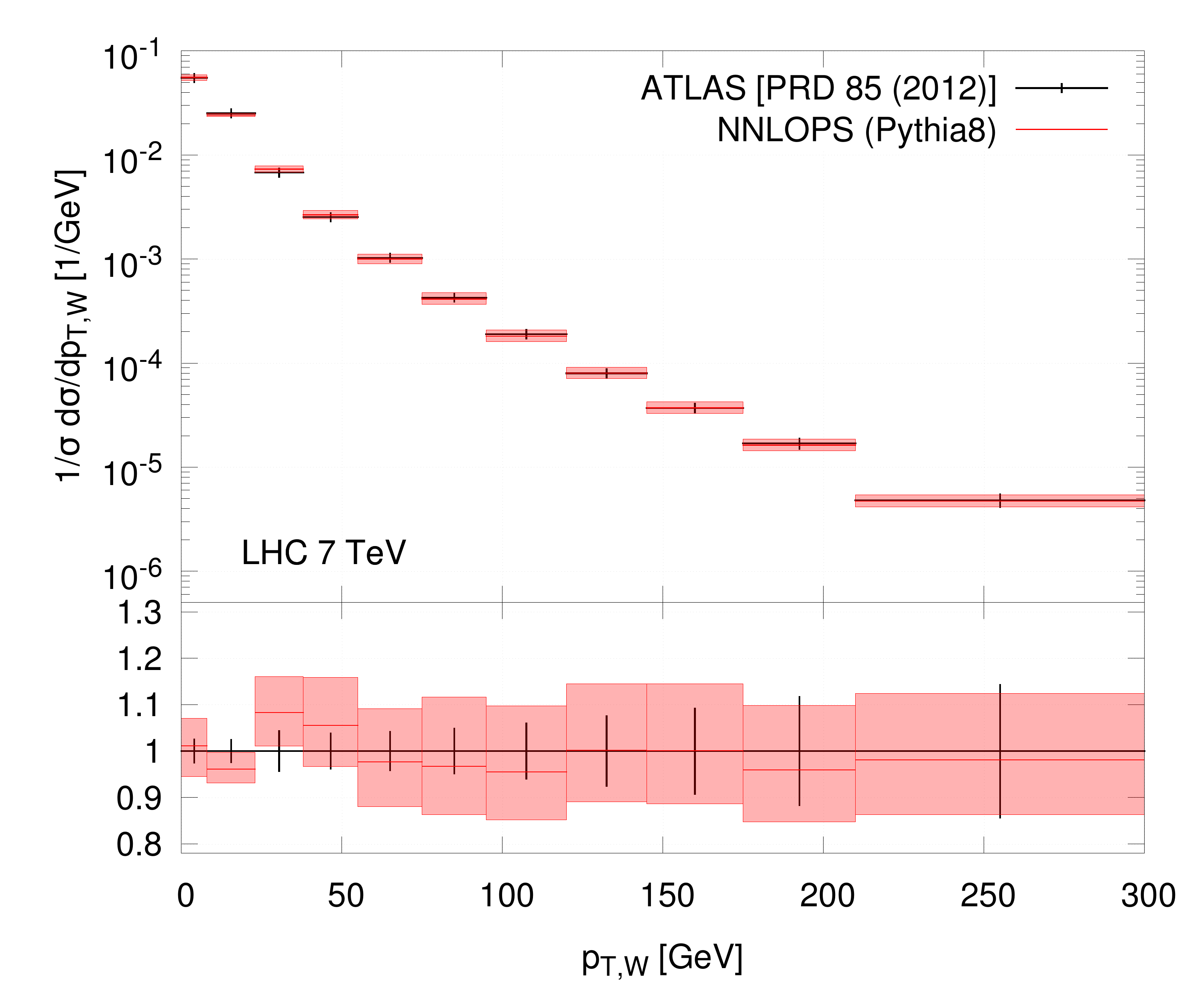}
\includegraphics[clip,width=0.49\textwidth]{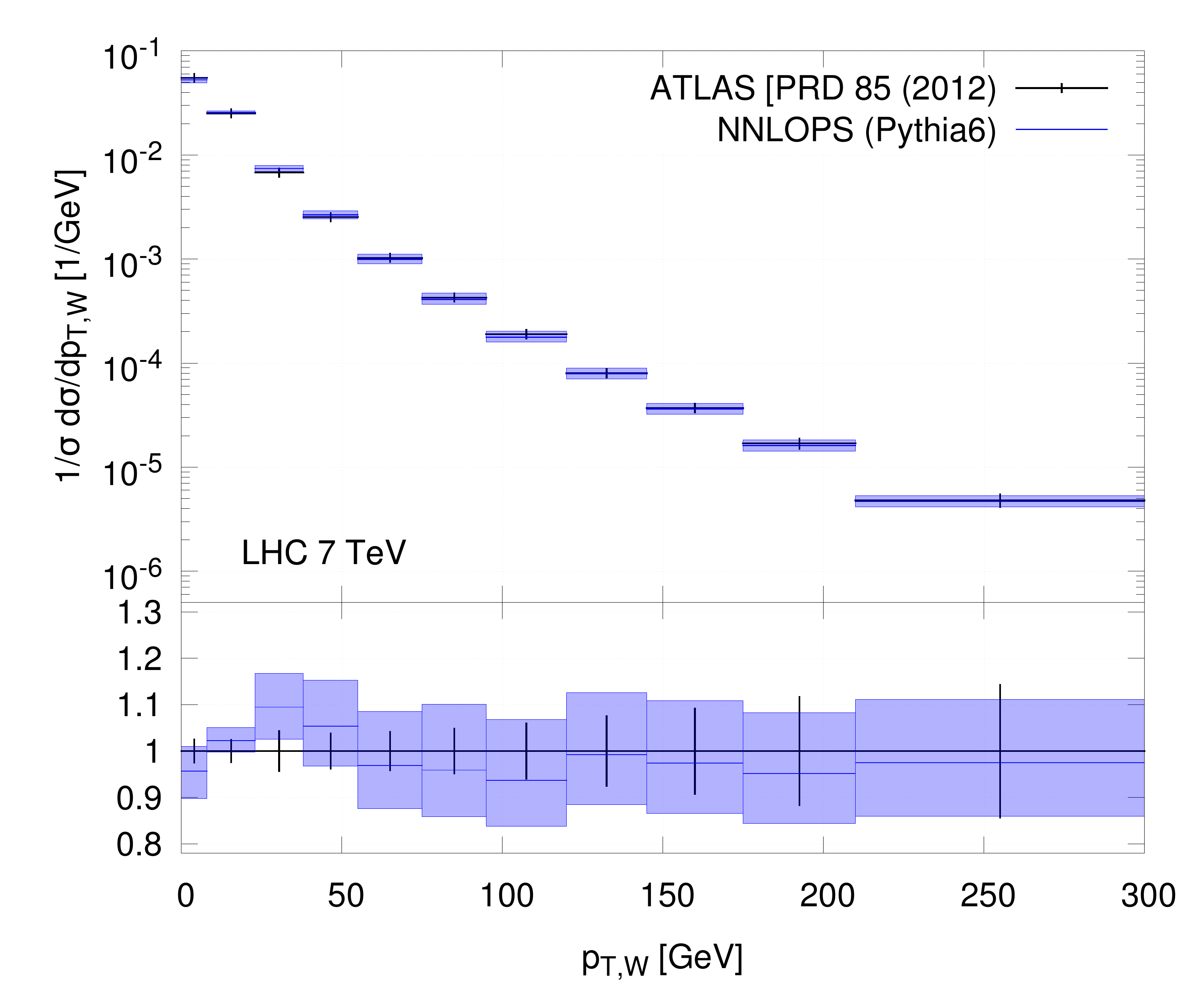}
\par\end{centering}
\caption{Comparison of \NNLOPS{} prediction (red) to data (black) from
  ref.~\cite{Aad:2011fp} for the $W$ boson $\ptw$ distribution, using
  \PYTHIA{8} (left) and \PYTHIA{6} (right).}
\label{fig:data-ptw}
\end{figure}

We start by showing in Fig.~\ref{fig:data-ptw} the transverse momentum
of the $W$ boson $\ptw$. As expected, differences in the parton shower
algorithm only play a visible role in the small $\pt$ region, where
minor differences between \PYTHIA{6} and \PYTHIA{8} can be
observed. For high $\ptw$, the two predictions are consistent with
each other, and agree quite well with data.

We also compared our predictions with the analysis performed by the
ATLAS collaboration in ref.~\cite{Aad:2013ueu}. We show results for
$\kt$-splitting scales in $W+$jets events. These observables are
defined as the smallest distances found by the (inclusive)
$\kt$-algorithm at each step in the clustering sequence. The splitting
scale $d_k$ is the smallest among all distances found by the algorithm
when going from $(k+1)$ to $k$ objects. Therefore $\sqrt{d_0}$
corresponds to the transverse momentum of the leading jet, whereas
$\sqrt{d_1}$ is the smallest distance among pseudo-jets when
clustering from $2$ to $1$ jet:
\begin{equation}
  \label{eq:d0}
  d_1=\min(d_{1B},d_{2B},d_{12})\,,
\end{equation}
where
\begin{eqnarray}
  d_{iB} & = & p^2_{\scriptscriptstyle \mathrm{T,i}}\,, \nonumber \\
  d_{ij} & = &\min(p^2_{\scriptscriptstyle \mathrm{T,i}},p^2_{\scriptscriptstyle \mathrm{T,j}})\frac{(\Delta R_{ij})^2}{R^2}\,,
\end{eqnarray}
are the usual distances used in the $\kt$-algorithm. Among other
reasons, these observables are interesting because they can be used as
a probe of the details of matching and merging schemes. Due to the
underlying \ZJMINLO{} simulation, our \NNLOPS{} prediction is NLO
accurate for large values of $\ptjone$, and it is at least LL accurate in
describing the $1\to 0$ jet transition, which is measured in the $d_0$
distribution. The second jet spectrum and the $2\to 1$ jet transition
(which is encapsulated in $d_1$) are instead described at LO+LL, due to
the underlying \POWHEG{} simulation. Since the definition of $d_1$
contains $d_{12}$, this observable is a measure of the internal
structure of the first jet, and not only of the second jet transverse
momentum.

\begin{figure}[htb]
\begin{centering}
\includegraphics[clip,width=0.49\textwidth]{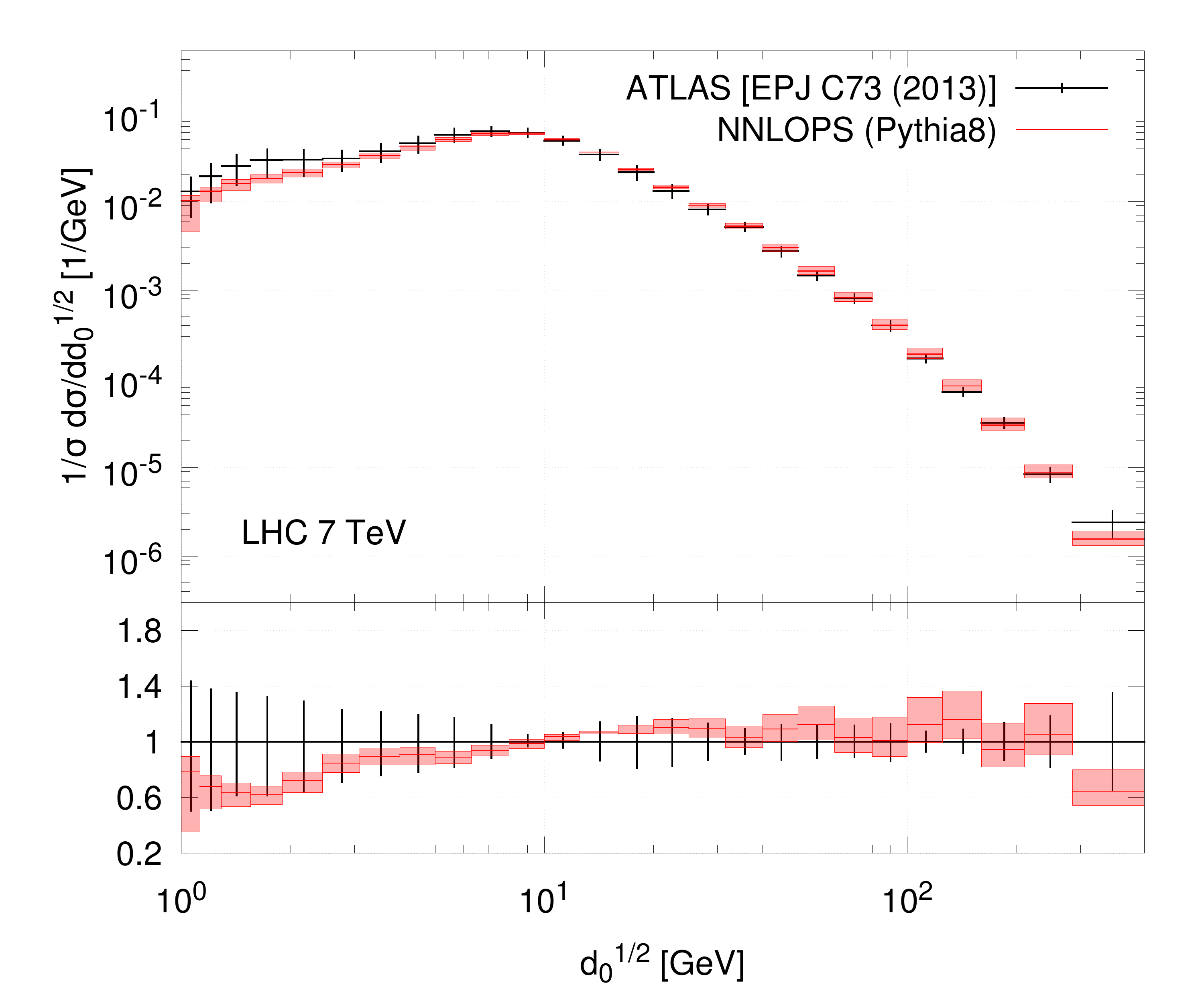}
\includegraphics[clip,width=0.49\textwidth]{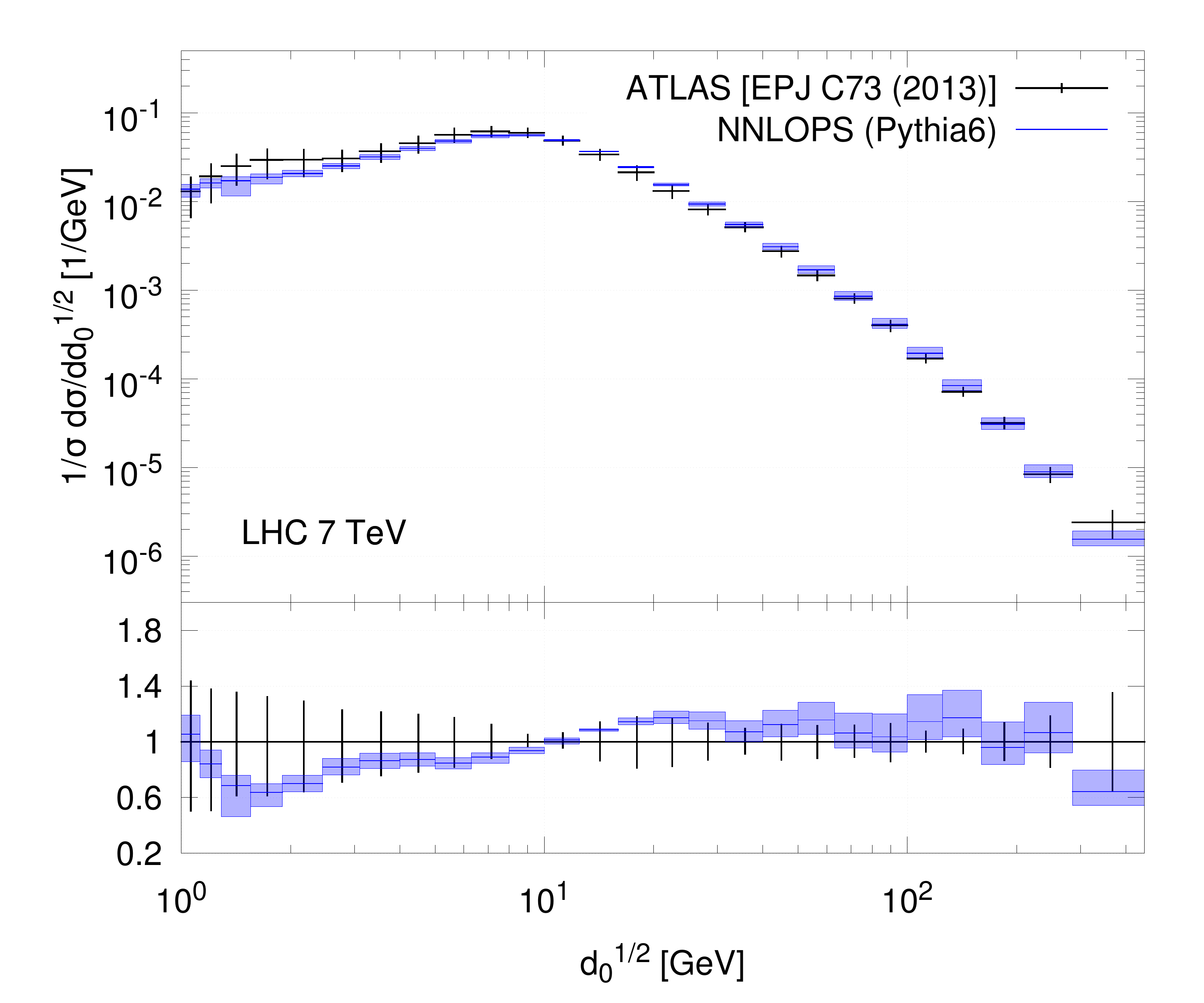}
\par\end{centering}
\caption{Comparison of \NNLOPS{} prediction (red) to 7 TeV LHC data (black) from
  ref.~\cite{Aad:2013ueu} for the $W$ boson $\kt$ splitting scale
  $\sqrt{d_0}$ as defined in the text using \PYTHIA{8} (left) and
  \PYTHIA{6} (right).}
\label{fig:data-d0}
\end{figure}

\begin{figure}[htb]
\begin{centering}
\includegraphics[clip,width=0.49\textwidth]{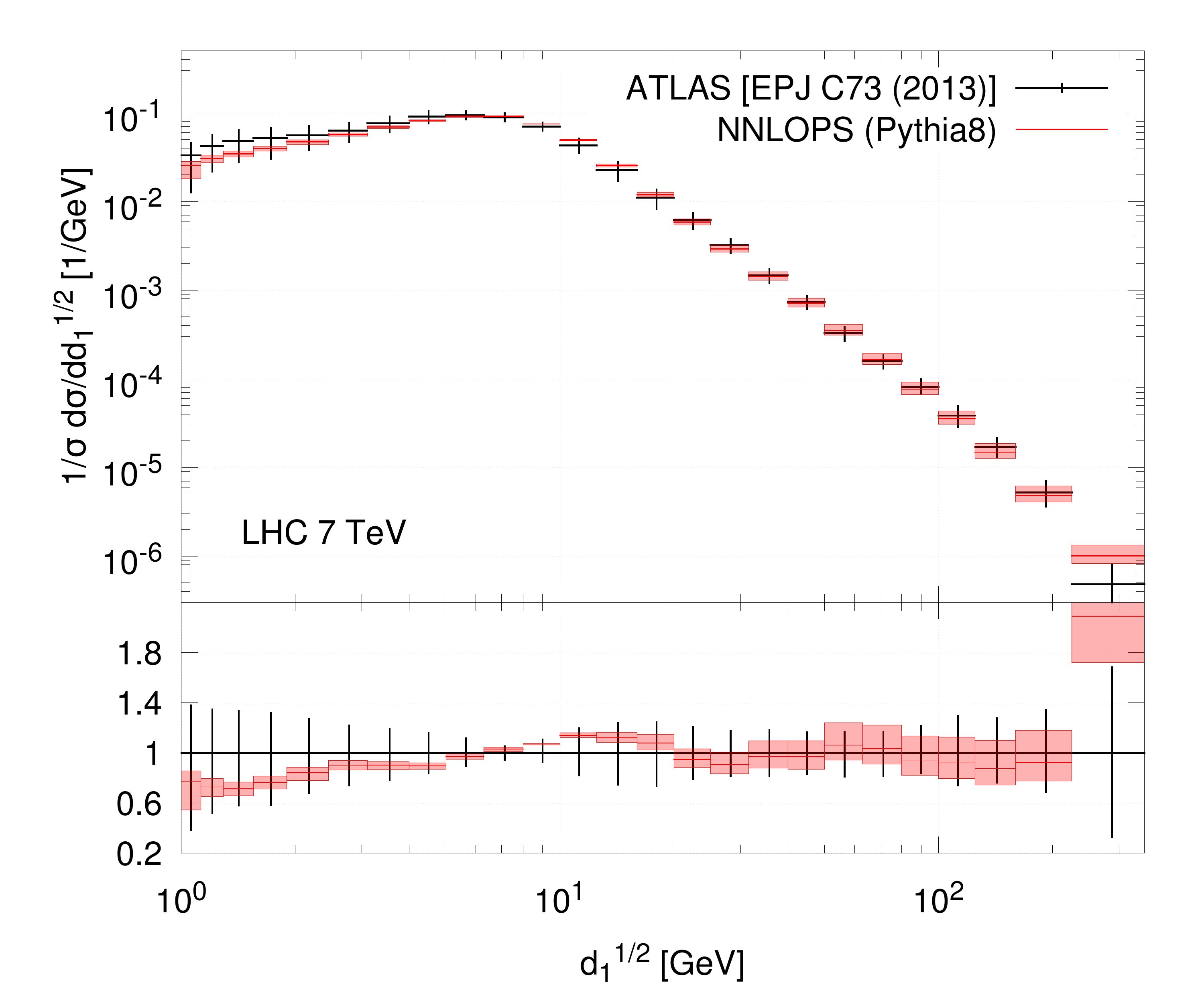}
\includegraphics[clip,width=0.49\textwidth]{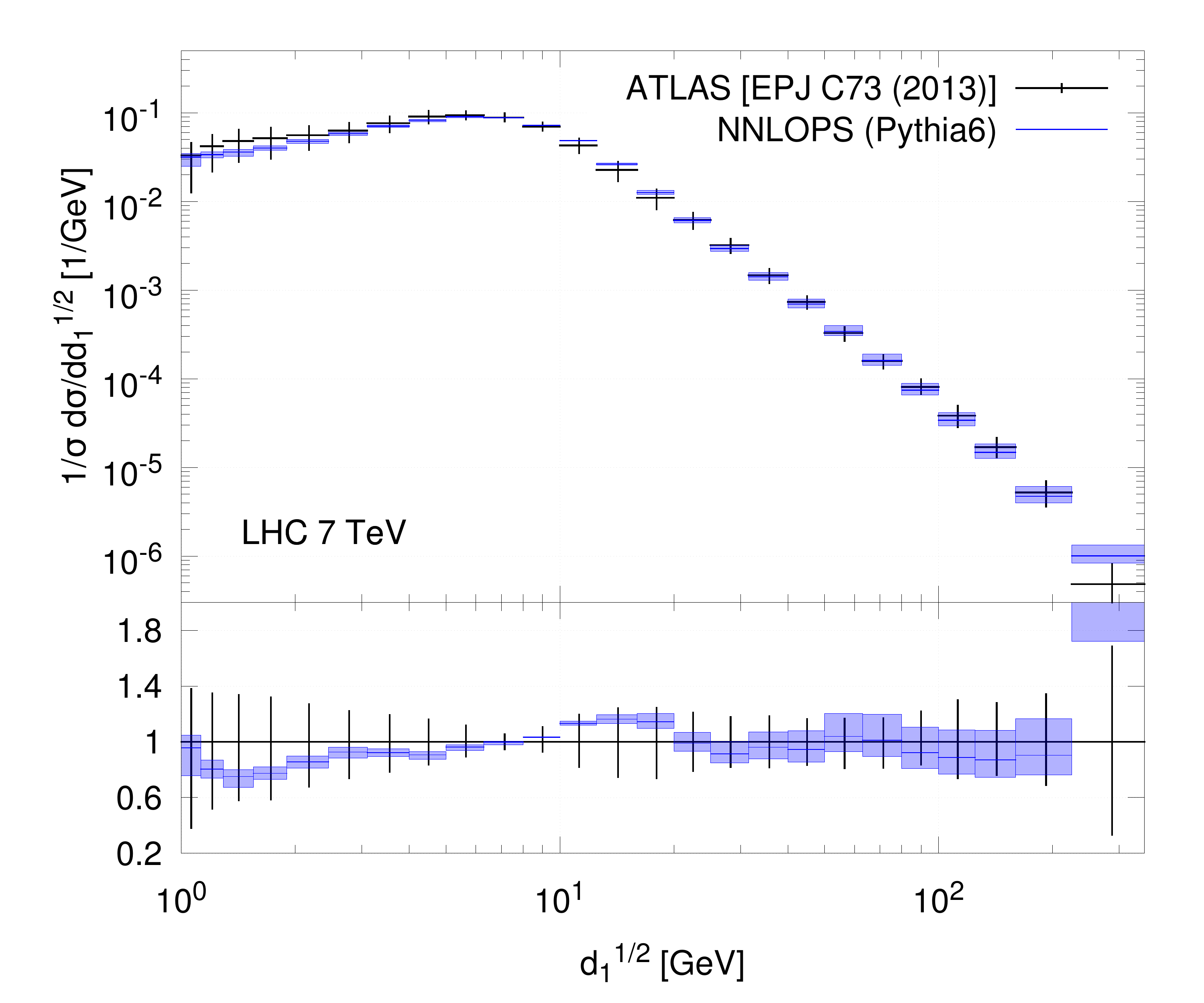}
\par\end{centering}
\caption{Comparison to 7 TeV LHC data from ref.~\cite{Aad:2013ueu} for the $W$
  boson $\kt$ splitting scale $\sqrt{d_1}$ as defined in
  eq.~\eqref{eq:d0} using \PYTHIA{8} (left) and \PYTHIA{6} (right).}
\label{fig:data-d1}
\end{figure}

In Figs.~\ref{fig:data-d0} and~\ref{fig:data-d1} 
we show our \NNLOPS{}
predictions against ATLAS data, using as jet radius $R=0.6$. We find good
agreement, especially when $\sqrt{d_i}>10$ GeV. Below this value, we
are still compatible with the experimental uncertainty bands, although
we are systematically lower than data. Once more, one should consider
that the region below $5-10$ GeV will be affected also by
non-perturbative effects. For large values of $d_i$ we are instead
sensitive to the level of accuracy that we reach in describing hard
emissions. In this respect, it is no surprise that we have a better
agreement with data than the \POWHEG{} results shown in
ref.~\cite{Aad:2013ueu}, where $d_1$ is poorly described since the
second emission is only described in the shower approximation. NLO
corrections to the $W+1$ jet region are included in the \NNLOPS{}
simulation, and are very likely the reason why we have a description
of $d_0$ that is better than what was observed in
ref.~\cite{Aad:2013ueu}.

\begin{figure}[htb]
\begin{centering}
\includegraphics[clip,width=0.49\textwidth]{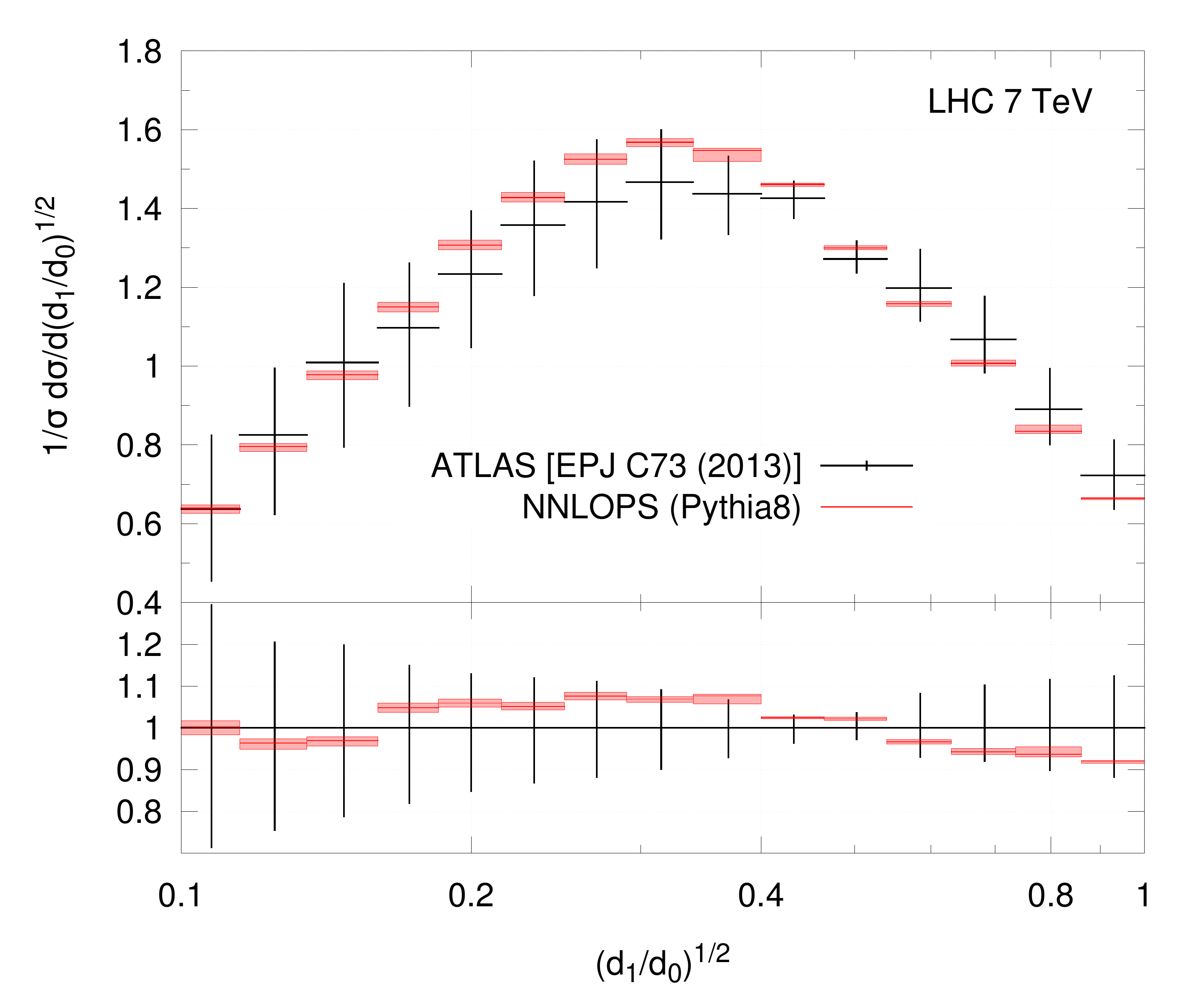}
\includegraphics[clip,width=0.49\textwidth]{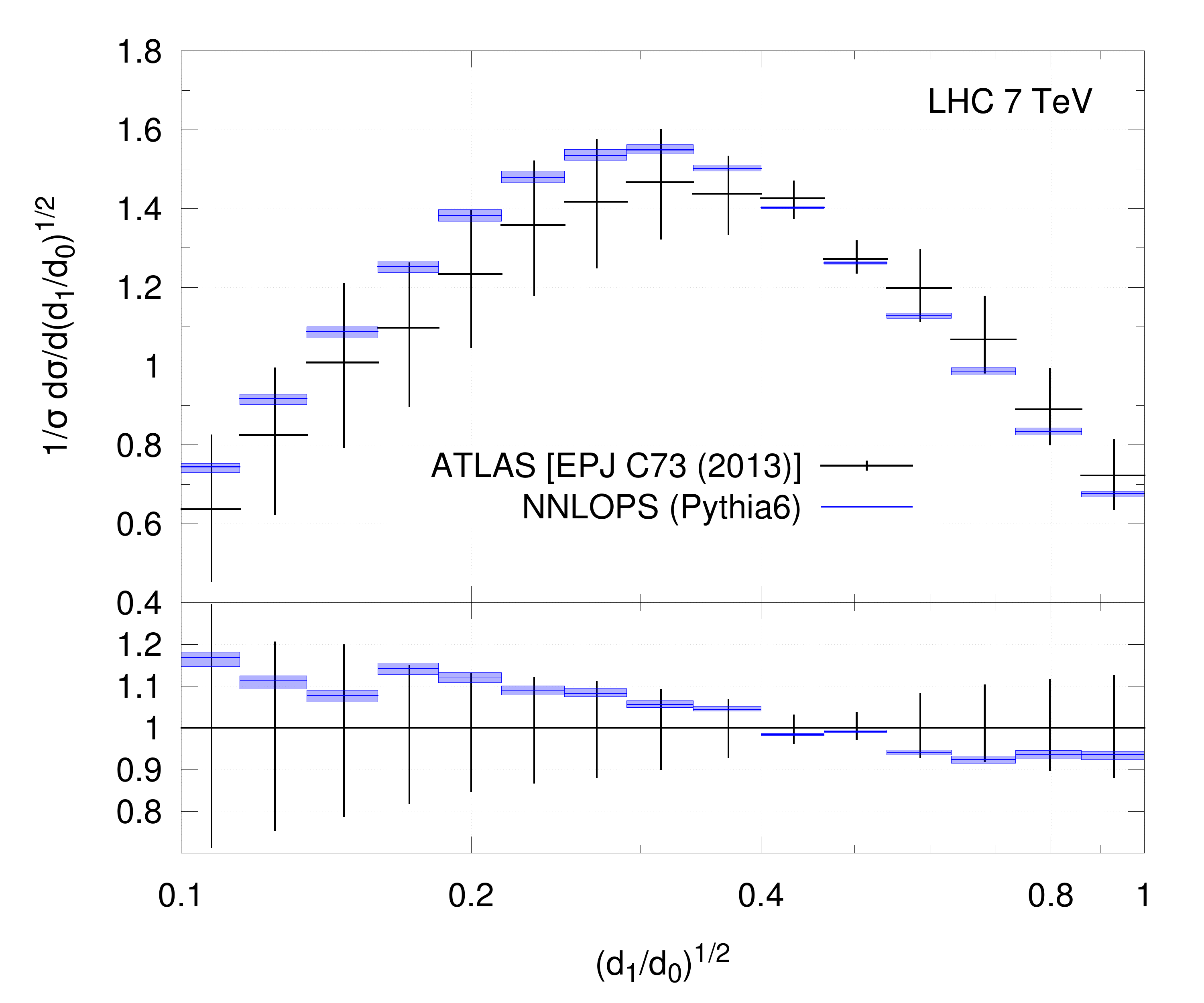}
\par\end{centering}
\caption{Comparison to 7 TeV LHC data from ref.~\cite{Aad:2013ueu} for
  the $W$ boson ratio of the $\kt$ splitting scales $\sqrt{d_0}$ and
  $\sqrt{d_1}$ using \PYTHIA{8} (left) and \PYTHIA{6} (right).}
\label{fig:data-d1d0}
\end{figure}

Finally in Fig.~\ref{fig:data-d1d0} 
we show the distribution for the
ratio $d_1/d_0$, for events with $\sqrt{d_0}>20 \mbox{ GeV}$.  Due to
the ratio nature of this quantity, a simultaneous over- or
underestimation in predicting $d_1$ and $d_0$ should be partially
compensated when plotting $d_1/d_0$. It is therefore no surprise that
the agreement with data is better than in Figs.~\ref{fig:data-d0}
and~\ref{fig:data-d1}.

\subsection{W and Z polarization}
\label{subsec:polarization}
Recently both ATLAS~\cite{ATLAS:2012au} and
CMS~\cite{Chatrchyan:2011ig} have published results on the
polarization of the $W$ boson at 7 TeV confirming the Standard Model
prediction, that $W$ bosons are mostly left-handed in $pp$ collisions
at large transverse momenta~\cite{Bern:2011ie}. Knowledge about the
$W$ boson polarization is important, as it provides a discriminant in
searches for new physics.

We first very briefly review how to measure the polarization in terms of
angular coefficients but refer the reader to the literature for a
complete description of the
topic~\cite{Bern:2011ie,Collins:1977iv,Lam:1980uc,Hagiwara:1984hi,Mirkes:1992hu,Mirkes:1994eb,Mirkes:1994dp,Hagiwara:2006qe}. Here
we will follow the derivation of~\cite{Bern:2011ie}. We then continue
to compare ATLAS data~\cite{ATLAS:2012au} to our \NNLOPS{} results for the
$W$ boson polarization and present predictions for the angular
coefficients for the $Z$ boson at 8 TeV.

The vector boson cross-section can be expanded in terms of $\cos{\theta^*}$ and $\phi^*$ as 
\begin{align}
  \frac{1}{\sigma}\frac{d\sigma}{d(\cos{\theta^*})d\phi^*} = &\frac{3}{16\pi}\Bigl[(1+\cos^2\theta^*)+A_0 \frac{1}{2}(1-3\cos^2{\theta^*})+A_1\sin 2\theta^*\cos\phi^* \notag \\
& + A_2\frac{1}{2}\sin^2\theta^*\cos 2\phi^* + A_3\sin\theta^*\cos\phi^* +A_4\cos\theta^* \notag \\ 
& + A_5\sin\theta^*\sin\phi^* + A_6\sin 2\theta^*\sin\phi^* + A_7\sin^2\theta^*\sin 2\phi^*\Bigr]\,, 
\label{eq:polexp}
\end{align}
where the polar angle $\theta^*$ and the azimuthal angle $\phi^*$ are
defined in some particular rest frame of the dilepton system. Here we
will make use of two different frames, the Collins-Soper
frame\footnote{One defines in the laboratory frame the two beam
  directions by $\vec{b}_{+}=(0,0,1;1)$ and
  $\vec{b}_{-}=(0,0,-1;1)$. After boosting to the dilepton centre of
  mass frame, $\mathcal{O}^{'}$, one defines the z-axis as the
  bisector of $\vec{b}^{'}_{+}$ and $-\vec{b}^{'}_{-}$ such that the
  z-axis points into the hemisphere of the $Z$ boson direction (in the
  lab frame). One then defines a q-axis lying in the plane spanned by
  the $\vec{b}^{'}_{+}$ and $\vec{b}^{'}_{-}$ vectors, orthogonal to
  the z-axis and pointing in the direction opposite to
  $\vec{b}^{'}_{+}+\vec{b}^{'}_{-}$.  $\theta^*$ is now defined with
  respect to the z-axis and $\phi^*$ with respect to the q-axis.}
\cite{Collins:1977iv} for the $Z$ boson and the helicity frame for the
$W$ boson defined as the dilepton rest frame with the z-axis pointing
along the direction of flight of the $W$ boson in the lab frame. The
cross-section can be differential in any quantity that does not depend
on the individual lepton kinematics.  The angular coefficients can
then be expressed in terms of expectation values defined as
\begin{align}
<f(\theta^*,\phi^*)> = \int_{-1}^{1}d(\cos\theta^*)\int_0^{2\pi}d\phi^*\frac{1}{\sigma}\frac{d\sigma}{d(\cos{\theta^*})d\phi^*}f(\theta^*,\phi^*)
\label{eq:expval}
\end{align} 
 by
\begin{align}
A_0&=4-<10\cos^2\theta^*>, &\quad A_1=&<5\sin 2\theta^*\cos\phi^*>, &\quad  A_2=&<10\sin^2\theta^*\cos 2\phi^*>\,, \notag \\
A_3&=<4\sin\theta^*\cos\phi^*>, &  A_4=&<4\cos\theta^*>,\phantom{\sin\theta^*} & A_5=&<5\sin^2\theta^*\sin 2\phi^*>\,, \notag \\
A_6&=<5\sin 2\theta^*\sin\phi^*>, & A_7=&<4\sin\theta^*\sin\phi^*>. &
\label{eq:angcoe}
\end{align}
For both $W$ and $Z$ production it is known that at
$\mathcal{O}(\alpha_s)$ $A_5=A_6=A_7=0$ and that the higher-order
corrections are small. We have checked that with 20 million events
processed the coefficients do not deviate significantly from zero and
we have therefore chosen not to show them here. It is also interesting
to notice that as a consequence of the spin-1 structure of the gluon,
the Lam-Tung relation $A_0=A_2$\footnote{For $q\bar{q}$ initiated
  production the relation is exact to all orders. For $qg$ initiated
  production it is violated at the NLO level~\cite{Bern:2011ie}. This
  is independent of the frame.} holds at LO~\cite{Lam:1978zr}. As we
will see, the deviations from this lowest-order result can become
quite large, $\mathcal{O}(20\%)$, in the $\ptz>10 \mbox{ GeV}$ region.
From the angular coefficients we then define the left, $f_L$, right,
$f_R$, and longitudinal, $f_0$, polarization fractions of the $W^\pm$
as
\begin{align}
f_L=\frac{1}{4}(2-A_0\mp A_4), \quad f_R=\frac{1}{4}(2-A_0\pm A_4), \quad f_0=\frac{1}{2}A_0\,.
\label{eq:polfra}
\end{align}
It is clear from these equations that the polarization fractions are
normalised such that $f_L+f_R+f_0=1$ and it is therefore sufficient to
show results for $f_L-f_R$ and $f_0$. In this way we also separate the
$A_0$ and $A_4$ dependence. In Table \ref{table:Wpol} we show our
prediction for $f_L-f_R$ and $f_0$ for combined $W^+$ and $W^-$
production at 7 TeV compared to ATLAS data \cite{ATLAS:2012au} for two
different $\pt$ regions. We find good agreement between the data and
predictions - three of them at the $1\sigma$ level and at the
$2\sigma$ level for $f_L-f_R$ in the low $\pt$ range. It should be
noted that the measurements are dominated by systematics
uncertainties.
  \begin{table}[htb]
    \centering    
    \scalebox{0.8}{\begin{tabular} {l c c | c c}
          &  \multicolumn{2}{c}{$35\mbox{ GeV} < \ptw <50\mbox{ GeV} $} & \multicolumn{2}{c}{$\ptw >50\mbox{ GeV} $} \\
        \toprule
        & $f_L - f_R$ & $f_0$ &$f_L - f_R$ & $f_0$ \\
        \midrule
        Data &$0.238\pm 0.02 \pm 0.034$ & $0.219 \pm 0.033\pm 0.134$ & $0.252\pm 0.017\pm 0.034$&$0.127\pm 0.03\pm 0.108 $\\
        \NNLOPS{} &$0.317\pm0.002$ & $0.198\pm 0.004$&$0.289\pm 0.004$ &$0.214\pm 0.009$ \\
        \bottomrule
    \end{tabular}}
  \caption{A comparison between combined $W^+$ and $W^-$ at 7 TeV
    ATLAS data \cite{ATLAS:2012au} and our \NNLOPS{} prediction for
    $f_L-f_R$ and $f_0$ as defined in eq.~\eqref{eq:polfra} and. For
    data the first uncertainty is statistical and the second one
    systematic. For the theoretical prediction the error is purely
    statistical. Except for the $\ptw$ cut the only cut imposed on the
    theoretical prediction is a transverse mass cut of $\mtw > 40
    \mbox{ GeV}$. The data and predictions are in good agreement.}
    \label{table:Wpol}
  \end{table}

\begin{figure}[htb]
\begin{centering}
\includegraphics[clip,width=0.49\textwidth]{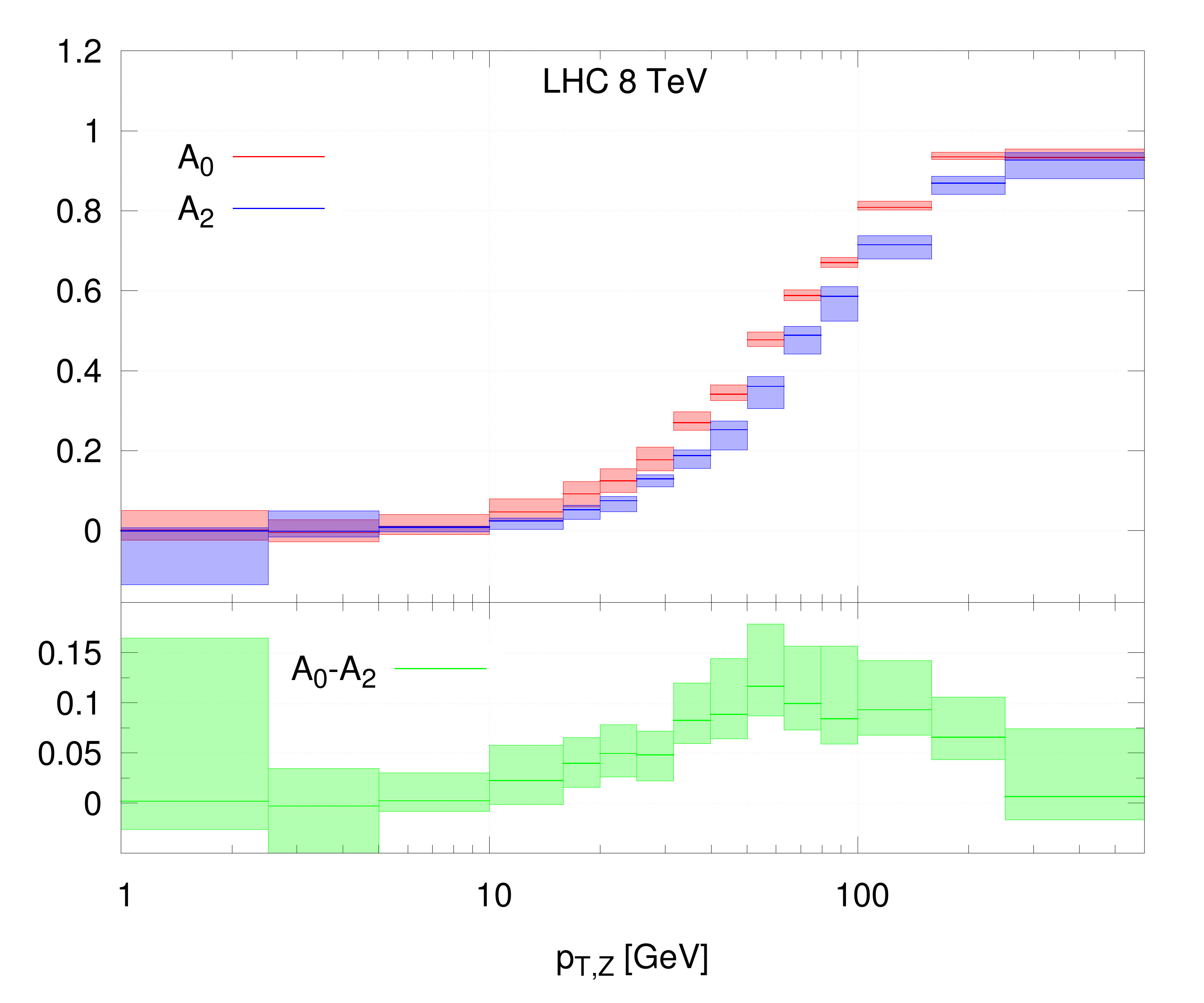}
\includegraphics[clip,width=0.49\textwidth]{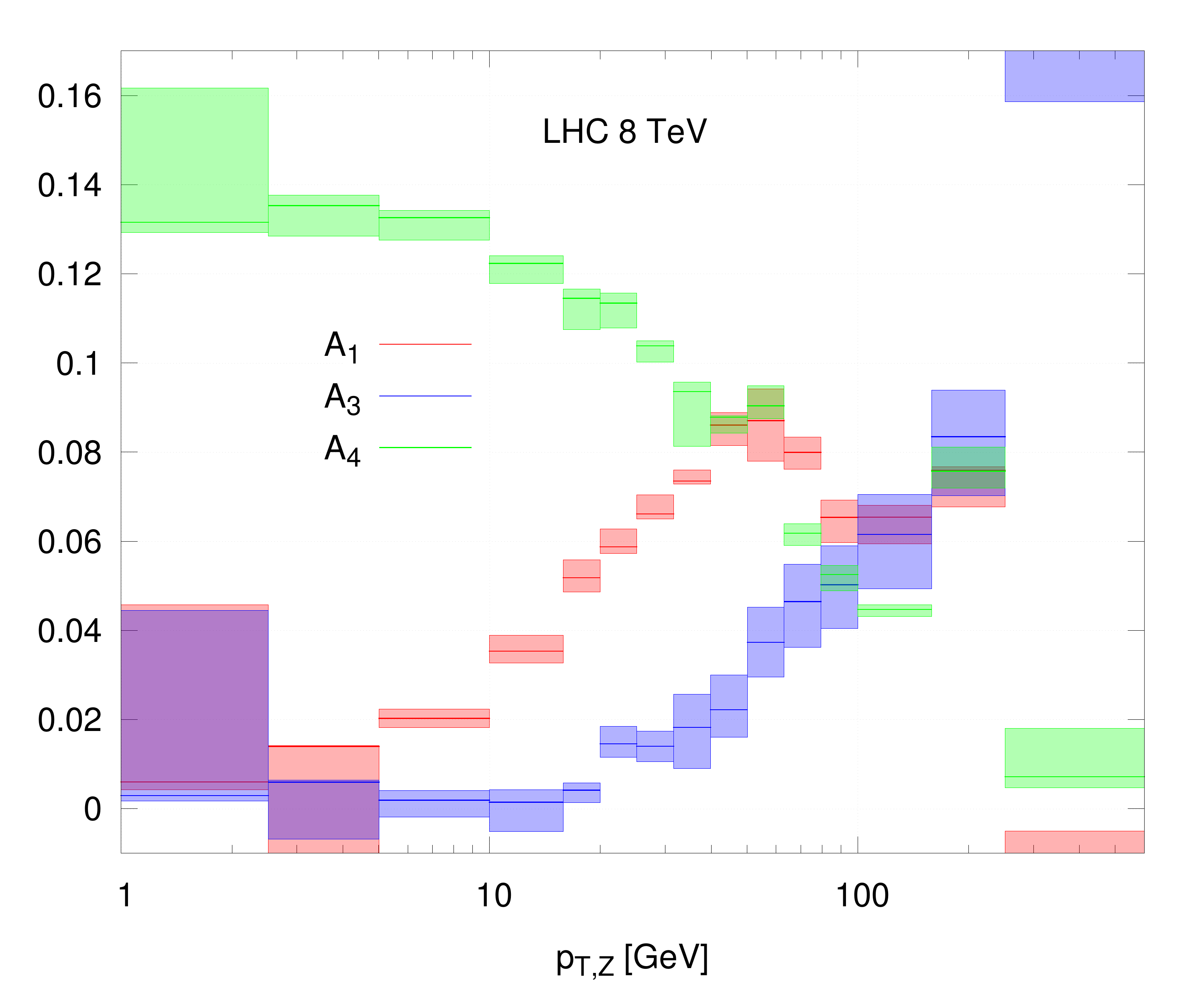}
\par\end{centering}
\caption{Predictions for angular coefficients eq.~\eqref{eq:angcoe} in
  the Collins-Soper frame for $Z$ production at 8 TeV. In the left
  panel we show $A_0$ (red) and $A_2$ (blue) along with the difference
  $A_0-A_2$ (green) and in the right panel we plot $A_1$ (red), $A_3$
  (blue) and $A_4$ (green).}
\label{fig:angcoe}
\end{figure}

  There is currently no public measurement for the angular
  coefficients for the $Z$ boson in $pp$ collisions. Previously
  measurements of $A_0,A_2,A_3$ and $A_4$ in $p\bar{p}$ collisions
  have been published by the CDF collaboration~\cite{Aaltonen:2011nr}
  and anticipating that such an analysis might be carried out at the
  LHC at 8 TeV we present here our predictions for the angular
  coefficients as defined in eq.~\eqref{eq:angcoe}, as a function of
  the boson transverse momentum. In the left panel of
  Fig.~\ref{fig:angcoe} we plot $A_0$ (red) and $A_2$ (blue) along
  with the difference $A_0-A_2$,
while in the right panel we plot $A_1,A_3$ and $A_4$.
For the two coefficients $A_0$ and $A_2$ we find that they are in very
good agreement for low transverse momentum, $\ptz < 10 \mbox{ GeV}$.
For higher transverse momenta deviations start to appear peaking
around 60 GeV.  For the three other coefficients, we observe that the
$\pt$ dependence is strongest for larger transverse momenta. We can
define polarization fractions for the $Z$ boson analogously to those
defined in eq.~\eqref{eq:polfra} by~\cite{Stirling:2012zt}
\begin{align}
f_L=\frac{1}{4}(2-A_0-\alpha A_4)\,, \quad f_R=\frac{1}{4}(2-A_0+\alpha A_4)\,, \quad f_0=\frac{1}{2}A_0\,,
\label{eq:polfraZ}
\end{align}
where $\alpha=\frac{c_L^2-c_R^2}{c_L^2+c_R^2}$, $c_L$ is the coupling
to the left-handed lepton and $c_R$ is the coupling to the
right-handed lepton. As for the $W$ boson these are normalised such
that $f_L+f_R+f_0=1$ and  we have in this case that $f_L-f_R=-\frac{1}{2}\alpha A_4$. Using
$c_L=\sin^2\theta_W-\frac{1}{2}$ and $c_R=\sin^2\theta_W$ we get
$\alpha\approx 0.032$. Comparing with Fig.~\ref{fig:angcoe} we see
that at low $\pt$ the longitudinal polarization of the $Z$ boson is
highly suppressed and it is produced almost equally between the left-
and right-handed polarization. For high $\pt$ the longitudinal
polarization starts to dominate.

\section{Conclusions} 
\label{sec:Conclusions}
In this paper we presented an implementation for $Z\to l^+l^-$ and
$W^\pm\to l^\pm \nu $ production at NNLO level including parton shower
effects. At the core of this method lies the fact that \ZJMINLO{} and
\WJMINLO{} simulations achieve NLO accuracy also for fully inclusive
distributions, \emph{i.e.}~once the jet is integrated out. In the case
of Higgs production considered recently in
ref.~\cite{Hamilton:2013fea}, it was enough to rescale the NLOPS
results to reproduce the NNLO Higgs rapidity spectrum, thereby
achieving NNLOPS accuracy. In the present case instead, to properly
take into account the decay of the boson to leptons, NNLOPS accuracy
is reached by performing a three-dimensional rescaling of the events
generated by \ZJMINLO{} and \WJMINLO{} using NNLO distributions as
computed, for instance, with \DYNNLO{}. This implies that the
calculation is numerically more intensive.

We have validated our procedure by considering observables typically
used to study Drell-Yan processes. We have found extremely good
agreement with NNLO results for observables fully inclusive over QCD
radiation, and all the features expected in a computation matched with
parton showers for more exclusive observables. We have also compared
our \NNLOPS{} predictions to state-of-the-art analytic resummation for
observables sensitive to soft-collinear QCD radiation, which are not
described accurately by a fixed-order NNLO calculation. Despite the
logarithmic accuracy of our simulation cannot be claimed to reach the
same precision as these analytic resummations, we have found
reasonably good agreement for the vector boson transverse momentum.
Slightly more pronounced discrepancies were instead observed for
$\phi^*$ and for the jet-veto efficiency for large values of the jet
radius $R$.

We have also compared with a number of available experimental results,
not only for fully inclusive observables but also for $\phi^*$, the
transverse momentum of the vector boson, as well as $\kt$ splitting
scales. The successful outcome of these comparisons is an indication
that our computation can be used as a state-of-the art prediction for
future studies where a simultaneous inclusion of NNLO corrections and
parton shower effects are needed. We have also illustrated how our
generator can be used to study the polarization of $W$ and $Z$ bosons
produced in Drell-Yan events.

We finally remark that we did not include in this work electroweak
corrections. These corrections are known exactly at one
loop~\cite{Dittmaier:2001ay,Dittmaier:2009cr} and can be included, at
${\cal O}(\alpha_{ew})$ either additively or multiplicative on top of
the QCD corrections included here. Version 3 of the program
\FEWZ{}~\cite{Li:2012wna} includes directly both EW and QCD
corrections, however we could not use it to generate the three
dimensional distributions needed here.

The code developed for this paper will soon be made available within
the \POWHEGBOX{}-{\ttfamily V2} public repository. By parallelising on
a cluster the runs needed to obtain the necessary inputs for the
``NLO-to-NNLO'' reweighter, results can be obtained in a reasonably
short timescale. Nevertheless, as mentioned above, running this
simulation from scratch requires more computing power than what was
needed for the Higgs case. Therefore, together with the code that
allows one to compute all the ingredients to achieve NNLOPS accuracy,
we find it useful to release also the distributions $d\sigma^{\rm
  NNLO}/d\Phi_B$ for $W$ and $Z$ production for three reference scales
$\mur=\muf= K M_V$ with $K=0.5,1,2.0$ ($M_V$ denotes the mass of the
vector boson) and also provide the $d\sigma^{\rm MINLO}_A/d\Phi_B$ and
$d\sigma^{\rm MINLO}_B/d\Phi_B$ distributions for 7 reference scales,
as discussed in the text. In these cases, NNLOPS accuracy can then be
reached by just running \POWHEG{} with \MINLO{} switched on, and
processing the generated events with the ``NLO-to-NNLO'' reweighter,
using as input the aforementioned distributions.

\section*{Acknowledgements}
We thank Andrea Banfi, Keith Hamilton, Pier Francesco Monni, Paolo Nason, 
Elzbieta Richter-Was, Gavin Salam and Peter Skands for
useful discussions. We are also grateful to Giancarlo Ferrera and
Massimiliano Grazzini for providing us with a preliminary version of
\DYQT{} and to Andrea Banfi and Lee Tomlinson for providing us with
their resummed results for $\phi^*$. AK and ER thank CERN for
hospitality while this work was being finalized.
This research was supported by the ERC grant 614577 ``HICCUP -- High
Impact Cross Section Calculations for Ultimate Precision''.  AK is
supported by the British Science and Technology Facilities Council and
by the Buckee Scholarship at Merton College.

\bibliographystyle{JHEP} \bibliography{dynnlops}

\providecommand{\href}[2]{#2}\begingroup\raggedright\begin{thebibliography}{10}

\bibitem{Ball:2013hta}
{\bf NNPDF} Collaboration, R.~D. Ball et~al., {\it {Parton distributions with
  QED corrections}},  {\em Nucl.Phys.} {\bf B877} (2013), no.~2 290--320,
  [\href{http://xxx.lanl.gov/abs/1308.0598}{{\tt arXiv:1308.0598}}].

\bibitem{Anastasiou:2003yy}
C.~Anastasiou, L.~J. Dixon, K.~Melnikov, and F.~Petriello, {\it {Dilepton
  rapidity distribution in the Drell-Yan process at NNLO in QCD}},  {\em
  Phys.Rev.Lett.} {\bf 91} (2003) 182002,
  [\href{http://xxx.lanl.gov/abs/hep-ph/0306192}{{\tt hep-ph/0306192}}].

\bibitem{Melnikov:2006kv}
K.~Melnikov and F.~Petriello, {\it {Electroweak gauge boson production at
  hadron colliders through O(alpha(s)**2)}},  {\em Phys.Rev.} {\bf D74} (2006)
  114017, [\href{http://xxx.lanl.gov/abs/hep-ph/0609070}{{\tt
  hep-ph/0609070}}].

\bibitem{Catani:2009sm}
S.~Catani, L.~Cieri, G.~Ferrera, D.~de~Florian, and M.~Grazzini, {\it {Vector
  boson production at hadron colliders: A Fully exclusive QCD calculation at
  NNLO}},  {\em Phys.Rev.Lett.} {\bf 103} (2009) 082001,
  [\href{http://xxx.lanl.gov/abs/0903.2120}{{\tt arXiv:0903.2120}}].

\bibitem{Gavin:2010az}
R.~Gavin, Y.~Li, F.~Petriello, and S.~Quackenbush, {\it {FEWZ 2.0: A code for
  hadronic Z production at next-to-next-to-leading order}},  {\em
  Comput.Phys.Commun.} {\bf 182} (2011) 2388--2403,
  [\href{http://xxx.lanl.gov/abs/1011.3540}{{\tt arXiv:1011.3540}}].

\bibitem{Baur:1997wa}
U.~Baur, S.~Keller, and W.~Sakumoto, {\it {QED radiative corrections to $Z$
  boson production and the forward backward asymmetry at hadron colliders}},
  {\em Phys.Rev.} {\bf D57} (1998) 199--215,
  [\href{http://xxx.lanl.gov/abs/hep-ph/9707301}{{\tt hep-ph/9707301}}].

\bibitem{Baur:2001ze}
U.~Baur, O.~Brein, W.~Hollik, C.~Schappacher, and D.~Wackeroth, {\it
  {Electroweak radiative corrections to neutral current Drell-Yan processes at
  hadron colliders}},  {\em Phys.Rev.} {\bf D65} (2002) 033007,
  [\href{http://xxx.lanl.gov/abs/hep-ph/0108274}{{\tt hep-ph/0108274}}].

\bibitem{Dittmaier:2001ay}
S.~Dittmaier and .~Kramer, Michael, {\it {Electroweak radiative corrections to
  W boson production at hadron colliders}},  {\em Phys.Rev.} {\bf D65} (2002)
  073007, [\href{http://xxx.lanl.gov/abs/hep-ph/0109062}{{\tt
  hep-ph/0109062}}].

\bibitem{CarloniCalame:2007cd}
C.~Carloni~Calame, G.~Montagna, O.~Nicrosini, and A.~Vicini, {\it {Precision
  electroweak calculation of the production of a high transverse-momentum
  lepton pair at hadron colliders}},  {\em JHEP} {\bf 0710} (2007) 109,
  [\href{http://xxx.lanl.gov/abs/0710.1722}{{\tt arXiv:0710.1722}}].

\bibitem{Arbuzov:2007db}
A.~Arbuzov, D.~Bardin, S.~Bondarenko, P.~Christova, L.~Kalinovskaya, et~al.,
  {\it {One-loop corrections to the Drell--Yan process in SANC. (II). The
  Neutral current case}},  {\em Eur.Phys.J.} {\bf C54} (2008) 451--460,
  [\href{http://xxx.lanl.gov/abs/0711.0625}{{\tt arXiv:0711.0625}}].

\bibitem{Dittmaier:2009cr}
S.~Dittmaier and M.~Huber, {\it {Radiative corrections to the neutral-current
  Drell-Yan process in the Standard Model and its minimal supersymmetric
  extension}},  {\em JHEP} {\bf 1001} (2010) 060,
  [\href{http://xxx.lanl.gov/abs/0911.2329}{{\tt arXiv:0911.2329}}].

\bibitem{Denner:2000jv}
A.~Denner and S.~Pozzorini, {\it {One loop leading logarithms in electroweak
  radiative corrections. 1. Results}},  {\em Eur.Phys.J.} {\bf C18} (2001)
  461--480, [\href{http://xxx.lanl.gov/abs/hep-ph/0010201}{{\tt
  hep-ph/0010201}}].

\bibitem{Kuhn:2001hz}
J.~H. Kuhn, S.~Moch, A.~Penin, and V.~A. Smirnov, {\it {Next-to-next-to-leading
  logarithms in four fermion electroweak processes at high-energy}},  {\em
  Nucl.Phys.} {\bf B616} (2001) 286--306,
  [\href{http://xxx.lanl.gov/abs/hep-ph/0106298}{{\tt hep-ph/0106298}}].

\bibitem{Campbell:2013qaa}
J.~Campbell, K.~Hatakeyama, J.~Huston, F.~Petriello, J.~R. Andersen, et~al.,
  {\it {Report of the Snowmass 2013 energy frontier QCD working group}},
  \href{http://xxx.lanl.gov/abs/1310.5189}{{\tt arXiv:1310.5189}}.

\bibitem{Li:2012wna}
Y.~Li and F.~Petriello, {\it {Combining QCD and electroweak corrections to
  dilepton production in FEWZ}},  {\em Phys.Rev.} {\bf D86} (2012) 094034,
  [\href{http://xxx.lanl.gov/abs/1208.5967}{{\tt arXiv:1208.5967}}].

\bibitem{Dittmaier:2014qza}
S.~Dittmaier, A.~Huss, and C.~Schwinn, {\it {Mixed QCD-electroweak
  $O(\alpha_s\alpha)$ corrections to Drell-Yan processes in the resonance
  region: pole approximation and non-factorizable corrections}},
  \href{http://xxx.lanl.gov/abs/1403.3216}{{\tt arXiv:1403.3216}}.

\bibitem{Nason:2004rx}
P.~Nason, {\it {A New method for combining NLO QCD with shower Monte Carlo
  algorithms}},  {\em JHEP} {\bf 0411} (2004) 040,
  [\href{http://xxx.lanl.gov/abs/hep-ph/0409146}{{\tt hep-ph/0409146}}].

\bibitem{Alioli:2010xd}
S.~Alioli, P.~Nason, C.~Oleari, and E.~Re, {\it {A general framework for
  implementing NLO calculations in shower Monte Carlo programs: the POWHEG
  BOX}},  {\em JHEP} {\bf 1006} (2010) 043,
  [\href{http://xxx.lanl.gov/abs/1002.2581}{{\tt arXiv:1002.2581}}].

\bibitem{Frixione:2002ik}
S.~Frixione and B.~R. Webber, {\it {Matching NLO QCD computations and parton
  shower simulations}},  {\em JHEP} {\bf 0206} (2002) 029,
  [\href{http://xxx.lanl.gov/abs/hep-ph/0204244}{{\tt hep-ph/0204244}}].

\bibitem{Bernaciak:2012hj}
C.~Bernaciak and D.~Wackeroth, {\it {Combining NLO QCD and Electroweak
  Radiative Corrections to W boson Production at Hadron Colliders in the POWHEG
  Framework}},  {\em Phys.Rev.} {\bf D85} (2012) 093003,
  [\href{http://xxx.lanl.gov/abs/1201.4804}{{\tt arXiv:1201.4804}}].

\bibitem{Barze:2012tt}
L.~Barze, G.~Montagna, P.~Nason, O.~Nicrosini, and F.~Piccinini, {\it
  {Implementation of electroweak corrections in the POWHEG BOX: single W
  production}},  {\em JHEP} {\bf 1204} (2012) 037,
  [\href{http://xxx.lanl.gov/abs/1202.0465}{{\tt arXiv:1202.0465}}].

\bibitem{Barze':2013yca}
L.~Barze, G.~Montagna, P.~Nason, O.~Nicrosini, F.~Piccinini, et~al., {\it
  {Neutral current Drell-Yan with combined QCD and electroweak corrections in
  the POWHEG BOX}},  {\em Eur.Phys.J.} {\bf C73} (2013) 2474,
  [\href{http://xxx.lanl.gov/abs/1302.4606}{{\tt arXiv:1302.4606}}].

\bibitem{Alioli:2010qp}
S.~Alioli, P.~Nason, C.~Oleari, and E.~Re, {\it {Vector boson plus one jet
  production in POWHEG}},  {\em JHEP} {\bf 1101} (2011) 095,
  [\href{http://xxx.lanl.gov/abs/1009.5594}{{\tt arXiv:1009.5594}}].

\bibitem{Hamilton:2012rf}
K.~Hamilton, P.~Nason, C.~Oleari, and G.~Zanderighi, {\it {Merging H/W/Z + 0
  and 1 jet at NLO with no merging scale: a path to parton shower + NNLO
  matching}},  {\em JHEP} {\bf 1305} (2013) 082,
  [\href{http://xxx.lanl.gov/abs/1212.4504}{{\tt arXiv:1212.4504}}].

\bibitem{Nason:2012pr}
P.~Nason and B.~Webber, {\it {Next-to-Leading-Order Event Generators}},  {\em
  Ann.Rev.Nucl.Part.Sci.} {\bf 62} (2012) 187--213,
  [\href{http://xxx.lanl.gov/abs/1202.1251}{{\tt arXiv:1202.1251}}].

\bibitem{Bozzi:2010xn}
G.~Bozzi, S.~Catani, G.~Ferrera, D.~de~Florian, and M.~Grazzini, {\it
  {Production of Drell-Yan lepton pairs in hadron collisions:
  Transverse-momentum resummation at next-to-next-to-leading logarithmic
  accuracy}},  {\em Phys.Lett.} {\bf B696} (2011) 207--213,
  [\href{http://xxx.lanl.gov/abs/1007.2351}{{\tt arXiv:1007.2351}}].

\bibitem{Banfi:2011dx}
A.~Banfi, M.~Dasgupta, and S.~Marzani, {\it {QCD predictions for new variables
  to study dilepton transverse momenta at hadron colliders}},  {\em Phys.Lett.}
  {\bf B701} (2011) 75--81, [\href{http://xxx.lanl.gov/abs/1102.3594}{{\tt
  arXiv:1102.3594}}].

\bibitem{Banfi:2012jm}
A.~Banfi, P.~F. Monni, G.~P. Salam, and G.~Zanderighi, {\it {Higgs and Z-boson
  production with a jet veto}},  {\em Phys.Rev.Lett.} {\bf 109} (2012) 202001,
  [\href{http://xxx.lanl.gov/abs/1206.4998}{{\tt arXiv:1206.4998}}].

\bibitem{Hoeche:2014aia}
S.~Hoeche, Y.~Li, and S.~Prestel, {\it {Drell-Yan lepton pair production at
  NNLO QCD with parton showers}},
  \href{http://xxx.lanl.gov/abs/1405.3607}{{\tt arXiv:1405.3607}}.

\bibitem{Alioli:2013hqa}
S.~Alioli, C.~W. Bauer, C.~Berggren, F.~J. Tackmann, J.~R. Walsh, et~al., {\it
  {Matching Fully Differential NNLO Calculations and Parton Showers}},
  \href{http://xxx.lanl.gov/abs/1311.0286}{{\tt arXiv:1311.0286}}.

\bibitem{Hamilton:2013fea}
K.~Hamilton, P.~Nason, E.~Re, and G.~Zanderighi, {\it {NNLOPS simulation of
  Higgs boson production}},  {\em JHEP} {\bf 1310} (2013) 222,
  [\href{http://xxx.lanl.gov/abs/1309.0017}{{\tt arXiv:1309.0017}}].

\bibitem{Frixione:2007vw}
S.~Frixione, P.~Nason, and C.~Oleari, {\it {Matching NLO QCD computations with
  Parton Shower simulations: the POWHEG method}},  {\em JHEP} {\bf 0711} (2007)
  070, [\href{http://xxx.lanl.gov/abs/0709.2092}{{\tt arXiv:0709.2092}}].

\bibitem{Banfi:2012yh}
A.~Banfi, G.~P. Salam, and G.~Zanderighi, {\it {NLL+NNLO predictions for
  jet-veto efficiencies in Higgs-boson and Drell-Yan production}},  {\em JHEP}
  {\bf 1206} (2012) 159, [\href{http://xxx.lanl.gov/abs/1203.5773}{{\tt
  arXiv:1203.5773}}].

\bibitem{Catani:1993hr}
S.~Catani, Y.~L. Dokshitzer, M.~Seymour, and B.~Webber, {\it {Longitudinally
  invariant $K_t$ clustering algorithms for hadron hadron collisions}},  {\em
  Nucl.Phys.} {\bf B406} (1993) 187--224.

\bibitem{Martin:2009iq}
A.~Martin, W.~Stirling, R.~Thorne, and G.~Watt, {\it {Parton distributions for
  the LHC}},  {\em Eur.Phys.J.} {\bf C63} (2009) 189--285,
  [\href{http://xxx.lanl.gov/abs/0901.0002}{{\tt arXiv:0901.0002}}].

\bibitem{Cacciari:2005hq}
M.~Cacciari and G.~P. Salam, {\it {Dispelling the $N^{3}$ myth for the $k_t$
  jet-finder}},  {\em Phys.Lett.} {\bf B641} (2006) 57--61,
  [\href{http://xxx.lanl.gov/abs/hep-ph/0512210}{{\tt hep-ph/0512210}}].

\bibitem{Cacciari:2011ma}
M.~Cacciari, G.~P. Salam, and G.~Soyez, {\it {FastJet User Manual}},  {\em
  Eur.Phys.J.} {\bf C72} (2012) 1896,
  [\href{http://xxx.lanl.gov/abs/1111.6097}{{\tt arXiv:1111.6097}}].

\bibitem{Ellis:1993tq}
S.~D. Ellis and D.~E. Soper, {\it {Successive combination jet algorithm for
  hadron collisions}},  {\em Phys.Rev.} {\bf D48} (1993) 3160--3166,
  [\href{http://xxx.lanl.gov/abs/hep-ph/9305266}{{\tt hep-ph/9305266}}].

\bibitem{Sjostrand:2007gs}
T.~Sjostrand, S.~Mrenna, and P.~Z. Skands, {\it {A Brief Introduction to PYTHIA
  8.1}},  {\em Comput.Phys.Commun.} {\bf 178} (2008) 852--867,
  [\href{http://xxx.lanl.gov/abs/0710.3820}{{\tt arXiv:0710.3820}}].

\bibitem{Skands:2014pea}
P.~Skands, S.~Carrazza, and J.~Rojo, {\it {Tuning PYTHIA 8.1: the Monash 2013
  Tune}},  \href{http://xxx.lanl.gov/abs/1404.5630}{{\tt arXiv:1404.5630}}.

\bibitem{Nason:2013uba}
P.~Nason and C.~Oleari, {\it {Generation cuts and Born suppression in POWHEG}},
   \href{http://xxx.lanl.gov/abs/1303.3922}{{\tt arXiv:1303.3922}}.

\bibitem{Becher:2010tm}
T.~Becher and M.~Neubert, {\it {Drell-Yan production at small $q_T$, transverse
  parton distributions and the collinear anomaly}},  {\em Eur.Phys.J.} {\bf
  C71} (2011) 1665, [\href{http://xxx.lanl.gov/abs/1007.4005}{{\tt
  arXiv:1007.4005}}].

\bibitem{Banfi:2012du}
A.~Banfi, M.~Dasgupta, S.~Marzani, and L.~Tomlinson, {\it {Predictions for
  Drell-Yan $\phi^*$ and $Q_T$ observables at the LHC}},  {\em Phys.Lett.} {\bf
  B715} (2012) 152--156, [\href{http://xxx.lanl.gov/abs/1205.4760}{{\tt
  arXiv:1205.4760}}].

\bibitem{Cacciari:2008gp}
M.~Cacciari, G.~P. Salam, and G.~Soyez, {\it {The Anti-k(t) jet clustering
  algorithm}},  {\em JHEP} {\bf 0804} (2008) 063,
  [\href{http://xxx.lanl.gov/abs/0802.1189}{{\tt arXiv:0802.1189}}].

\bibitem{Banfi:2010cf}
A.~Banfi, S.~Redford, M.~Vesterinen, P.~Waller, and T.~Wyatt, {\it
  {Optimisation of variables for studying dilepton transverse momentum
  distributions at hadron colliders}},  {\em Eur.Phys.J.} {\bf C71} (2011)
  1600, [\href{http://xxx.lanl.gov/abs/1009.1580}{{\tt arXiv:1009.1580}}].

\bibitem{Aad:2011dm}
{\bf ATLAS Collaboration} Collaboration, G.~Aad et~al., {\it {Measurement of
  the inclusive $W^\pm$ and Z/gamma cross sections in the electron and muon
  decay channels in $pp$ collisions at $\sqrt{s}=7$ TeV with the ATLAS
  detector}},  {\em Phys.Rev.} {\bf D85} (2012) 072004,
  [\href{http://xxx.lanl.gov/abs/1109.5141}{{\tt arXiv:1109.5141}}].

\bibitem{Aad:2011gj}
{\bf ATLAS Collaboration} Collaboration, G.~Aad et~al., {\it {Measurement of
  the transverse momentum distribution of Z/gamma* bosons in proton-proton
  collisions at $\sqrt{s}=7$ TeV with the ATLAS detector}},  {\em Phys.Lett.}
  {\bf B705} (2011) 415--434, [\href{http://xxx.lanl.gov/abs/1107.2381}{{\tt
  arXiv:1107.2381}}].

\bibitem{Aad:2012wfa}
{\bf ATLAS Collaboration} Collaboration, G.~Aad et~al., {\it {Measurement of
  angular correlations in Drell-Yan lepton pairs to probe Z/gamma* boson
  transverse momentum at sqrt(s)=7 TeV with the ATLAS detector}},  {\em
  Phys.Lett.} {\bf B720} (2013) 32--51,
  [\href{http://xxx.lanl.gov/abs/1211.6899}{{\tt arXiv:1211.6899}}].

\bibitem{Aad:2014xaa}
{\bf ATLAS Collaboration} Collaboration, G.~Aad et~al., {\it {Measurement of
  the $Z/\gamma^*$ boson transverse momentum distribution in $pp$ collisions at
  $\sqrt{s}$ = 7 TeV with the ATLAS detector}},
  \href{http://xxx.lanl.gov/abs/1406.3660}{{\tt arXiv:1406.3660}}.

\bibitem{Balazs:1997xd}
C.~Balazs and C.~Yuan, {\it {Soft gluon effects on lepton pairs at hadron
  colliders}},  {\em Phys.Rev.} {\bf D56} (1997) 5558--5583,
  [\href{http://xxx.lanl.gov/abs/hep-ph/9704258}{{\tt hep-ph/9704258}}].

\bibitem{Aad:2011fp}
{\bf ATLAS Collaboration} Collaboration, G.~Aad et~al., {\it {Measurement of
  the Transverse Momentum Distribution of $W$ Bosons in $pp$ Collisions at
  $\sqrt{s}=7$ TeV with the ATLAS Detector}},  {\em Phys.Rev.} {\bf D85} (2012)
  012005, [\href{http://xxx.lanl.gov/abs/1108.6308}{{\tt arXiv:1108.6308}}].

\bibitem{Aad:2013ueu}
{\bf ATLAS Collaboration} Collaboration, G.~Aad et~al., {\it {Measurement of kT
  splitting scales in W->lv events at sqrt(s)=7 TeV with the ATLAS detector}},
  {\em Eur.Phys.J.} {\bf C73} (2013) 2432,
  [\href{http://xxx.lanl.gov/abs/1302.1415}{{\tt arXiv:1302.1415}}].

\bibitem{ATLAS:2012au}
{\bf ATLAS Collaboration} Collaboration, G.~Aad et~al., {\it {Measurement of
  the polarisation of $W$ bosons produced with large transverse momentum in
  $pp$ collisions at $\sqrt{s}=7$ TeV with the ATLAS experiment}},  {\em
  Eur.Phys.J.} {\bf C72} (2012) 2001,
  [\href{http://xxx.lanl.gov/abs/1203.2165}{{\tt arXiv:1203.2165}}].

\bibitem{Chatrchyan:2011ig}
{\bf CMS Collaboration} Collaboration, S.~Chatrchyan et~al., {\it {Measurement
  of the Polarization of W Bosons with Large Transverse Momenta in W+Jets
  Events at the LHC}},  {\em Phys.Rev.Lett.} {\bf 107} (2011) 021802,
  [\href{http://xxx.lanl.gov/abs/1104.3829}{{\tt arXiv:1104.3829}}].

\bibitem{Bern:2011ie}
Z.~Bern, G.~Diana, L.~Dixon, F.~Febres~Cordero, D.~Forde, et~al., {\it
  {Left-Handed W Bosons at the LHC}},  {\em Phys.Rev.} {\bf D84} (2011) 034008,
  [\href{http://xxx.lanl.gov/abs/1103.5445}{{\tt arXiv:1103.5445}}].

\bibitem{Collins:1977iv}
J.~C. Collins and D.~E. Soper, {\it {Angular Distribution of Dileptons in
  High-Energy Hadron Collisions}},  {\em Phys.Rev.} {\bf D16} (1977) 2219.

\bibitem{Lam:1980uc}
C.~Lam and W.-K. Tung, {\it {A Parton Model Relation Sans {QCD} Modifications
  in Lepton Pair Productions}},  {\em Phys.Rev.} {\bf D21} (1980) 2712.

\bibitem{Hagiwara:1984hi}
K.~Hagiwara, K.-i. Hikasa, and N.~Kai, {\it {Parity Odd Asymmetries in $W$ Jet
  Events at Hadron Colliders}},  {\em Phys.Rev.Lett.} {\bf 52} (1984) 1076.

\bibitem{Mirkes:1992hu}
E.~Mirkes, {\it {Angular decay distribution of leptons from W bosons at NLO in
  hadronic collisions}},  {\em Nucl.Phys.} {\bf B387} (1992) 3--85.

\bibitem{Mirkes:1994eb}
E.~Mirkes and J.~Ohnemus, {\it {$W$ and $Z$ polarization effects in hadronic
  collisions}},  {\em Phys.Rev.} {\bf D50} (1994) 5692--5703,
  [\href{http://xxx.lanl.gov/abs/hep-ph/9406381}{{\tt hep-ph/9406381}}].

\bibitem{Mirkes:1994dp}
E.~Mirkes and J.~Ohnemus, {\it {Angular distributions of Drell-Yan lepton pairs
  at the Tevatron: Order $\alpha-s^{2}$ corrections and Monte Carlo studies}},
  {\em Phys.Rev.} {\bf D51} (1995) 4891--4904,
  [\href{http://xxx.lanl.gov/abs/hep-ph/9412289}{{\tt hep-ph/9412289}}].

\bibitem{Hagiwara:2006qe}
K.~Hagiwara, K.-i. Hikasa, and H.~Yokoya, {\it {Parity-Odd Asymmetries in $W^-$
  Jet Events at the Tevatron}},  {\em Phys.Rev.Lett.} {\bf 97} (2006) 221802,
  [\href{http://xxx.lanl.gov/abs/hep-ph/0604208}{{\tt hep-ph/0604208}}].

\bibitem{Lam:1978zr}
C.~Lam and W.-K. Tung, {\it {Structure Function Relations at Large Transverse
  Momenta in Lepton Pair Production Processes}},  {\em Phys.Lett.} {\bf B80}
  (1979) 228.

\bibitem{Aaltonen:2011nr}
{\bf CDF Collaboration} Collaboration, T.~Aaltonen et~al., {\it {First
  Measurement of the Angular Coefficients of Drell-Yan $e^{+}e^{-}$ pairs in
  the Z Mass Region from $p\bar{p}$ Collisions at $\sqrt{s}$ = 1.96 TeV}},
  {\em Phys.Rev.Lett.} {\bf 106} (2011) 241801,
  [\href{http://xxx.lanl.gov/abs/1103.5699}{{\tt arXiv:1103.5699}}].

\bibitem{Stirling:2012zt}
W.~Stirling and E.~Vryonidou, {\it {Electroweak gauge boson polarisation at the
  LHC}},  {\em JHEP} {\bf 1207} (2012) 124,
  [\href{http://xxx.lanl.gov/abs/1204.6427}{{\tt arXiv:1204.6427}}].

\end{thebibliography}\endgroup

\end{document}